\documentclass[a4paper,10pt]{article}
\usepackage[dvipsnames]{xcolor}
\usepackage{jheppub}
\usepackage{epsfig,amsmath,amssymb,slashed}
\usepackage{graphicx}
\usepackage{color}
\usepackage{bm}
\usepackage[dvipsnames]{xcolor}
\usepackage{tikz}
\usetikzlibrary{arrows.meta,decorations.markings,decorations.pathmorphing}
\usepackage{dirtytalk}

\usepackage{newunicodechar}
\newunicodechar{ζ}{\zeta}
\newunicodechar{η}{\eta}
\usepackage[utf8]{inputenc}
\usepackage{booktabs}

\def\wT{{\widehat T}}
\def\wj{{\widehat j}}

\def\wP{{\widehat P}}
\def\wA{{\widehat A}}
\def\wB{{\widehat B}}
\def\wQ{{\widehat Q}}

\def\wpsi{{\widehat{\psi}}}
 
\def\wrho{{\widehat{\rho}}}

\def\wpi{{\widehat\pi}}

\def\wA{\widehat A}

\def\wB{\widehat B}

\def\wW{\widehat W}

\def\wO{\widehat O}
\def\wQ{\widehat Q}

\def\wphi{\widehat \phi}
\def\wpsi{\widehat \psi}

\def\wpi{\widehat \pi}

\def\wpsi{\widehat \psi}

\def\mm{\mathfrak{m}}
\def\nn{\mathfrak{n}}

\def\rr{\mathfrak{r}}
\def\wvrphi{\widehat{\varphi}}

\makeatletter
\newcommand{\compactdouble}[1]{%
  \mathord{\mathpalette\compactdouble@{#1}}%
}
\newcommand{\compactdouble@}[2]{%
  \smash{%
    \widehat{%
      \smash{\overline{#2}}%
      \vphantom{\raisebox{.4ex}{$\m@th#1#2$}}%
    }%
  }%
  \vphantom{\widehat{#2}}%
}
\makeatother

\newcommand{\compactpsi}{\compactdouble{\psi}}
\newcommand{\compactPhi}{\compactdouble{\phi}}

\newcommand{\compactphi}{%
  \mathord{\smash{\widehat{\overline{\varphi}}}%
  \vphantom{\widehat{\varphi}}}%
}
\newcommand{\compactV}{%
  \mathord{\smash{\widehat{\overline{V}}}%
  \vphantom{\widehat{V}}}%
}
\newcommand{\wGamma}{\widehat{\Gamma}}

\newcommand{\tr}{{\rm tr}}  
\newcommand{\Tr}{{\rm Tr}}  
\newcommand{\e}{{\rm e}}

\newcommand{\di}{{\rm d}}
\newcommand{\ii}{{\rm i}}

\newcommand{\be}{\begin{equation}}
\newcommand{\ee}{\end{equation}}   

\newcommand{\wbmV}{\widehat{\boldsymbol{\mathcal{V}}}}
\newcommand{\bZZ}{\boldsymbol{\mathfrak{Z}}}

\newcommand{\ZZ}{\mathfrak{Z}}

\newcommand{\mrr}{\mathfrak{r}}
\newcommand{\mss}{\mathfrak{s}}

\def\checkmark{\tikz\fill[scale=0.4](0,.35) -- (.25,0) -- (1,.7) -- (.25,.15) -- cycle;}

\begin{document}


\title{Quadratic response of the  Wigner function at local thermodynamic equilibrium:\\ spin polarization and pion spectra}  

\author[a]{Andrea Olzi}
\author[b]{\,\,Daniele Roselli}
\affiliation[a]{Universit\'e Paris Cit\'e, CNRS, Astroparticule et Cosmologie, F-75006 Paris, France}
\affiliation[b]{School of Science and Engineering, The Chinese University of Hong Kong (Shenzhen),\\ Longgang, Shenzhen, Guangdong, 518172, China}

\emailAdd{aolzi@apc.in2p3.fr, danieleroselli@cuhk.edu.cn}

\abstract{We derive a quadratic-response expansion of the covariant Wigner function for free relativistic fields of arbitrary spin and internal symmetry at local thermodynamic equilibrium, retaining the geometry of the decoupling hypersurface. Combining quadratic response with second-order linear response, we find that the local second-order gradient correction to the spin-polarization vector of massive Dirac fermions vanishes in the Belinfante and canonical pseudo-gauges, independently of hypersurface orientation and curvature. We trace this cancellation to a microscopic gradient-parity selection rule applicable to a broader class of pseudo-gauges. For the pion isospin triplet, we obtain non-vanishing second-order Wigner corrections, including contributions from transverse non-Abelian isospin gradients. For a fixed isospin direction, hypersurface integration yields a compact correction to the pion momentum spectra involving products of first thermodynamic gradients but no explicit extrinsic-curvature terms. We also derive corrections to the charged-pion asymmetry and neutral-to-charged imbalance.}


\maketitle
\flushbottom

\section{Introduction}
The quark gluon plasma (QGP) produced in heavy ion collisions is the natural environment in which to study strongly interacting matter under extreme rotation and acceleration \cite{Shuryak:2003xe}. One of the most striking features of the QGP is that it behaves like an almost perfect fluid \cite{Gustafsson:1984ka, Gutbrod:1988hh, Rischke:1995ir, Rischke:1995mt, Luzum:2008cw, Song:2010mg, Heinz:2013th, Gale:2013da, McDonald:2016vlt, Jaiswal:2016hex, Romatschke:2017ejr, Jaiswal:2020hvk} and, as such, it is assumed to reach local thermodynamic equilibrium at some time after its production. After thermalization the QGP evolves according to hydrodynamic equations up to the decoupling, which is the moment in space and time where the quark and the gluons recombine together to produce bound strong states, i.e. they hadronize. The produced weakly interacting hadrons subsequently evolve driving the system out of equilibrium until they are eventually observed as asymptotic free states in the infinite future.

The properties of the QGP can be inferred by measuring various observables related with the hadrons produced at the decoupling such as, for instance, the energy distribution in momentum space or the spin polarization of strange hadrons \cite{ALICE:2019onw, ALICE:2020iev, ALICE:2022dyy, Alzhrani:2022dpi, STAR:2017ckg, STAR:2018gyt, STAR:2019erd, STAR:2020xbm, STAR:2021beb, STAR:2022fan, CMS:2025nqr}. In particular, in the past twenty years, spin polarization of strange hadrons \cite{Liang:2004ph, Becattini:2007nd, Becattini:2013fla}, and spin alignment of vector mesons \cite{Sheng:2022wsy,Dong:2023cng,Chen:2024afy,Zhang:2024mhs, Yang:2024fkn}, have been considered as powerful probes of the QGP being extremely sensitive to the gradients of the thermodynamic fields \cite{Becattini:2021iol, Becattini:2021suc, Fukushima:2020ucl, Fu:2021pok, Liu:2021uhn, Weickgenannt:2020aaf, Liu:2020dxg, Fu:2022myl, Buzzegoli:2021wlg}, rotation, acceleration and also bulk viscosity \cite{Palermo:2024tza, Singh:2026iew} present in the plasma.

The fundamental theoretical tool to bridge the macroscopic observables with the microscopic hadronic degrees of freedom at the decoupling is the covariant Wigner function \cite{Becattini:2020sww, Li:2025pef}. In particular, the spectra, spin density matrix and the expectation values of local operators such as the stress-energy tensor can be expressed in terms of opportune integrals of it. Assuming that hadrons at decoupling are in local thermodynamic equilibrium the standard Cooper–Frye formula \cite{Cooper:1974mv} is recovered at leading order, while gradients of the thermodynamic fields generate corrections to the Wigner function and, consequently, to the corresponding observables.

These corrections can be derived systematically within relativistic quantum statistical mechanics, starting from the Zubarev non-equilibrium statistical operator \cite{Zubarev:1979afm, Hosoya:1983id, Becattini:2019dxoZUB, Liu:2021nyg}. In this framework, the local-equilibrium expectation values are given in terms of quantum-field operators, and their gradient expansion is determined by microscopic thermal correlation functions leading to a field-theoretical derivation of the gradient corrections without having to rely on a phenomenological ansatz for the particle distribution function.

Recently \cite{Sheng:2025cjk} a new expansion scheme was developed to obtain the gradient expansion of the Wigner function for Dirac fermions at any order in linear response theory. The formula was later applied to the case of Rarita-Schwinger field at finite spin density \cite{Buzzegoli:2026sge}. The new expansion scheme allows for an exact integration, order by order in the gradient expansion, on the decoupling hypersurface allowing to compute the corrections without having to resort to any geometrical approximation for the shape of the hypersurface which is generally expected to be non-flat and to even possess time-like branches.
In this approach the Wigner function of the hadrons, and the physical quantities derived from it, receive corrections depending on the gradients of the thermodynamic fields, on the curvature of the decoupling hypersurface and combinations of the two. The two contributions are related with two independent  parameters with dimension of a length: $l_G\equiv|\beta/\partial\beta|$, which is the typical length over which the thermodynamic fields vary, and $l_C\equiv|\hat{n}/\partial \hat{n}|$, which measures the typical scale of  the curvature of the hypersurface, with $\hat{n}$ being its normal vector. Both lengths are assumed to be sufficiently large compared with microscopic length scales for a gradient expansion to apply. 

From numerical simulations
some components of the gradient of the thermal vorticity  can be comparable with the thermal vorticity itself \cite{Karpenko:2016jyx} whereas the estimation of $l_C$ is not easy and in general one expects the decoupling curvature to be different from a simple flat hyperplane. In particular curvature of the decoupling surface can couple with gradients of the thermodynamic fields leading to the appearance of otherwise absent contributions, for example shear induced spin alignment \cite{Zhang:2025vlk}.

In this work we extend the computation performed in \cite{Sheng:2025cjk} obtaining the general quadratic response expansion for the Wigner function of free fields \emph{of any spin} and general decoupling hypersurface valid in \emph{any pseudo-gauge}, which is explicitly given in equation \eqref{eq: QUAD FINAL}. The employed formalism is not restricted to Abelian charges only. Hadrons can be organized in multiplets
of the isospin group, and, more generally, the fields can belong to any representation of a non-Abelian internal
symmetry group. The local equilibrium state can be extended to contain a chemical potential which is a vector in the internal
space, whose direction can, in principle, also change from point to point. The resulting final
formulae then apply to multiplets of any spin and any representation.

We use the newly obtained formula to compute the spin polarization vector
of the $\Lambda$ hyperons and the local equilibrium corrections to the Cooper-Frye formula for the pion spectra up to the
second order in the gradients of the thermodynamic fields and curvature of the decoupling hypersurface. For the
former we obtain that the second order gradient correction to the spin polarization vector \emph{exactly vanishes} in a vast
class of pseudo-gauges once the decoupling hypersurface is properly retained. For the latter we find that the first
order correction vanishes and we obtain the second order correction in closed form.

The paper is structured as follows: In section \ref{SEC: LTE} we introduce the local equilibrium statistical operator and the quadratic response expansion for a generic observable. In section \ref{SEC: Wigner function} we introduce the covariant Wigner operator for free hadronic fields  of generic spin. In section \ref{SEC: linear and quadratic response} we compute the linear \eqref{eq: linwigFIN} and quadratic \eqref{eq: QUAD FINAL} response correction to the local equilibrium Wigner function. In section \ref{SEC: Dirac} we apply the newly obtained formula to compute the spin polarization vector up to the second order in the gradients. We extend the formula obtained in \cite{Sheng:2025cjk} to the case of canonical pseudo-gauge \eqref{eq: S canonical} and we obtain that the total correction at order $\mathcal{O}(\partial^2)$ is \emph{exactly vanishing} for a vast class of pseudo-gauges. In section \ref{SEC: pions} we compute the quadratic local equilibrium correction to the spectra of the pions given in \eqref{eq: pion spectrum final}. We finally collect our results and our concluding remarks in section \ref{SEC: conclusions}. The technical details regarding the computations are collected in the appendices.

\subsection*{Notations}
Throughout this paper we use the mostly minus signature convention for the flat metric $g^{\mu\nu}=\mbox{diag}\left(+,-,-,-\right)$. 
We use natural units, $\hbar=k_B=c=1$. The spatial part of a four-vector $k^\mu$ is denoted in boldface, ${\bf k}$, 
and a dot denotes both the four-vector scalar product, $k\cdot p=k_\mu p^\mu$, and the three-vector scalar product, ${\bf k}\cdot{\bf p}=k_jp^j$. 
We adopt the Einstein summation convention, with Greek indices running over $0,1,2,3$ and Latin indices over $1,2,3$. 
The Heaviside $\theta$ function is defined as: $\theta(x) = 1$ if $x > 0$ and $0$ if $ x \le 0$. Isospin indices for the representation are denoted with a mathfrak Latin letter such as $\mm$ or $\mrr$. For a field in the representation of total isospin ${\rm I}$ the index is $\mm=-{\rm I},-{\rm I}+1,\ldots,{\rm I}$. The convention for the total antisymmetric Levi-Civita symbol is $\epsilon^{0123}=+1$.

\section{Local thermodynamic equilibrium and response expansion}\label{SEC: LTE}
If the QGP achieves local thermodynamic equilibrium on an initial, space-like, hypersurface $\Sigma_0$, the quantum state is assumed to be given by the \emph{non-equilibrium} statistical operator $\wrho$, which, for a system described by a stress-energy tensor $\wT^{\mu\nu}$ a set of Abelian vector currents $\wj^\mu_\ell$, a spin tensor $\widehat{\mathcal{S}}^{\mu,\lambda\rho}$ and non-Abelian vector current $\wbmV=\widehat{\mathcal{V}}_\rr$, with $\rr$ the isospin index, reads \cite{Becattini:2019dxoZUB}:\footnote{Note that the generalization to several non-Abelian currents, associated with mutually commuting non-Abelian symmetry groups, is straightforward and amounts to dress both the current $\wbmV$ and the corresponding chemical potential $\bZZ$ with an index $l$.}
\begin{equation}\label{def: rho}
    \wrho\equiv\frac{1}{Z}\exp\left[-\int_{\Sigma_{0}}\di\Sigma_\mu(y)\left(\wT^{\mu\nu}(y)\beta_\nu(y)-\begin{matrix} \sum_{\ell}\end{matrix}\wj^\mu_\ell(y)\zeta_\ell(y)-\frac{1}{2}\mathfrak{S}_{\lambda\rho}(y)\widehat{\mathcal{S}}^{\mu,\lambda\rho}(y)-\wbmV(y)\cdot\bZZ(y)\right)\right]\;,
\end{equation}
where $\beta^\mu=u^\mu/T$ is the four-temperature, $\zeta_\ell=\mu_\ell/T$ are the reduced chemical potentials, $\mathfrak{S}_{\lambda\rho}$ is the reduced spin potential and $\ZZ_\rr=\mu_\rr/T$ is the reduced \textit{iso-vector} chemical potential. We refer to $\mathfrak{r}$ as the isospin index and to $\bZZ$ as the reduced iso-vector chemical potential because we have in mind as underlying non-Abelian symmetry the isospin one under which the hadronic states are organized in multiplets.

\begin{figure}[h]
    \centering
    \includegraphics[width=0.5\linewidth]{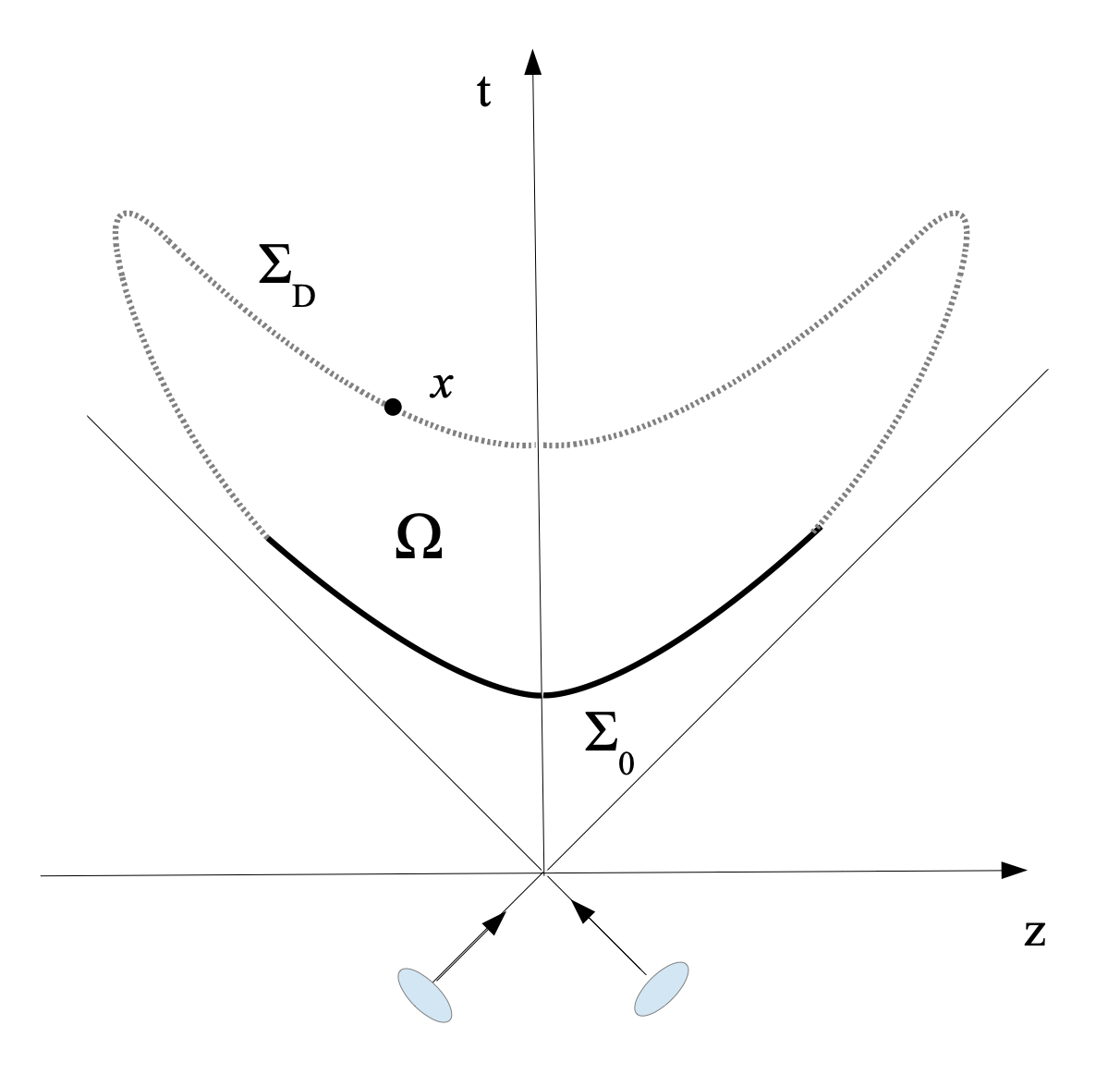}
    \caption{Schematic illustration of a nuclear collision at high energy. The initial local equilibrium hypersurface 
    is $\Sigma_0$ (solid line) and the decoupling hypersurface $\Sigma_{\rm D}$ (finely dotted line); the quark-gluon
    plasma as a fluid lives in the encompassed region $\Omega$. A point $x$ on $\Sigma_{\rm D}$ is the most suitable place 
    where an approximate expression of hadronic observables such as the spectrum or the spin polarization vector can be obtained from the Wigner operator. Figure from \cite{Becattini:2025rqp}. }
    \label{fig:Gauss Theorem Region}
\end{figure}

The non-equilibrium state \eqref{def: rho}, stationary in Heisenberg representation, can be expressed in terms of the decoupling hypersurface $\Sigma_{\rm D}$ by applying the Gauss theorem in the region $\Omega$ enclosed by $\Sigma_0$ and $\Sigma_{\rm D}$ (see figure \ref{fig:Gauss Theorem Region}) so that:

\begin{equation}\label{def: rho gauss}
    \begin{split}
        \wrho\equiv\frac{1}{Z}\exp\Bigg\{&-\int_{\Sigma_{\rm D}}\di\Sigma_\mu(y)\bigg(\wT^{\mu\nu}(y)\beta_\nu(y)-\begin{matrix}\sum_\ell\end{matrix}\wj^\mu_\ell(y)\zeta_\ell(y)-\frac{1}{2}\mathfrak{S}_{\lambda\rho}(y)\widehat{\mathcal{S}}^{\mu,\lambda\rho}(y)+\\
        &-\wbmV^\mu(y)\cdot\bZZ(y)\bigg)+\int_\Omega\di^4y\bigg[\wT^{\mu\nu}(y)\partial_\mu\beta_\nu(y)-\begin{matrix}\sum_\ell\end{matrix}\wj^\mu_\ell(y)\partial_\mu\zeta_\ell(y)+\\
        &-\frac{1}{2}\widehat{\mathcal{S}}^{\mu,\lambda\rho}(y)\partial_\mu\mathfrak{S}_{\lambda\rho}(y)++\frac{1}{2}\mathfrak{S}_{\lambda\rho}(y)\left(\wT^{\lambda\rho}(y)-\wT^{\rho\lambda}(y)\right)-\wbmV^\mu(y)\cdot\partial_\mu\ZZ(y)\bigg]\Bigg\}\;,
    \end{split}
\end{equation}
where we took advantage of the continuity equation of the densities:
\begin{equation*}
    \partial_\mu\wT^{\mu\nu}=0\;,\qquad\partial_\mu\wj_\ell^\mu=0\;,\qquad\partial_\mu\widehat{\mathcal{S}}^{\mu,\lambda\rho}=\widehat{T}^{\rho\lambda}-\widehat{T}^{\lambda\rho}\;,\qquad\partial_\mu\wbmV^\mu=0\;.
\end{equation*}
The first term in the exponent of \eqref{def: rho gauss} identifies the proper \emph{local equilibrium} term while the second term contains dissipative as well as non-local and memory effects.

In the following we concentrate only on the local equilibrium term:
\begin{equation}\label{def: rhoLTE}
    \wrho_{\rm LTE}\equiv\frac{1}{Z}\exp\left[-\int_{\Sigma_{\rm D}}\di\Sigma_\mu(y)\left(\wT^{\mu\nu}(y)\beta_\nu(y)-\begin{matrix}\sum_\ell  \end{matrix}\wj^\mu_\ell(y)\zeta_\ell(y)-\frac{1}{2}\mathfrak{S}_{\lambda\rho}(y)\widehat{\mathcal{S}}^{\mu,\lambda\rho}(y)-\wbmV^\mu(y)\cdot\bZZ(y)\right)\right]
\end{equation}
where, given that at thermodynamic equilibrium $\mathfrak{S}=\varpi\sim \partial\beta$, being $\varpi$ the \emph{thermal vorticity}, we considered the reduced spin potential of the same order of the variations. The dissipative correction given by the integral over the volume $\Omega$, is usually treated separately. Being dependent on the gradients of the thermodynamic fields, in hydrodynamic regime it is supposedly smaller than the local equilibrium part and has been often considered a correction. However we do not claim here that this is certainly the case. The description of dissipative corrections requires the inclusion of interactions leading to the appearance of unknown dynamical coefficients which in principle may be as important as the local equilibrium contribution (see \cite{Wang:2025mfz, Buzzegoli:2025zud, Becattini:2025rqp}) for recent discussions in the framework of heavy ion collisions and Wigner function approach). 

For a generic observable $\wO(x)$ we define the \emph{local equilibium} expectation value as:
\begin{equation}\label{def: LTE expval}
    \langle\wO\rangle_{\rm LTE}(x)\equiv\Tr\left[\wrho_{\rm LTE}\,\wO(x)\right]\;.
\end{equation}
For an observable defined in the space-time point $x$ the leading contribution to the expectation value \eqref{def: LTE expval} comes from the values of the thermodynamic fields around $y\simeq x$.
As such we introduce the variation of the thermodynamic fields:
\begin{equation}\label{def: DELTAs}
    \Delta\beta_\nu(y,x)\equiv\beta_\nu(y)-\beta_\nu(x)\;,\qquad\Delta\zeta_\ell(y,x)\equiv\zeta_\ell(y)-\zeta_\ell(x)\;,\qquad\Delta\bZZ(y,x)\equiv\bZZ(y)-\bZZ(x)\;.
\end{equation}
The above can in principle be expanded in Taylor series around $x=y$ so that \eqref{def: DELTAs} contain all the deviation from the values of the thermodynamic fields around $x=y$ induced by the corresponding gradients, where the reduced spin chemical potential $\mathfrak{S}$ itself is considered a deviation from the equilibrium state.
Plugging \eqref{def: DELTAs} in \eqref{def: rhoLTE} and using:
\begin{equation*}
    \int_{\Sigma}\di\Sigma_\mu(y)\wT^{\mu\nu}(y)=\wP^\nu\;,\qquad\int_{\Sigma}\di\Sigma_\mu(y)\wj^{\mu}_\ell(y)=\wQ_\ell\;,\qquad\int_{\Sigma}\di\Sigma_\mu(y)\wbmV^{\mu}(y)=\widehat{\boldsymbol{\mathcal{Q}}}\;,
\end{equation*}
the local equilibrium operator \eqref{def: rhoLTE} can be equivalently written as:
\begin{equation}\label{def: rho AB}
    \wrho_{\rm LTE}=\frac{1}{Z_{A+B}}\e^{\wA+\wB}\;,\qquad Z_{A+B}\equiv \Tr\left[\e^{\wA+\wB}\right]\;,
\end{equation}
where we defined:
\begin{subequations}\label{def: A e B}
    \begin{align}
        \wA&=-\beta(x)\cdot\wP+\sum_\ell\zeta_\ell(x)\wQ_\ell+\mathfrak{Z}(x)\widehat{\mathcal{Q}}_3\;,\label{def: A}\\
    \widehat{B}&=-\int_{\Sigma_{\rm D}}\di\Sigma_\mu(y)\left[\wT^{\mu\nu}(y)\Delta\beta_\nu(y,x)-\sum_\ell\wj^\mu_\ell(y)\Delta\zeta_\ell(y,x)-\frac{1}{2}\mathfrak{S}_{\lambda\rho}(y)\widehat{\mathcal{S}}^{\mu,\lambda\rho}(y)-\widehat{\boldsymbol{\mathcal{V}}}^\mu(y)\cdot\Delta\boldsymbol{\mathfrak{Z}}(y,x)\right].\label{def: B}
    \end{align}
\end{subequations}
Note that in \eqref{def: A} we performed a local rotation at $x$ in the internal space aligning the non-Abelian charge along the \emph{third} direction. However the same rotation, which is associated with a global symmetry group, cannot, in general, align the local current density at every point $y$ so that the term in \eqref{def: B} must retain all the components of the internal (isospin) product. We will further discuss this point in section \ref{SEC: pions}.

All the terms in \eqref{def: B} have the same structure i.e. a local operator convoluted on the decoupling hypersurface $\Sigma_{\rm D}$ with a corresponding thermodynamic field. It is convenient to write:
\begin{equation}\label{def: A e B collective}
    \widehat{B}=\sum_a\int_{\Sigma_{\rm D}}\di\Sigma_\mu(y)\,\vartheta_a(y,x)\,\widehat{\Gamma}^\mu_a(y)\;,
\end{equation}
where the collective index $a$ contains the kind of operator, its free Lorentz index (e.g.\ $\nu$ for $\wT^{\mu\nu}$) and its isospin component:
\begin{equation}\label{Table: B table}
    \begin{array}{c|cccc}
        \widehat{\Gamma}^\mu_a(y) & \wT^{\mu\nu}(y) & \wj^\mu_\ell(y) &\widehat{\mathcal{S}}^{\mu,\lambda\rho}(y) &\widehat{\mathcal{V}}^\mu_i(y) \\ \hline
        \vartheta_a(y,x) & -\Delta\beta_\nu(y,x) & \Delta\zeta_\ell(y,x)&\frac{1}{2}\mathfrak{S}_{\lambda\rho}(y) & \Delta\mathfrak{Z}_i(y,x) 
    \end{array}\;.
\end{equation}
For the considered operators $\wGamma_a(y)$ is local and it commutes with the conserved Abelian charges $\wQ_\ell$, then:
\begin{equation*}
    \e^{\ii\wP\cdot c}\,\widehat{\Gamma}_a(y)\,\e^{-\ii\wP\cdot c}=\widehat{\Gamma}_a(y+c)\;,\qquad\e^{\ii\theta\widehat{\mathcal{Q}}_3}\,\widehat{\Gamma}_a(y)\,\e^{-\ii\theta\widehat{\mathcal{Q}}_3}=\e^{-\ii\mathfrak{q}_a\theta}\,\widehat{\Gamma}_a(y)\;,\qquad[\wQ_\ell,\widehat{\Gamma}_a(y)]=0\;,
\end{equation*}
where $\mathfrak{q}_a$ is the isospin charge of the operator ($\mathfrak{q}_a=0$ for $\wT$, $\widehat{\mathcal{S}}$ and $\wj_\ell$). 

If the deviations from the local values \eqref{def: DELTAs} are small, namely if the thermodynamic fields are sufficiently slowly varying over the decoupling hypersurface, it is possible to expand the local equilibrium operator \eqref{def: rho AB} in powers of $\wB$. 

Introducing:
\begin{equation*}
    \widehat{B}(\lambda)\equiv\e^{\lambda\wA}\widehat{B}\,\e^{-\lambda\wA}\;,
\end{equation*}
the operator $\exp(\wA+\wB)$ can be expanded as:
\begin{equation*}
    \e^{\wA+\widehat{B}}=\left[\widehat{\mathbb{I}}+\int_0^1\di\lambda\,\widehat{B}(\lambda)+\int_0^1\di\lambda_1\int_0^{\lambda_1}\di\lambda_2\,\widehat{B}(\lambda_2)\widehat{B}(\lambda_1)+\mathcal{O}(\widehat{B}^3)\right]\e^{\wA}\;.
\end{equation*}
By taking the expectation value with $\wO(x)$ and using the cyclicity of the trace we get:
\begin{equation*}
    \begin{split}
        \frac{\Tr[\e^{\wA+\widehat{B}}\wO(x)]}{Z_A}&=\langle\wO(x)\rangle_{\rm GE}+\int_0^1\di\lambda\,\langle\wO(x)\,\widehat{B}(\lambda)\rangle_{\rm GE}+\int_0^1\di\lambda_1\int_0^{\lambda_1}\di\lambda_2\,\langle\wO(x)\,\widehat{B}(\lambda_2)\widehat{B}(\lambda_1)\rangle_{\rm GE}\;,\\
        \frac{\Tr[\e^{\wA+\widehat{B}}]}{Z_A}&=1+\langle\widehat{B}\rangle_{\rm GE}+\int_0^1\di\lambda_1\int_0^{\lambda_1}\di\lambda_2\,\langle\widehat{B}(\lambda_2)\widehat{B}(\lambda_1)\rangle_{\rm GE}\;,
    \end{split}
\end{equation*}
where the \emph{global homogeneous} equilibrium expectation value is defined as:
\begin{equation*}
    \langle\wO(x)\rangle_{\rm GE}\equiv\Tr\left(\wrho_{\rm GE}\,\wO(x)\right)=\frac{1}{Z_A}\Tr\left(\e^{\wA}\,\wO(x)\right)\;,
\end{equation*}
with $\wrho_{\rm GE}$ reference global equilibrium state in $x$:
\begin{equation}\label{eq: GE}
    \wrho_{\rm GE}=\frac{1}{Z_{\rm GE}}\exp\left[-\beta(x)\cdot\wP+\sum_\ell\zeta_\ell(x)\wQ_\ell+\mathfrak{Z}(x)\widehat{\mathcal{Q}}_3\right]\;.
\end{equation}
Note that, due to homogeneity of \eqref{eq: GE} $\langle\widehat{B}(\lambda)\rangle_{\rm GE}=\langle\widehat{B}\rangle_{\rm GE}$. Dividing by $Z_{A+B}$ and using $1/(1+\epsilon)=1-\epsilon+\epsilon^2+\mathcal{O}(\epsilon^3)$ the local equilibrium expectation value turns out to be:
\begin{equation}\label{eq: OLTE1}
    \begin{split}
        \langle\wO(x)\rangle_{\rm LTE}&=\langle\wO(x)\rangle_{\rm GE}+\int_0^1\di\lambda\left[\langle\wO(x)\,\widehat{B}(\lambda)\rangle_{\rm GE}-\langle\wO(x)\rangle_{\rm GE}\langle\widehat{B}\rangle_{\rm GE}\right]\\
        &+\int_0^1\di\lambda_1\int_0^{\lambda_1}\di\lambda_2\Big[\langle\wO(x)\,\widehat{B}(\lambda_2)\widehat{B}(\lambda_1)\rangle_{\rm GE}-\langle\wO(x)\rangle_{\rm GE}\langle\widehat{B}(\lambda_2)\widehat{B}(\lambda_1)\rangle_{\rm GE}\Big]\\
        &-\langle\widehat{B}\rangle_{\rm GE}\int_0^1\di\lambda\,\langle\wO(x)\,\widehat{B}(\lambda)\rangle_{\rm GE}+\langle\wO(x)\rangle_{\rm GE}\langle\widehat{B}\rangle^2_{\rm GE}+\mathcal{O}(\widehat{B}^3)\;.
    \end{split}
\end{equation}
The correction at each order in $\wB$ can be expressed in terms of the global equilibrium connected correlators:
\begin{subequations}
    \begin{align}
    \langle\wO,\widehat{X}\rangle_{c,\rm GE}&\equiv\langle\wO\widehat{X}\rangle_{\rm GE}-\langle\wO\rangle_{\rm GE}\langle\widehat{X}\rangle_{\rm GE}\;,\label{A: cumulant2}\\ \nonumber
    \langle\wO,\widehat{X},\widehat{Y}\rangle_{c,{\rm GE}}&\equiv\langle\wO\widehat{X}\widehat{Y}\rangle_{\rm GE}-\langle\wO\widehat{X}\rangle_{\rm GE}\langle\widehat{Y}\rangle_{\rm GE}-\langle\wO\widehat{Y}\rangle_{\rm GE}\langle\widehat{X}\rangle_{\rm GE}-\langle\wO\rangle_{\rm GE}\langle\widehat{X}\widehat{Y}\rangle_{\rm GE}\\
    &\quad\;+2\langle\wO\rangle_{\rm GE}\langle\widehat{X}\rangle_{\rm GE}\langle\widehat{Y}\rangle_{\rm GE}\;.\label{A: cumulant3}
\end{align}    
\end{subequations}
Setting $\widehat{X}=\widehat{B}(\lambda_2)$, $\widehat{Y}=\widehat{B}(\lambda_1)$ in \eqref{eq: OLTE1} and integrating over the triangle $T\equiv\{0\le\lambda_2\le\lambda_1\le1\}$, we finally obtain the local equilibrium expectation value of $\wO(x)$ at quadratic order in $\wB$:
\begin{equation}\label{eq: OLTE2}
\begin{split}
    \langle\wO(x)\rangle_{\rm LTE}=&\langle\wO(x)\rangle_{\rm GE}+\int_0^1\di\lambda\,\langle\wO(x),\widehat{B}(\lambda)\rangle_{c,\rm GE}\\ 
    &+\int_0^1\di\lambda_1\int_0^{\lambda_1}\di\lambda_2\,\langle\wO(x),\widehat{B}(\lambda_2),\widehat{B}(\lambda_1)\rangle_{c,\rm {GE}}+\mathcal{O}(\widehat{B}^3)\;.
    \end{split}
\end{equation}
Plugging \eqref{def: A e B collective} in equation \eqref{eq: OLTE2} and separating the linear and quadratic order in $\wB$ we thus define, respectively, the \emph{linear response}:
\begin{equation}\label{def: lin O}
    \Delta_{\rm lin}O(x)\equiv\sum_a\int_{\Sigma_{\rm D}}\di\Sigma_\mu(y)\,\vartheta_a(y,x)\int_0^1\di\lambda\,\langle\wO(x),\widehat{\Gamma}^\mu_a(y;\lambda)\rangle_{c,\rm GE}\;,
\end{equation}
and \emph{quadratic response} correction:
\begin{equation}\label{def: quad O}
\begin{split}
    \Delta_{\rm quad}O(x)\equiv&\sum_{a,b}\int_{\Sigma_{\rm D}}\di\Sigma_\mu(y_1)\vartheta_a(y_1,x)\int_{\Sigma_{\rm D}}\di\Sigma_\nu(y_2)\vartheta_b(y_2,x)\times\\
    &\times\int_0^1\!\!\!\di\lambda_1\int_0^{\lambda_1}\!\!\!\di\lambda_2\,\langle\wO(x),\widehat{\Gamma}^\nu_b(y_2;\lambda_2),\widehat{\Gamma}^\mu_a(y_1;\lambda_1)\rangle_{c,\rm GE}\;,
\end{split}
\end{equation}
where $\widehat{\Gamma}(y;\lambda)\equiv\e^{\lambda\wA}\widehat{\Gamma}(y)\e^{-\lambda\wA}$ and the double sum runs over all ordered pairs $(a,b)$, including $a=b$. 

Note that \eqref{def: lin O} and \eqref{def: quad O} are exact functional of the thermodynamic fields on $\Sigma_{\rm D}$: no gradient expansion has been performed so far, and each of them contains contributions of arbitrarily high order in the gradients of such thermodynamic fields.


\section{Covariant Wigner operator for free hadronic fields}\label{SEC: Wigner function}
The covariant Wigner operator is defined as the Fourier transform of the two point function with respect to the relative coordinate $s$:
\begin{equation}\label{def: Wig op}
    \wW^{\mm\nn}_{\rho\gamma}(x,k)\equiv\frac{1}{(2\pi)^4}\int\di^4s\,\e^{-\ii  s\cdot k}\,\compactphi^{\mm}_\rho\left(x+\frac{s}{2}\right)\wvrphi^\nn_\gamma\left(x-\frac{s}{2}\right)\;,
\end{equation}
where $\wvrphi_\gamma$ is generic iso-multiplet hadronic field, with spin/representation index $\gamma$ and isospin index $\nn={\rm I},{\rm I-1},\ldots,-{\rm I}$. $\compactphi^\mm_\rho$ denotes instead the complex adjoint of the field with $\gamma\to\rho$ and $\nn\to \mm$. The covariant Wigner operator is a ${\rm (2I+1)}\times{\rm (2I+1)}$ matrix in the isospin space transforming under the mixed representation of the fields.

The indexes $\gamma,\rho$ denote all the possible internal indexes pertaining the Lorentz representation of the field. For instance for scalar fields it is absent, for Dirac field is a spinorial index $\gamma\equiv A=1,2,3,4$ while for Proca fields is a proper four-vector index $\gamma=0,1,2,3$.

For scalar $\wvrphi_\gamma\equiv\wphi$ and Proca $\wvrphi_\gamma\equiv \widehat{V}_\gamma$ the covariant adjoint is simply the dagger of the field:
\begin{equation*}
    \compactPhi^\mm=\wphi^{\mm\dagger}\;,\qquad\compactV^\mm_\gamma=\widehat{V}^{\mm\dagger}_\gamma\;,
\end{equation*}
while for Dirac fermions $\wvrphi_\gamma\equiv\wpsi_A$ it corresponds to the Dirac adjoint:
\begin{equation*}
    \compactpsi^\mm_A=\wpsi^{\mm\dagger}_B(\gamma^0)_{AB}\;.
\end{equation*}
The conjugate of a generic hadron multiplet in the charged isospin base can either describe a different multiplet or the same multiplet. In the first case (complex multiplet) the field and its conjugate are differentiated by other Abelian charges, such as strangeness or electric charge, hence $\wvrphi^\mm_\rho$ and $\compactphi^\mm_\rho$ are independent fields. Instead the components of \textit{self-conjugate} hadrons multiplets are related by conjugation:
\begin{equation*}
    \compactphi^{\mm}(x)=C^{\mm\nn}\wvrphi^{\nn}(x)\;,\qquad C^{\mm\nn}=\delta^{\mm,-\nn}\;.
\end{equation*}
A very important example of self-conjugate fields is the pion field. The Cartesian components are real, $\widehat{\boldsymbol{\pi}}=(\wpi_1,\wpi_2,\wpi_3)$, with $\wpi_i\in\mathbb{R}$. However in the base that diagonalizes $\widehat{\mathcal{Q}}_3$ the components are complex $\wpi^\pm=(1/\sqrt{2})(\wpi_1\pm\ii\wpi_2)$ and $\wpi^0=\wpi_3$ such that:
\begin{equation*}
    \wpi^{\pm\dagger}(x)=\wpi^\mp(x)\;,\qquad\wpi^{0\dagger}(x)=\wpi^0(x)\;.
\end{equation*}

A more useful form of the Wigner operator \eqref{def: Wig op} can be obtained by working in Fourier space. We start from the Fourier transform of the fields:
\begin{equation}\label{def: FT field}
    \begin{split}
        \widehat{\varphi}^\mm_\lambda(x)=\frac{1}{(2\pi)^4}\int\di^4p\,\e^{-\ii p\cdot x}\,\widehat{\varphi}^\mm_\lambda(p)\;,\qquad\compactphi^\mm_\lambda(x)=\frac{1}{(2\pi)^4}\int\di^4p\,\e^{\ii p\cdot x}\,\compactphi^\mm_\lambda(p)\;.
    \end{split}
\end{equation}
Note that, for self-conjugate multiplets, the field mode and its adjoint are related by:
\begin{equation}\label{eq: mode self-adjoint}
    \compactphi^\mm_\lambda(p)=C^{\mm\nn}\wvrphi^\nn_\lambda(-p)\;.
\end{equation}
We restrict our attention to the case of free fields for which
 momentum $p$ is on shell, i.e. $p^2=m^2$ with $m$ \emph{mass} of the hadron. For free fields commutation (anticommutation) relations completely determine the spectral function $\varrho$:
\begin{equation}\label{eq: spectral kernel}
    \left[\wvrphi^\nn_\rho(k),\compactphi^{\mm}_\lambda(k')\right]_\eta=(2\pi)^4\delta^{\mm\nn}\delta^4(k-k')\varrho_{\rho\lambda}(k)\;,
\end{equation}
which in turn is independent from the quantum state,
where $[\wA,\wB]_\eta\equiv\wA\wB+\eta\wB\wA$ denotes the commutator and anticommutator respectively with
$\eta=-1$ for bosons and $+1$ for fermions. Based on the spin of the hadron the free spectral function can be written as:
\begin{equation}\label{eq: free spectral}
    \varrho_{\rho\lambda}(k)=2\pi\,{\rm sgn}(k^0)\delta(k^2-m^2_h)\Pi_{\rho\lambda}(k)\;,
\end{equation}
where $\Pi^{}_{\rho\lambda}$ is the on-shell polarization projector and contains all the dependence on the Lorentz/spin representation of the field. For instance, for scalars, Dirac fermions and Proca vector mesons it is given by:
\begin{equation}\label{eq: Pi 0 1/2 1}
   \Pi^{(0)}_{\rho\lambda}(k)=1\;,\qquad\Pi^{(1/2)}_{BA}(k)=\left(\slashed{k}+m\right)_{BA}\;,\qquad\Pi^{(1)}_{\rho\lambda}(k)=-g_{\rho\lambda}+\frac{k_\rho k_\lambda}{m^2}\;. 
\end{equation}
By plugging equation \eqref{eq: free spectral} in \eqref{eq: spectral kernel} we obtain the following useful expression for the commutator of two free Fourier modes:
\begin{equation*}
    \left[\wvrphi^\nn_\rho(k),\compactphi^{\mm}_\lambda(k')\right]_\eta=(2\pi)^5\delta^{\mm\nn}\,{\rm sgn}(k^0)\delta^4(k-k')\delta(k^2-m^2_h)\Pi_{\rho\lambda}(k)\;.
\end{equation*}
For the subsequent discussion it is important to determine the transformation properties of the Fourier modes under translation, Abelian rotation and iso-vector rotation. The fields themselves satisfy:
\begin{equation*}
    \begin{split}
        \e^{\ii\wP\cdot y}\wvrphi^{\mss}_\alpha(x)\e^{-\ii\wP\cdot y}&=\wvrphi^{\mss}_\alpha(x+y)\;,\\
        \e^{\ii\wQ_{\ell}\phi}\wvrphi^\mss_\alpha(x)\e^{-\ii\wQ_{\ell}\phi}&=\e^{-\ii\phi q_\ell}\wvrphi^\mss_\alpha(x)\;,\\
        \e^{\ii\widehat{\mathcal{Q}}_3\theta}\wvrphi^\mss_\alpha(x)\e^{-\ii\widehat{\mathcal{Q}}_3\theta}&=\e^{-\ii\mss\theta}\wvrphi^\mss_\alpha(x)\;.
    \end{split}
\end{equation*}
By plugging \eqref{def: FT field} in the above expression we thus obtain the transformation of the Fourier modes under translation, Abelian rotation and iso-vector rotation:
\begin{subequations}\label{eq: Transformation mode}
    \begin{align}
        \e^{\ii\wP\cdot y}\wvrphi^{\mss}_\alpha(p)\e^{-\ii\wP\cdot y}&=\e^{-\ii p\cdot y}\wvrphi^\mss_\alpha(p)\;,\;\qquad \e^{\ii\wP\cdot y}\compactphi^{\mss}_\alpha(p)\e^{-\ii\wP\cdot y}=\e^{\ii p\cdot y}\compactphi^{\mss}_\alpha(p)\;,\label{eq: transP}\\
        \e^{\ii\wQ_{\ell}\phi}\wvrphi^{\mss}_\alpha(p)\e^{-\ii\wQ_{\ell}\phi}&=\e^{-\ii q_\ell\phi}\wvrphi^\mss_\alpha(p)\;,\;\qquad \e^{\ii\wQ_{\ell}\phi}\compactphi^{\mss}_\alpha(p)\e^{-\ii\wQ_{\ell} \phi}=\e^{\ii q_\ell \phi}\compactphi^{\mss}_\alpha(p)\;,\label{eq: transQ}\\
        \e^{\ii\widehat{\mathcal{Q}}_3\theta}\wvrphi^{\mss}_\alpha(p)\e^{-\ii\widehat{\mathcal{Q}}_3\theta}&=\e^{-\ii\mss\theta}\wvrphi^\mss_\alpha(p)\;,\qquad \e^{\ii\widehat{\mathcal{Q}}_3\theta}\compactphi^{\mss}_\alpha(p)\e^{-\ii\widehat{\mathcal{Q}}_3\theta}=\e^{\ii\mss\theta}\compactphi^{\mss}_\alpha(p)\;.\label{eq: transI}
    \end{align}
\end{subequations}
From \eqref{eq: transP}, \eqref{eq: transQ} and \eqref{eq: transI}, writing $\e^{-\lambda\beta\cdot\wP}=\e^{\ii\wP\cdot c}$ with $c=\ii\lambda\beta$ and $\e^{\lambda\mathfrak{Z}\widehat{\mathcal{Q}}_3}=\e^{\ii\theta\widehat{\mathcal{Q}}_3}$ with $\theta=-\ii\lambda\mathfrak{Z}$, we get:
\begin{equation}\label{eq: A on bilinear}
    \e^{\lambda\wA}\compactphi^\mrr_\alpha(p)\wvrphi^\mss_\beta(p')\e^{-\lambda\wA}=\e^{-\lambda\left[\beta\cdot(p-p')-(\mu_\mrr-\mu_\mss)\right]}\compactphi^\mrr_\alpha(p)\wvrphi^\mss_\beta(p')\;,
\end{equation}
where $\wA$ is the global equilibrium exponent \eqref{def: A} and $\mu_\mrr\equiv\sum_\ell\zeta_\ell q_\ell+\mathfrak{Z}\mrr$ is the reduced chemical potential of the component $\mrr$, $q_\ell$ being the Abelian charges of the multiplet. 
 
Plugging the Fourier transform of the field \eqref{def: FT field} in equation \eqref{def: Wig op} and integrating over $\di^4s$ we get:
\begin{equation*}
    \begin{split}
        \wW^{\mm\nn}_{\lambda\rho}(x,k)=\frac{1}{(2\pi)^8}\int\di^4p\,\di^4p'\delta^4\left(k-\frac{p+p'}{2}\right)\e^{\ii x\cdot(p-p')}\compactphi^\mm_\lambda(p)\widehat{\varphi}^\nn_\rho(p')\;.
    \end{split}
\end{equation*}
It is convenient to change integration variables introducing the two momenta $P$ and $q$:
\begin{equation*}
    \begin{split}
        &P=\frac{p+p'}{2}\;,\qquad q=p-p'\;,\qquad p=P+\frac{q}{2}\;,\qquad p'= P-\frac{q}{2}\;.
    \end{split}
\end{equation*}
Then, taking into account that $\di^4p\,\di^4p'=\di^4P\,\di^4q$ we can integrate over $\di^4P$ taking advantage of $\delta^4(P-k)$ leading to the following compact expression: 
\begin{equation}\label{eq: Wig comp}
    \wW^{\mm\nn}_{\lambda\rho}(x,k)=\frac{1}{(2\pi)^8}\int\di^4q\,\e^{\ii x\cdot q}\,\compactphi^\mm_\lambda(k_+)\widehat{\varphi}^\nn_\rho(k_-)\;,
\end{equation}
where the momenta of the two free fields are now:
\begin{equation}\label{def: k q}
    k\equiv\frac{k_++k_-}{2}\;,\qquad q\equiv k_+-k_-\;.
\end{equation}
Note that, being the fields on-shell, i.e. $k^2_\pm=m^2$, the momenta \eqref{def: k q} satisfy:
\begin{equation}\label{eq: rel k q}
    k^2=m^2-\frac{q^2}{4}\;,\qquad k\cdot q=0\;.
\end{equation}
An important property of the Wigner operator for free fields is that:
\begin{equation*}
    k\cdot\partial\,\wW^{\mm\nn}_{\lambda\rho}(x,k)=0\;,
\end{equation*}
implying that the integral $\int_{\Sigma}\di\Sigma\cdot k\,\wW$ is \emph{independent} of the hypersurface $\Sigma$ \cite{Becattini:2020sww}.

The Wigner function is then defined as the expectation value, on the state of the system, of the Wigner operator.
In the local equilibrium state $\wrho_{\rm LTE}$ \eqref{def: rhoLTE}, the Wigner function is then:
\begin{equation}\label{def: Wig function}
    \langle \wW^{\mm\nn}_{\lambda\rho}(x,k)\rangle_{\rm LTE}\equiv\Tr\left[\wrho_{\rm LTE}\,\wW^{\mm\nn}_{\lambda\rho}(x,k)\right]-\left\langle0\right|\wW^{\mm\nn}_{\lambda\rho}(x,k)\left|0\right\rangle\;,
\end{equation}
where the vacuum expectation value has been subtracted in order to properly normalize the local equilibrium expectation value.

We end this section by computing the expectation value of a quadratic combination of fields mode in the global homogeneous equilibrium state. Using the cyclicity of the trace and \eqref{eq: Transformation mode} we obtain:
\begin{equation*}
    \begin{split}
        \langle\compactphi^\mm_\lambda(p)\wvrphi^\nn_\rho(p')\rangle_{\rm GE}&=\frac{\Tr\left[\wvrphi^\nn_\rho(p')\,\e^{\wA}\compactphi^\mm_\lambda(p)\right]}{Z_{\rm GE}}=\e^{-(\beta\cdot p'-\mu_\nn)}\langle\wvrphi^\nn_\rho(p')\compactphi^\mm_\lambda(p)\rangle_{\rm GE}\\
        &=\e^{-(\beta\cdot p'-\mu_\nn)}\Big(\left[\wvrphi^\nn_\rho(p'),\compactphi^\mm_\lambda(p)\right]_\eta-\eta\langle\compactphi^\mm_\lambda(p)\wvrphi^\nn_\rho(p')\rangle_{\rm GE}\Big)\;.
    \end{split}
\end{equation*}
Solving for $\langle\compactphi\wvrphi\rangle_{\rm GE}$ and using \eqref{eq: spectral kernel}:
\begin{subequations}\label{eq: 2pt thermal exp}
    \begin{align}
        \langle\compactphi^\mm_\lambda(p)\wvrphi^\nn_\rho(p')\rangle_{\rm GE}&=(2\pi)^4\delta^{\mm\nn}\delta^4(p-p')\,n_\eta(p,\mm)\,\varrho_{\rho\lambda}(p)\;,\label{eq: 2pt a}\\
    \langle\wvrphi^\nn_\rho(p')\compactphi^\mm_\lambda(p)\rangle_{\rm GE}&=(2\pi)^4\delta^{\mm\nn}\delta^4(p-p')\left[1-\eta\,n_\eta(p,\mm)\right]\varrho_{\rho\lambda}(p)\;,\label{eq: 2pt b}
    \end{align}
\end{subequations}
where:
\begin{equation}\label{def: n}
    n_\eta(p,\mm)\equiv\frac{1}{\e^{\beta(x)\cdot p-\mu_\mm(x)}+\eta}\;,
\end{equation}
is the Bose--Einstein for $\eta=-1$ and Fermi--Dirac for $\eta=+1$ distributions respectively.

\section{Linear and quadratic response of the Wigner function}\label{SEC: linear and quadratic response}
The local equilibrium expansion in quadratic response of the Wigner function is simply obtained by replacing $\wO(x)$ in equations \eqref{def: lin O} and \eqref{def: quad O}, with the Wigner operator $\wW(x,k)$ such that:
\begin{equation}\label{eq: WLTE exp}
    \langle\wW^{\mm\nn}_{\lambda\rho}(x,k)\rangle_{\rm LTE}\simeq \langle\wW^{\mm\nn}_{\lambda\rho}(x,k)\rangle_{\rm GE}+\Delta_{\rm lin}W^{\mm\nn}_{\lambda\rho}(x,k)+\Delta_{\rm quad}W^{\mm\nn}_{\lambda\rho}(x,k)\;,
\end{equation}
where the linear response correction is:
\begin{equation}\label{def: linW}
    \Delta_{\rm lin}W^{\mm\nn}_{\lambda\rho}(x,k)\equiv\sum_a\int_{\Sigma_{\rm D}}\di\Sigma_\mu(y)\,\vartheta_a(y,x)\int_0^1\di\lambda\,\langle\wW^{\mm\nn}_{\lambda\rho}(x,k),\widehat{\Gamma}^\mu_a(y;\lambda)\rangle_{c,\rm GE}\;,
\end{equation}
while the quadratic response correction is:
\begin{equation}\label{def: quadW}
    \begin{split}
        \Delta_{\rm quad}W^{\mm\nn}_{\lambda\rho}(x,k)&\equiv\sum_{a,b}\int_{\Sigma_{\rm D}}\!\!\!\di\Sigma_\mu(y_1)\vartheta_a(y_1,x)\int_{\Sigma_{\rm D}}\!\!\!\di\Sigma_\nu(y_2)\vartheta_b(y_2,x)\\
        &\times\int_0^1\!\!\!\di\lambda_1\int_0^{\lambda_1}\!\!\!\di\lambda_2\,\langle\wW^{\mm\nn}_{\lambda\rho}(x,k),\widehat{\Gamma}^\nu_b(y_2;\lambda_2),\widehat{\Gamma}^\mu_a(y_1;\lambda_1)\rangle_{c,\rm GE}\;.
    \end{split}
\end{equation}
The leading order term in the local equilibrium expansion \eqref{eq: WLTE exp} is the Wigner function at homogeneous global equilibrium obtained 
plugging \eqref{eq: Wig comp} in \eqref{def: Wig function} and using \eqref{eq: 2pt a}:
\begin{equation}\label{eq: W GE}
    \langle\wW^{\mm\nn}_{\lambda\rho}(x,k)\rangle_{\rm GE}=\frac{\delta^{\mm\nn}}{(2\pi)^3}{\rm sgn}(k^0)\delta(k^2-m^2)n_\eta(k,\mm)\Pi_{\rho\lambda}(k)\;.
\end{equation}
%

\subsection{Linear response}

We now consider the linear response correction to the Wigner function. This result has already been obtained, for the particular case of Dirac field in the Belinfante pseudo-gauge, in \cite{Sheng:2025cjk} and later extended to Rarita-Schwinger field in canonical, Belinfante and HW pseudo-gauges \cite{Buzzegoli:2026sge}. In the following we report the major steps and we extend to the case of finite iso-vector chemical potentials, arbitrary pseudo-gauge and for generic spin field.

We start by plugging equation \eqref{eq: Wig comp} in equation \eqref{def: linW} so that the linear response correction can be written as:
\begin{equation}\label{eq: linwig1}
    \Delta_{\rm lin}W^{\mm\nn}_{\lambda\rho}(x,k)=\sum_a\int_{\Sigma_{\rm D}}\di\Sigma_\mu(y)\vartheta_a(y,x)\,C^{\mu}_{W\Gamma_a}(x-y,k)^{\mm\nn}_{\lambda\rho}\;,
\end{equation}
where $C^{\mu,\mm\nn}_{W\Gamma_a\lambda\rho}$ is the Kubo-Mori correlator between the Wigner operator and the local density $\wGamma^\mu_a$:
\begin{equation}\label{def: KM correlator}
    C^{\mu}_{W\Gamma_a}(x-y,k)^{\mm\nn}_{\lambda\rho}\equiv\int^1_0\di\lambda\,\langle\wW^{\mm\nn}_{\lambda\rho}(x,k),\,\wGamma^\mu_a(y;\lambda)\rangle_{c,\rm GE}\;.
\end{equation}
All the local densities $\wGamma^\mu_a$ \eqref{Table: B table} are, for free fields, bilinear in the fields of each hadronic species. It is thus convenient to express them in terms of the field Fourier modes \eqref{def: FT field}:
\begin{equation}\label{eq: vertex}
    \widehat{\Gamma}^\mu_a(y)=\frac{1}{(2\pi)^8}\int\di^4p\,\di^4p'\,\e^{\ii(p-p')\cdot y}\left[\mathsf{f}^\mu_a(p,p')\right]^{\mrr\mss}_{\alpha\beta}\,\compactphi^\mrr_\alpha(p)\wvrphi^\mss_\beta(p')\;,
\end{equation}
which defines the \emph{vertex} $\mathsf{f}^\mu_a$, a matrix in the isospin and spin indices, polynomial in the momenta, and depending on the particular density considered.

Combining \eqref{eq: vertex} with \eqref{eq: A on bilinear} it follows:
\begin{equation}\label{eq: Gamma lambda}
    \widehat{\Gamma}^\mu_a(y;\lambda)=\frac{1}{(2\pi)^8}\int\di^4p\,\di^4p'\,\e^{\ii(p-p')\cdot y}\,\e^{-\lambda\left[\beta\cdot(p-p')-(\mu_\mrr-\mu_\mss)\right]}\left[\mathsf{f}^\mu_a(p,p')\right]^{\mrr\mss}_{\alpha\beta}\,\compactphi^\mrr_\alpha(p)\wvrphi^\mss_\beta(p')\;.
\end{equation}
For an operator of isospin charge $\mathfrak{q}_a$ only the components with $\mss-\mrr=\mathfrak{q}_a$ are non-vanishing, so that $\mu_\mrr-\mu_\mss=-\mathfrak{Z}\mathfrak{q}_a$.
By plugging \eqref{eq: Wig comp} and \eqref{eq: Gamma lambda} in \eqref{def: KM correlator}, and performing the integration in $\lambda$ we thus obtain: 
\begin{equation}\label{eq: KM1}
   \begin{split}
        C^{\mu}_{W\Gamma_a}(x-y,k)^{\mm\nn}_{\lambda\rho}&=\frac{1}{(2\pi)^{16}}\int\di^4q\int\di^4p\di^4p'\,\e^{\ii q\cdot x}\,\Lambda_1(z)\times\\
        &\times\e^{\ii(p-p')\cdot y}\left[\mathsf{f}^\mu_a(p,p')\right]^{\mrr\mss}_{\alpha\beta}\langle\compactphi^\mm_\lambda(k_+)\wvrphi^\nn_\rho(k_-),\,\compactphi^\mrr_\alpha(p)\wvrphi^\mss_\beta(p')\rangle_{c,\rm GE}\;,
   \end{split}
\end{equation}
where:
\begin{equation}\label{def: z lin}
    z\equiv\beta(x)\cdot (p'-p)-\mathfrak{Z}(x)(\mss-\rr)\;,
\end{equation}
and the $\lambda$-integral $\Lambda_1(z)$ is given by:
\begin{equation}\label{eq: Lambda1}
    \Lambda_1(z)\equiv\int_0^1\di\lambda\,\e^{\lambda z}=\frac{\e^z-1}{z}\;.
\end{equation}
The Kubo-Mori correlator \eqref{eq: KM1} is worked out by computing the four-field connected expectation value (see appendix \ref{APP: wick}), in particular, by plugging \eqref{eq: RISULTATO C4} in \eqref{eq: KM1} and integrating over $\di^4p\,\di^4p'$ we eventually obtain:
\begin{align*}
    C^{\mu}_{W\Gamma_a}(x-y,k)^{\mm\nn}_{\lambda\rho}=&\frac{1}{(2\pi)^8}\int\di^4q\,\e^{\ii q\cdot(x-y)}\,\Lambda_1(z)\,n_\eta(k_+,\mm)\left[1-\eta n_\eta(k_-,\nn)\right]\times\\
    &\times\left[\varrho(k_-)\,\mathsf{f}^{\mu,\nn\mm}_a(k_-,k_+)\,\varrho(k_+)\right]_{\rho\lambda}\;,
\end{align*}
where $n_\eta$ is the Bose-Einstein ($\eta=-1$) or Fermi-Dirac ($\eta=+1$) distribution \eqref{def: n}.
The above can be further simplified by considering that the two spectral function insertions are given by:
\begin{equation*}
    \varrho_{\beta\lambda}(k_+)\varrho_{\rho\alpha}(k_-)=(2\pi)^2{\rm sgn}(k^0_+){\rm sgn}(k^0_-)\delta(k^2_+-m^2)\delta(k^2_--m^2)\Pi_{\beta\lambda}(k_+)\Pi_{\rho\alpha}(k_-)\;.
\end{equation*}
 Being the two four-vectors on-shell $k^2_\pm=m^2$ using \eqref{eq: rel k q} the two on-shell deltas can be written as:
\begin{equation*}
    \delta(k^2_+-m^2)\delta(k^2_--m^2)=\frac{1}{2}\delta(k\cdot q)\delta\left(k^2+\frac{q^2}{4}-m^2\right)\;.
\end{equation*}
By means of these identity we finally obtain that the Kubo-Mori correlator can be written as:
\begin{equation}\label{eq: KMfinal}
    \begin{split}
        C^{\mu}_{W\Gamma_a}(x-y,k)^{\mm\nn}_{\lambda\rho}&=\frac{1}{2(2\pi)^6}\int\di^4q\,\delta(k\cdot q)\e^{\ii q\cdot(x-y)}\,\Lambda_1(z)\,n_\eta(k_+,\mm)\left[1-\eta n_\eta(k_-,\nn)\right]\\
        &\times{\rm sgn}(k^0_+){\rm sgn}(k^0_-)\delta\left(k^2+\frac{q^2}{4}-m^2\right)\left[\Pi(k_-)\,\mathsf{f}^{\mu,\nn\mm}_a(k_-,k_+)\,\Pi(k_+)\right]_{\rho\lambda}\;.
    \end{split}
\end{equation}
Finally, plugging equation \eqref{eq: KMfinal} in equation \eqref{def: linW} we obtain that the linear response correction to the local equilibrium Wigner function can be written compactly as:
\begin{equation}\label{eq: LIN FG}
    \Delta_{\rm lin}W^{\mm\nn}_{\lambda\rho}(x,k)=\sum_a\int\di^4q\,\delta(k\cdot q)\,F^a_\mu(x,q)\left[G^\mu_a(k,q)\right]^{\mm\nn}_{\lambda\rho}\;,
\end{equation}
where we defined the microscopic kernel correlator:
\begin{equation}\label{def: G lin}
    \begin{split}
        \left[G^\mu_a(k,q)\right]^{\mm\nn}_{\lambda\rho}&\equiv\frac{1}{2(2\pi)^6}\,\delta\!\left(k^2-m^2+\frac{q^2}{4}\right){\rm sgn}(k^0_+){\rm sgn}(k^0_-)\,\Lambda_1\big(\beta\cdot q-\mathfrak{Z}(\mm-\nn)\big)\\
        &\times n_\eta(k_+,\mm)\left[1-\eta n_\eta(k_-,\nn)\right]\left[\Pi(k_-)\,\mathsf{f}^{\mu,\nn\mm}_a(k_-,k_+)\,\Pi(k_+)\right]_{\rho\lambda}\;,
    \end{split}
\end{equation}
and the thermodynamic field spatial Fourier transform:
\begin{equation}\label{def: FT tdfield}
    F^a_\mu(x,q)\equiv\int_{\Sigma_{\rm D}}\di\Sigma_\mu(y)\,\e^{\ii q\cdot(x-y)}\,\vartheta_a(y,x)\;.
\end{equation}
Equation \eqref{eq: LIN FG} makes the application of the hydrodynamic limit particularly clear and is the starting point of the subsequent gradient expansion.

The linear response correction to the Wigner function  is given by the integral convolution, in the momentum $q$, between the Fourier transform in the relative distance $x-y$ of the variation of the thermodynamic field \eqref{def: FT tdfield}, with a microscopic kernel \eqref{def: G lin}. In the hydrodynamic limit, when the thermodynamic fields are slowly varying compared to the microscopic scales of the system, the former is an extremely peaked function for $q=0$ while the latter is assumed to be finite in the limit $q\to0$\footnote{Being the Wigner function a distribution in $k$ the kernel is assumed to be finite modulo an overall $\delta$ function in $k$.}.  
This allows to evaluate the linear response theory by expanding the function $G$ in power of $q$:
\begin{equation*}
    \begin{split}
        \Delta_{\rm lin} W^{\mm\nn}_{\lambda\rho}(x,k)&=\sum\limits_a\sum_{N=0}^{+\infty}\frac{1}{N!}\partial^q_{\nu_1}\ldots\partial^q_{\nu_N}\left[G^\mu_a(k,q)\right]^{\mm\nn}_{\lambda\rho}\Big|_{q=0}\\
        &\times\int_{\Sigma_{\rm D}}\di\Sigma_\mu(y)\vartheta_a(y,x)\int\di^4q\,\delta(k\cdot q)\,q^{\nu_1}\ldots q^{\nu_N}\,\e^{\ii q\cdot(x-y)}\;.
    \end{split}
\end{equation*}
 The integration over $\di^4q$ and the subsequent integration of the decoupling hypersurface can then be performed exactly at any order in $q$. Appendix \ref{APP: integration over Sigma} describes in detail the computation that leads to the following final results:
\begin{equation}\label{eq: linwigFIN}
\boxed{
   \begin{split}
        \Delta_{\rm lin} W^{\mm\nn}_{\lambda\rho}(x,k)&=(2\pi)^3\sum_a\sum_{N=0}^\infty\frac{(-\ii)^N}{N!}\,\partial^q_{\alpha_1}\cdots\partial^q_{\alpha_N}\left[G^\mu_a(k,q)\right]^{\mm\nn}_{\lambda\rho}\Big|_{q=0}\\
        &\times\;\sum_{\overline{y}_{i,k}(x)}\Bigg[\mathfrak{D}^{\alpha_1}_y(y)\circ\ldots\circ\mathfrak{D}^{\alpha_N}_y(y)\left(\frac{\hat{n}_\mu(y)}{|k\cdot \hat{n}(y)|}\,\vartheta_a(y,x)\right)\Bigg]\Big|_{y=\overline{y}_{i,k}(x)}
   \end{split}}\;,
\end{equation}
where the differential operator $\mathfrak{D}_y$ is defined as:
\begin{equation}\label{def: Dy}
    \mathfrak{D}^y_\alpha(y)\equiv\mathcal{P}^\beta_{\;\alpha}(y)\,\partial^y_\beta=\partial^y_\alpha-\frac{\hat{n}_\alpha(y)}{k\cdot \hat{n}(y)}\,k\cdot\partial^y\;,\qquad\mathcal{P}^\beta_{\;\alpha}(y)\equiv g^\beta_\alpha-\frac{k^\beta\hat{n}_\alpha(y)}{k\cdot\hat{n}(y)}\;,
\end{equation}
and the symbol $\circ$ denotes the composition of the differential operators: each $\mathfrak{D}$ acts on everything on its right, including the normal vectors contained in the operators $\mathfrak{D}$ on its right.
For instance, for $N=2$ and arbitrary function $H(y,x)$, since $\mathfrak{D}_{\alpha_2}H=\partial_{\alpha_2}H-\nu_{\alpha_2}\,k\cdot\partial H$:
\begin{equation*}
    \mathfrak{D}_{\alpha_1}\circ\mathfrak{D}_{\alpha_2}H=\mathcal{P}^{\beta_1}_{\,\alpha_1}\mathcal{P}^{\beta_2}_{\,\alpha_2}\,\partial_{\beta_1}\partial_{\beta_2}H-\mathcal{P}^{\beta_1}_{\,\alpha_1}\left(\partial_{\beta_1}\nu_{\alpha_2}\right)k\cdot\partial H\;,
\end{equation*}
where:
\begin{equation}\label{def: nu}
    \nu_\alpha(y)\equiv\frac{\hat{n}_\alpha(y)}{k\cdot\hat{n}(y)}\;,
\end{equation}
and where the second term comes from the derivative of the normal vector contained in $\mathfrak{D}_{\alpha_2}$ (see also appendix \ref{APP: integration over Sigma}). The absolute value $|k\cdot\hat{n}|$ appearing in \eqref{eq: linwigFIN} coincides with $k\cdot\hat{n}$ on the space-like part of $\Sigma_{\rm D}$ for $k^0>0$, and in particular at the trivial intersection considered in the following, while it differs from it at the intersections with the time-like branches.

$\mathcal{P}^\beta_{\,\alpha}$ is the (oblique) projector onto the tangent space of $\Sigma_{\rm D}$ along the direction of $k$:
 \begin{align}\label{eq: ortho P maintext}
     &\hat n_\beta(y)\mathcal{P}^\beta_\alpha(y) = 0\;,& &\mathcal{P}^\beta_\alpha(y)k^\alpha = 0\;,& &\mathcal{P}^\beta_\gamma(y)\mathcal{P}^\gamma_\alpha(y) = \mathcal{P}^\beta_\alpha(y)\;.&
 \end{align}
 The first relation implies that $\mathfrak{D}_\alpha$ only differentiates along $\Sigma_{\rm D}$, the second one that $k^\alpha\mathfrak{D}_\alpha=0$, consistently with $\delta(k\cdot q)$. Note that instead $k_\beta\mathcal{P}^\beta_\alpha = k_\alpha-k^2\nu_\alpha\neq 0$ and $\mathcal{P}^\beta_\alpha\hat n^\alpha\neq 0$.
  
\begin{figure}[h!]
    \centering
    \includegraphics[width=0.5\linewidth]{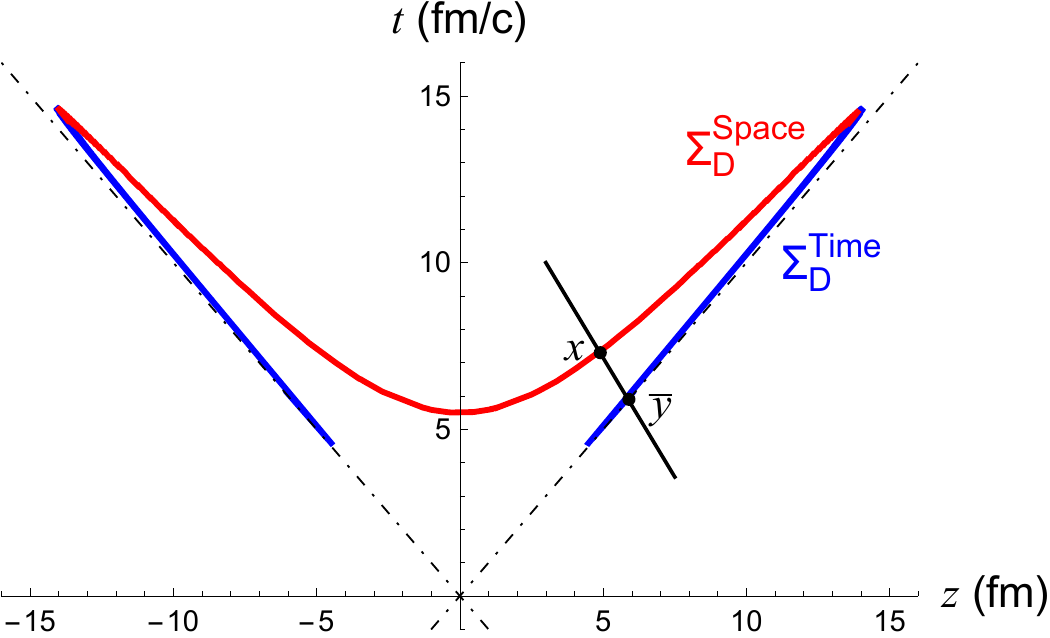}
\caption{\label{fig: intersection} A slice of the decoupling hypersurface $\Sigma_{\rm D}$
(red and blue lines) on the transverse plane $x=y=0$, obtained by 3+1D viscous hydrodynamic
simulations with the CLVisc model at $\sqrt{s_{\text{NN}}}=27$ GeV \citep{Pang:2018zzo,Wu:2021fjf,Wu:2022mkr}.
The black solid line represents the worldline \eqref{eq: worldline}, which intersects $\Sigma_{\rm D}$ at 
two points: one on the spacelike branch and the other one on the timelike branch. Figure taken from \cite{Sheng:2025cjk}.}
\end{figure}

In formula \eqref{eq: linwigFIN} the space-time point
 $\overline{y}_{i,k}(x)$ denotes an intersection between the decoupling hypersurface and the world line through $x$ with direction $k$, i.e. the trajectory of a free particle with momentum $k$ passing through $x$:
\begin{equation}\label{eq: worldline}
    (x-y)^\mu=\tau k^\mu\;,\qquad \tau\in\mathbb{R}\;.
\end{equation}
If the decoupling hypersurface also has time-like branches, there are in principle more possible intersections $\overline{y}_{i,k}(x)$, while the trivial one $\overline{y}_k(x)=x$ is always present (see figure \ref{fig: intersection}).

Equation \eqref{eq: linwigFIN} contains all the corrections, linear in the gradients of the thermodynamic fields and the curvature. If we are interested in considering corrections up to the quadratic order in the gradients we cut the series at $N=2$:
\begin{equation}\label{eq: tcimo}
    \begin{split}
        &\Delta_{\rm lin} W^{\mm\nn}_{\lambda\rho}(x,k)\simeq(2\pi)^3\sum_a\left[G^\mu_a(k,0)\right]^{\mm\nn}_{\lambda\rho}\sum_{\overline{y}_{i,k}(x)}\frac{\hat{n}_\mu(\overline{y}_{i,k}(x))}{|k\cdot\hat{n}(\overline{y}_{i,k}(x)|}\,\vartheta_a(\overline{y}_{i,k}(x),x)\\
        &-\ii(2\pi)^3\sum_a\,\partial^q_\alpha\left[G^\mu_a(k,q)\right]^{\mm\nn}_{\lambda\rho}\Big|_{q=0}\sum_{\overline{y}_{i,k}(x)}\mathfrak{D}^\alpha_y(y)\left[\frac{\hat{n}_\mu(y)}{|k\cdot \hat{n}(y)|}\,\vartheta_a(y,x)\right]\Big|_{y=\overline{y}_{i,k}(x)}\\
        &-\frac{(2\pi)^3}{2}\sum_a\,\partial^q_{\alpha_1}\partial^q_{\alpha_2}\left[G^\mu_a(k,q)\right]^{\mm\nn}_{\lambda\rho}\Big|_{q=0}\sum_{\overline{y}_{i,k}(x)}\left\{\mathfrak{D}^{\alpha_1}_y(y)\left[\mathfrak{D}^{\alpha_2}_y(y)\left(\frac{\hat{n}_\mu(y)}{|k\cdot \hat{n}(y)|}\,\vartheta_a(y,x)\right)\right]\right\}\bigg|_{y=\overline{y}_{i,k}(x)}\;.
    \end{split}
\end{equation}
The first term depends on the finite difference of the thermodynamic fields between the two points $\overline{y}_{i,k}(x)$ and $x$. As  observed in \cite{Becattini:2025rqp} these contributions are present also in a generic interacting theory and are related with memory of the initial state. For free fields these reduce to the free-streaming of a particle from the intersection $\overline{y}_{i,k}$ towards space-time point $x$. These terms are expected to be canceled once the volume term in \eqref{def: rho gauss} is also included and thus we neglect them given that they give rise to non-local corrections.
We then assume that the leading order local equilibrium correction is coming from the trivial intersection ${\overline{y}_{i,k}}(x)=x$ only, for which the first term in the formula above is zero. This point is further discussed in section \ref{SEC: discussion spin pol} for the specific case of spin polarization of Dirac fermions. Note that on the space-like patch of $\Sigma_{\rm D}$ $k\cdot\hat{n}>0$.

It is convenient to separate the corrections induced by the conserved densities $\wT,\,\wj_\ell$ and $\widehat{\boldsymbol{\mathcal{V}}}$ from those induced by the spin potential $\widehat{\mathcal{S}}$. For the first $\vartheta_a(x,x)=0$ while for the latter $\mathfrak{S}(x)\ne0$, which is already of order $\partial$:
\begin{equation*}
    \Delta_{\rm lin}W^{\mm\nn}_{\lambda\rho}(x,k)=\Delta^{\rm con}_{\rm lin}W^{\mm\nn}_{\lambda\rho}(x,k)+\Delta^{\mathfrak{S}}_{\rm lin}W^{\mm\nn}_{\lambda\rho}(x,k)\;,
\end{equation*}
where:
\begin{equation}\label{eq: W linear N2}
    \begin{split}
        \Delta^{\rm con}_{\rm lin} W^{\mm\nn}_{\lambda\rho}(x,k)&=-\ii(2\pi)^3\sum_{a\ne\mathfrak{S}}\,\partial^q_\alpha\left[G^\mu_a(k,q)\right]^{\mm\nn}_{\lambda\rho}\Big|_{q=0}\mathfrak{D}^\alpha_y\left[\mathcal{R}^a_\mu(y,x)\right]\Big|_{y=x}\\
        &-\frac{(2\pi)^3}{2}\sum_{a\ne\mathfrak{S}}\,\partial^q_{\alpha_1}\partial^q_{\alpha_2}\left[G^\mu_a(k,q)\right]^{\mm\nn}_{\lambda\rho}\Big|_{q=0}\left\{\mathfrak{D}^{\alpha_1}_y\left[\mathfrak{D}^{\alpha_2}_y\left(\mathcal{R}^a_\mu(y,x)\right)\right]\right\}\Big|_{y=x}+\mathcal{O}(\partial^3)\;,
    \end{split}
\end{equation}
is the correction induced by the gradient associated with conserved densities, while:
\begin{equation}\label{eq: W linear spin potential}
    \begin{split}
        \Delta^{\mathfrak{S}}_{\rm lin} W^{\mm\nn}_{\lambda\rho}(x,k)&=\frac{(2\pi)^3}{2}\left[G^{\mu,\sigma\gamma}_{\mathcal{S}}(k,0)\right]^{\mm\nn}_{\lambda\rho}\nu_\mu(x)\mathfrak{S}_{\sigma\gamma}(x)\\
        &-\ii\frac{(2\pi)^3}{2}\partial^q_\alpha\left[G^{\mu,\sigma\gamma}_{\mathcal{S}}(k,q)\right]^{\mm\nn}_{\lambda\rho}\Big|_{q=0}\mathfrak{D}^\alpha_y\left[\nu_\mu(y)\mathfrak{S}_{\sigma\gamma}(y)\right]\Big|_{y=x}+\mathcal{O}(\partial^3)\;,
    \end{split}
\end{equation}
is the correction induced by the spin chemical potential.
We introduced the following compact notation for the derivatives:
\begin{equation}\label{def: R e nu maintext}
    \mathcal{R}^a_\mu(y,x)\equiv\nu_\mu(y)\vartheta_a(y,x)\;,\qquad \nu_\mu(y)\equiv\frac{\hat{n}_\mu(y)}{k\cdot\hat{n}(y)}\;,\qquad \mathfrak{D}_y^\alpha=\partial^\alpha_y-\nu^\alpha(y)k^\beta\partial^y_\beta\;.
\end{equation}
Now, recalling the definition \eqref{def: Dy} the projector is $\mathcal{P}^{\beta}_{\,\alpha}=g^\beta_\alpha-k^\beta\nu_\alpha$ and, being $\mathcal{R}^{\Delta\beta}_{\mu\nu}(x,x)=0$,  $\mathcal{R}^{\Delta\zeta}_\mu(x,x)=0$  and $\mathcal{R}^{\Delta\ZZ}_\mu(x,x)=0$:

\begin{equation}\label{eq: DR 1}
\begin{split}
    \mathfrak{D}^\alpha_y\mathcal{R}^{\Delta\beta}_{\mu\nu}(y,x)\Big|_{y=x}=\nu_\mu(x)&\mathcal{P}^{\beta\alpha}(x)\partial_\beta\beta_\nu(x)\,,\quad \mathfrak{D}^\alpha_y\mathcal{R}^{\Delta\zeta}_{\mu}(y,x)\Big|_{y=x}=\nu_\mu(x)\mathcal{P}^{\beta\alpha}(x)\partial_\beta\zeta(x)\,, \\&\mathfrak{D}^\alpha_y\mathcal{R}^{\Delta\ZZ}_{\mu}(y,x)\Big|_{y=x}=\nu_\mu(x)\mathcal{P}^{\beta\alpha}(x)\partial_\beta\ZZ(x)\,.
\end{split}
\end{equation}
Similarly, for the second derivative, we obtain:
\begin{equation}\label{eq: DR 2}
    \begin{split}
        \mathfrak{D}^{\alpha_1}_y\left[\mathfrak{D}^{\alpha_2}_y\mathcal{R}^a_\mu(y,x)\right]\Big|_{y=x}&=\Big\{\nu_\mu\,\mathcal{P}^{\beta_1\alpha_1}\mathcal{P}^{\beta_2\alpha_2}\partial_{\beta_1}\partial_{\beta_2}\vartheta_a-\nu_\mu\,\mathcal{P}^{\beta_1\alpha_1}\left(\partial_{\beta_1}\nu^{\alpha_2}\right)k^\gamma\partial_\gamma\vartheta_a\\
        &+\left(\mathfrak{D}^{\alpha_1}\nu_\mu\right)\mathcal{P}^{\beta\alpha_2}\partial_\beta\vartheta_a+\left(\mathfrak{D}^{\alpha_2}\nu_\mu\right)\mathcal{P}^{\beta\alpha_1}\partial_\beta\vartheta_a\Big\}\Big|_{y=x}\;,
    \end{split}
\end{equation}
where all the derivatives act on $y$, $\vartheta_a\equiv\vartheta_a(y,x)$, and we used $\vartheta_{a\ne\mathfrak{S}}(x,x)=0$ and $\partial_{\beta_1}\mathcal{P}^{\beta_2\alpha_2}=-k^{\beta_2}\partial_{\beta_1}\nu^{\alpha_2}$.

As already observed in \cite{Sheng:2025cjk} the differential operator $\mathfrak{D}$ computes the variation of a function only along the hypersurface. In particular the second derivative \eqref{eq: DR 2} contains terms $\mathfrak{D}\nu$ which have the clear physical meaning of \emph{extrinsic curvature} of the hypersurface. Overall the derivatives \eqref{eq: DR 1} and \eqref{eq: DR 2} only compute the variations on the direction parallel to the hypersurface. Changes in the thermodynamic fields orthogonal to $\Sigma_{\rm D}$ are automatically eliminated.
\subsection{Quadratic response}

We now come to the computation of the quadratic response correction to the local equilibrium Wigner function which is the main theoretical result of this work. 

We start by plugging the Wigner operator $\wW(x,k)$ in place of $\wO(x)$ in equation \eqref{def: quad O}:
\begin{equation*}
    \begin{split}
        \Delta_{\rm quad}W^{\mm\nn}_{\lambda\rho}(x,k)&=\sum_{a,b}\int_{\Sigma_{\rm D}}\di\Sigma_{\mu_1}(y_1)\vartheta_a(y_1,x)\int_{\Sigma_{\rm D}}\di\Sigma_{\mu_2}(y_2)\vartheta_b(y_2,x)C^{\mu_2\mu_1}_{W\Gamma_b\Gamma_a}(x,y_1,y_2,k)^{\mm\nn}_{\lambda\rho}\;,
    \end{split}
\end{equation*}
where, in analogy with equations \eqref{eq: linwig1} and \eqref{def: KM correlator}, we introduced the three-point Kubo-Mori correlator: 
\begin{equation}\label{def: KuboMori3}
    C^{\mu_2\mu_1}_{W\Gamma_b\Gamma_a}(x,y_1,y_2,k)^{\mm\nn}_{\lambda\rho}\equiv\int^1_0\di\lambda_1\int^{\lambda_1}_0\di\lambda_2\langle\wW^{\mm\nn}_{\lambda\rho}(x,k),\,\wGamma^{\mu_2}_b(y_2;\lambda_2),\,\wGamma^{\mu_1}_a(y_1;\lambda_1)\rangle_{c,\rm GE}\;.
\end{equation}
Using  equation \eqref{eq: vertex}, the connected three body thermal expectation value can be written as:
\begin{equation*}
    \begin{split}
        &\langle\wW^{\mm\nn}_{\lambda\rho}(x,k),\,\wGamma^{\mu_2}_b(y_2;\lambda_2),\,\wGamma^{\mu_1}_a(y_1;\lambda_1)\rangle_{c,\rm GE}\\
        &=\frac{1}{(2\pi)^{24}}\int\di^4q\,\e^{\ii q\cdot x}\int\di^4p_1\di^4p'_1\int\di^4p_2\di^4p'_2\e^{-\lambda_2[\beta\cdot(p_2-p'_2)-(\mu_{\mrr_2}-\mu_{\mss_2})]}\e^{-\lambda_1[\beta\cdot(p_1-p'_1)-(\mu_{\mrr_1}-\mu_{\mss_1})]}\\
        &\times\e^{\ii(p_1-p'_1)\cdot y_1}\e^{\ii(p_2-p'_2)\cdot y_2}\langle\compactphi^\mm_\lambda(k_+)\wvrphi^\nn_\rho(k_-),\,\compactphi^{\rr_2}_{\alpha_2}(p_2)\mathsf{f}^{\rr_2\mss_2}_b\wvrphi^{\mss_2}_{\beta_2}(p'_2),\,\compactphi^{\rr_1}_{\alpha_1}(p_1)\mathsf{f}^{\rr_1\mss_1}_a\wvrphi^{\mss_1}_{\beta_1}(p'_1)\rangle_{c,\rm GE}\;,
    \end{split}
\end{equation*}
so that the double integral over $\lambda_1,\lambda_2$ can be performed and reads:
\begin{equation}\label{A: Lambda2}
    \Lambda_2(z_a,z_b)\equiv\int_0^1\di\lambda_1\int_0^{\lambda_1}\di\lambda_2\,\e^{\lambda_1z_a+\lambda_2z_b}=\frac{\Lambda_1(z_a+z_b)-\Lambda_1(z_a)}{z_b}\;,
\end{equation}
where $z_{a,b}$ are defined according to \eqref{def: z lin} as:
\begin{equation}\label{def: z quad}
    z_{a,b}\equiv\beta\cdot(p'_{1,2}-p_{1,2})+(\mu_{\rr_{1,2}}-\mu_{\mss_{1,2}})\;.
\end{equation}
With this the correlator turns out to be given by:
\begin{equation}\label{eq: 3KM 1}
   \begin{split}
        &C^{\mu_2\mu_1}_{W\Gamma_b\Gamma_a}(x,y_1,y_2,k)=\frac{1}{(2\pi)^{24}}\int\di^4q\,e^{\ii q\cdot x}\int\di^4p_1\di^4p'_1\int\di^4p_2\di^4p'_2\Lambda_2(z_a,z_b)\e^{\ii(p_1-p'_1)\cdot y_1}\e^{\ii(p_2-p'_2)\cdot y_2}\\
        &\times[\mathsf{f}^{\rr_2\mss_2}_b(p_2,p'_2)]_{\alpha_2\beta_2}[\mathsf{f}^{\rr_1\mss_1}_a(p_1,p'_1)]_{\alpha_1\beta_1}\langle\compactphi^\mm_\lambda(k_+)\wvrphi^\nn_\rho(k_-),\,\compactphi^{\rr_2}_{\alpha_2}(p_2)\wvrphi^{\mss_2}_{\beta_2}(p'_2),\,\compactphi^{\rr_1}_{\alpha_1}(p_1)\wvrphi^{\mss_1}_{\beta_1}(p'_1)\rangle_{c,\rm GE}\;.
   \end{split}
\end{equation}
The 6 field connected thermal expectation value has been computed explicitly in appendix \ref{APP: wick}. After introducing the internal momenta:
\begin{equation}\label{def: qab l}
    q_a=\ell-k_-\;,\qquad q_b=k_+-\ell\;,\qquad q=q_a+q_b\;,\qquad \ell=k+\frac{q_a-q_b}{2}\;,
\end{equation}
and performing the integration over $p_{1,2}$, $p'_{1,2}$ one ultimately finds:
\begin{equation*}
    \begin{split}
        &C^{\mu_2\mu_1}_{W\Gamma_b\Gamma_a}(x-y_1,x-y_2,k)=\frac{1}{(2\pi)^{12}}\int\di^4 q_ad^4q_b\e^{\ii q_a\cdot(x-y_1)}\e^{\ii q_b\cdot(x-y_2)}[1-\eta\,n_{\eta}(k_-,\nn)]n_\eta(k_+,\mm)\times\\
        &\times\left[\Lambda_2(z_b,z_a)-\eta n_\eta(\ell,\rr)\Lambda_1(z_a)\Lambda_1(z_b)\right]\left[\varrho(k_-)\mathsf{f}^{\mu_1,\nn\rr}_a(k_-,\ell)\varrho(\ell)\mathsf{f}^{\mu_2,\rr\mm}_b(\ell,k_+)\varrho(k_+)\right]_{\rho\lambda}\;,
    \end{split}
\end{equation*}
with a sum over the internal isospin index $\rr$ understood.
After the integration on the internal momenta and the sum over the contracted isospin indexes the $z_{a,b}$ vector \eqref{def: z quad} are:
\begin{equation}\label{eq: z quad}
\begin{split}
    &\mathfrak{q}_a=\rr-\nn\;, \qquad\mathfrak{q}_b=\mm-\rr\;,\\
    &z_a=\beta(x)\cdot q_a-\ZZ(x)\,\mathfrak{q}_a\;, \quad z_b=\beta(x)\cdot q_b-\ZZ(x)\,\mathfrak{q}_b\;,\quad z_a+z_b=\beta\cdot q-\ZZ(\mm-\nn)\;.
\end{split}
\end{equation}
Now the three spectral functions contain the combination of the on-shell conditions $k^2_+=k^2_-=\ell^2=m^2$ which, using \eqref{def: qab l} can be written as:
\begin{equation*}
    \delta(k_+^2-m^2)\,\delta(k_-^2-m^2)\,\delta(\ell^2-m^2)=\frac14\,\delta\!\left(k^2-m^2+\frac{q^2}{4}\right)\,\delta\!\left(q_a\cdot\Big(k-\frac{q_b}{2}\Big)\right)\,\delta\!\left(q_b\cdot\Big(k+\frac{q_a}{2}\Big)\right)\;.
\end{equation*}
Eventually, using the above expression, one can show that the quadratic response correction to the Wigner function can be put in the following form:
\begin{equation}\label{eq: QUAD FG}
\begin{split}
    \Delta_{\rm quad}W^{\mm\nn}_{\lambda\rho}(x,k)=&\sum_{a,b}\int\di^4q_a\,\di^4q_b\;\delta\!\left[q_a\cdot\Big(k-\frac{q_b}{2}\Big)\right]\delta\!\left[q_b\cdot\Big(k+\frac{q_a}{2}\Big)\right]\times\\
    &\times F^a_\mu(x,q_a)\,F^b_\nu(x,q_b)\left[\mathsf{G}^{\mu\nu}_{ab}(k,q_a,q_b)\right]^{\mm\nn}_{\lambda\rho}\;,
\end{split}
\end{equation}
where the two $F^{a,b}_\mu$ are again the Fourier transform of the thermodynamic field on the hypersurface defined in \eqref{def: FT tdfield}, while the microscopic kernel is now given by:
\begin{equation}\label{def: G quad}
\begin{split}
    \left[\mathsf{G}^{\mu\nu}_{ab}(k,q_a,q_b)\right]^{\mm\nn}_{\lambda\rho}\equiv&\frac{1}{4(2\pi)^9}\,\delta\!\left(k^2-m^2+\frac{q^2}{4}\right)\times\\
    & \times\sum_{\mrr}\mathcal{N}^{\mm\mrr\nn}_{ab}(k_\pm,\ell)\left[\Pi(k_-)\,\mathsf{f}^{\mu,\nn\mrr}_a(k_-,\ell)\,\Pi(\ell)\,\mathsf{f}^{\nu,\mrr\mm}_b(\ell,k_+)\,\Pi(k_+)\right]_{\rho\lambda}\;.
\end{split}
\end{equation}
The statistical factor $\mathcal{N}$ is independent from the inserted operators $\wGamma_a$, $\wGamma_b$ and reads:
\begin{equation}\label{def: Sstat}
\begin{split}
     \mathcal{N}^{\mm\mrr\nn}_{ab}(k_\pm,\ell)\equiv &n_\eta(k_+,\mm)\left[1-\eta n_\eta(k_-,\nn)\right]\Big[\Lambda_2(z_b,z_a)-\eta\,n_\eta(\ell,\mrr)\,\Lambda_1(z_a)\Lambda_1(z_b)\Big]\times\\
     &\times {\rm sgn}(k_-^0)\,{\rm sgn}(\ell^0)\,{\rm sgn}(k_+^0)\;,
\end{split}
\end{equation}
where $q=q_a+q_b$, $k_\pm=k\pm q/2$, $\ell=k+(q_a-q_b)/2$ and $z_{a,b}$ are given in \eqref{eq: z quad}. 

The equation \eqref{eq: QUAD FG} is the quadratic response equivalent of \eqref{eq: LIN FG} and will be subsequent starting point for the hydrodynamic expansion. The two relative momenta $q_a$ and $q_b$ are the conjugate Fourier variable of two independent thermodynamic fields Fourier transforms. Hence in the hydrodynamic regime $F^{a,b}_{\mu_{1,2}}$ are peaked functions at $q_{a},\, q_b=0$ whereas the microscopic kernel is a regular distribution-supported function of $q_a,\, q_b$. It then follows that the quadratic correction \eqref{eq: QUAD FG} is given by the following Taylor expansion:
\begin{equation*}
    \begin{split}
        \Delta_{\rm quad}W^{\mm\nn}_{\lambda\rho}(x,k)&=\sum_{a,b}\sum_{N,M\ge0}\frac{1}{N!M!}\partial^{q_a}_{\alpha_1}\ldots\partial^{q_a}_{\alpha_N}\partial^{q_b}_{\beta_1}\ldots\partial^{q_b}_{\beta_M}[\mathsf{G}^{\mu_1\mu_2}_{ab}(k,q_a,q_b)]^{\mm\nn}_{\lambda\rho}\Big|_{q_a=q_b=0}\\
        &\times\int_{\Sigma_{\rm D}}\di\Sigma_{\mu_1}(y_1)\vartheta_a(y_1,x)\int_{\Sigma_{\rm D}}\di \Sigma_{\mu_2}(y_2)\vartheta_b(y_2,x)\int\di^4q_a\,\di^4q_b\\
        &\times\delta\!\left[q_a\cdot\Big(k-\frac{q_b}{2}\right)\Big]\delta\!\left[q_b\cdot\Big(k+\frac{q_a}{2}\Big)\right] q^{\alpha_1}_a\ldots q^{\alpha_N}_a\e^{\ii q_a\cdot(x-y_1)}q^{\beta_1}_b\ldots q^{\beta_M}_b\e^{\ii q_b\cdot(x-y_2)}\;.
    \end{split}
\end{equation*}
The resulting double integration in the decoupling hypersurface and relative momenta share a similar structure with the one obtained at linear order. It is possible once again to perform, order by order in $q_a$ and $q_b$, the exact integration over the two hypersurfaces. The computation has been developed in appendix \ref{APP: integration over Sigma} and we report here the final result which reads:
\begin{equation}\label{eq: QUAD FINAL}
    \boxed{
    \begin{split}
        \Delta_{\rm quad}W(x,k)&=(2\pi)^6\sum_{a,b}\sum_{i,j}\sum_{N,M=0}^{\infty}\frac{(-\ii)^{N+M}}{N!M!}\partial^{q_a}_{\alpha_1}\ldots\partial^{q_a}_{\alpha_N}\partial^{q_b}_{\beta_1}\ldots\partial^{q_b}_{\beta_M}\,[\mathsf{G}^{\mu_1\mu_2}_{ab}(k,q_a,q_b)^{\mm\nn}_{\lambda\rho}]\Big|_{q_a=q_b=0}\\
        &\times\sum_{n,m\ge0}\frac{\ii^{n+m}(-1)^n}{n!m!2^{n+m}}\Bigg\{\left[\mathsf{D}^{\alpha_1}_y\circ\ldots\circ\mathsf{D}_y^{\alpha_N}\circ\mathsf{D}^y_{\gamma_1}\circ\ldots\mathsf{D}^y_{\gamma_{n+m}}\left(\tau^n\mathcal{R}^a_{\mu_1}(y,x)\right)\right]\Big|_{i}\\
        &\times \left[\mathsf{D}^{\beta_1}_y\circ\ldots\circ\mathsf{D}_y^{\beta_M}\circ\mathsf{D}_y^{\gamma_1}\circ\ldots\mathsf{D}_y^{\gamma_{n+m}}\left(\tau^m\mathcal{R}^b_{\mu_2}(y,x)\right)\right]\Big|_{j}\Bigg\}\;,
    \end{split}
    }
\end{equation}
where the differential operator $\mathsf{D}_y$ is defined as:
\begin{equation}\label{def: D second order}
    \mathsf{D}^\alpha(y)\equiv\mathfrak{D}^\alpha(y)+\nu^\alpha(y)\partial_\tau=\mathcal{P}^\beta_{\,\alpha}(y)\partial^y_{\beta}+\nu^\alpha(y)\partial_\tau\;,   
\end{equation}
 while $\nu_\alpha(y)$ and $\mathcal{R}^{a,b}_{\mu_{1,2}}(y,x)$ are the same geometrical and thermodynamic factors obtained in the linear response and given explicitly by \eqref{def: R e nu maintext}. Formula \eqref{eq: QUAD FINAL} is one of the main results of this paper; it expresses the quadratic response correction to the Wigner function, at any order in the derivative of the thermodynamic fields and curvature, and is valid for any geometry of the decoupling hypersurface.

\begin{figure}[h!]
\centering
\begin{tikzpicture}[scale=1.25]
    \draw[thick] (-3,0.2) .. controls (-1,-0.6) and (1,0.8) .. (3,0);
    \node[black] at (3.3,0.15) {$\Sigma_{\rm D}$};
    \coordinate (y1) at (-1.2,-0.045);
    \coordinate (y2) at (0.9,0.22);
    \coordinate (x1) at (-0.54,2.155);
    \coordinate (x2) at (1.38,1.82);
    \draw[dashed] (-1.365,-0.595) -- (-0.441,2.485);
    \draw[dashed] (0.756,-0.26) -- (1.476,2.14);
    \draw[-{Latex},thick] (y1) -- (-0.837,1.165) node[left] {$\tau_1k$};
    \draw[-{Latex},thick] (y2) -- (1.164,1.10) node[right] {$\tau_2k$};
    \fill[black] (y1) circle (1.6pt) node[below left] {$y_1$};
    \fill[black] (y2) circle (1.6pt) node[below right] {$y_2$};
    \fill[black] (x1) circle (1.6pt) node[above left] {$x_1$};
    \fill[black] (x2) circle (1.6pt) node[above right] {$x_2$};
    \draw[-{Latex}] (-2.6,1.4) -- (-2.3,2.4) node[above] {$k$};
\end{tikzpicture}
\caption{\label{fig: two worldlines} The two worldlines entering the quadratic response \eqref{eq: QUAD FINAL}. They have the same direction $k$ and start from the two auxiliary points $x_1$ and $x_2$, whose derivatives bring down the momenta $q_a$ and $q_b$; their intersections with $\Sigma_{\rm D}$ are $y_{1,2}=x_{1,2}-\tau_{1,2}k$. The operator $\mathsf{D}$ takes into account both the displacement of $y_{1,2}$ along $\Sigma_{\rm D}$ and the change of the affine parameters $\tau_{1,2}$ when $x_{1,2}$ are varied. At the end $x_1=x_2=x$ and, for the trivial intersection, $y_1=y_2=x$ and $\tau_1=\tau_2=0$.}
\end{figure}
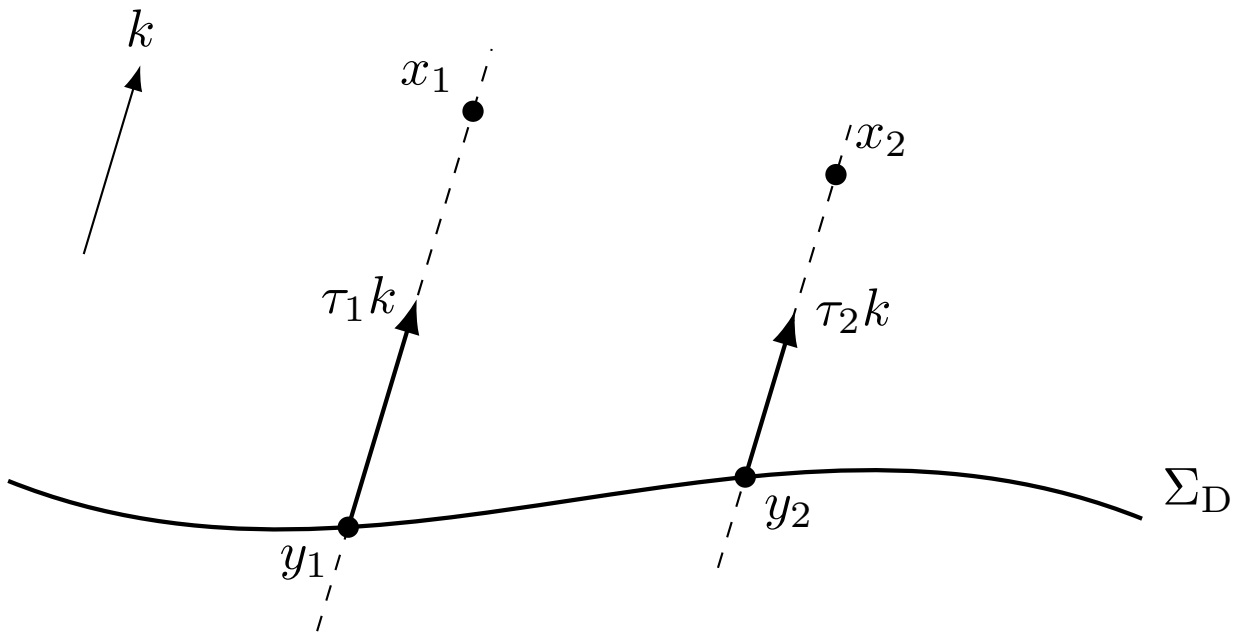

Note that, compared with the linear response formula \eqref{eq: linwigFIN}, the quadratic formula \eqref{eq: QUAD FINAL} \emph{is not} just the combination of two independent linear order formulae. The two hypersurface integrations are coupled through the affine parameters $\tau_{1,2}$ parametrizing the two world-lines with direction $k$ which start from the two auxiliary points $x_{1,2}$ introduced in appendix \ref{APP: integration over Sigma}, set equal to $x$ only at the end of the computation (see figure \ref{fig: two worldlines}):
\begin{equation*}
    (x_1-y_1)^\mu=k^\mu\,\tau_1(x_1)\;,\qquad (x_2-y_2)^\mu=k^\mu\,\tau_2(x_2)\;,
\end{equation*}
and the differentiation with respect to the operator $\mathsf{D}^y$ both take into account the variation along the tangential direction of $\Sigma$ and the variation along the affine parameters $\tau_{1,2}$.

The formula \eqref{eq: QUAD FINAL}, similarly to the linear response, contains non-local correction coming from the non-trivial intersections between the two independent world-lines and the decoupling hypersurface $\Sigma_{\rm D}$. In order to derive explicitly the quadratic contribution in the gradients we again only consider the trivial intersection $\overline{y}_{1,k}(x)=\overline{y}_{2,k}(x)=x$, for which $\tau_1(x)=\tau_2(x)=0$. Given that, from its definition \eqref{Table: B table}, $\vartheta_{a\ne\mathfrak{S}}(x,x)=0$ it holds $\mathcal{R}^{a,b}_{\mu_{1,2}}(x,x)=0$. This is not true for the case of the spin potential, for which $\mathcal{R}^\mathfrak{S}(x,x)\neq 0$, we will then discuss the terms involving the spin potential separately.

The generic term in the quadratic expansion \eqref{eq: Delta quad y=x} is labelled with $N,M,n,m$ and contains:
\begin{equation*}
    \begin{matrix}
        N+m+n\\
        {\rm factors}
    \end{matrix}\quad\mathsf{D}(y)\longrightarrow \tau^n\mathcal{R}^{a\ne\mathfrak{S}}_{\mu_1}(y,x)\;.
\end{equation*}
Given that, for $\overline{y}_{k}(x)=x$, $\tau$ is set to 0 after the derivation, $\partial_\tau$ must act exactly $n-$times to give a non-vanishing contribution. Hence it remains:
\begin{equation*}
    \begin{matrix}
        N+m\equiv j_1\\
        {\rm factors}
    \end{matrix}\quad \mathfrak{D}(y)\longrightarrow\mathcal{R}^{a\ne\mathfrak{S}}_{\mu_1}(y,x)\;.
\end{equation*}
Similarly for the other contribution we obtain:
\begin{equation*}
    \begin{matrix}
        M+n\equiv j_2\\
        {\rm factors}
    \end{matrix}\quad \mathfrak{D}(y)\longrightarrow\mathcal{R}^{b\ne\mathfrak{S}}_{\mu_2}(y,x)\;.
\end{equation*}
Now suppose each derivative acts as $\mathcal{O}(\epsilon)$ so that $\mathsf{D}^j\mathcal{R}=\mathcal{O}(\epsilon^j)$. It then follows that:
\begin{equation*}
    (N,M,n,m)\implies\mathcal{O}\left(\epsilon^{j_1+j_2}\right)=\mathcal{O}(\epsilon^{N+M+n+m})\;.
\end{equation*}
So, if we are interested in terms containing two powers of the derivative $\mathcal{O}(\epsilon^2)$ we need to compute:
\begin{equation*}
    j_1+j_2=2\implies j_1=j_2=1\implies N+m=1\;,\quad M+n=1\;.
\end{equation*}
Note that, for $a\ne\mathfrak{S}$, a factor with $j=0$ vanishes because $\mathcal{R}(x,x) = 0$, so $(j_1,j_2)= (2,0)$ and $(0,2)$ do not contribute. Labeling the general term with $(N,M,n,m)$, we need to compute the following four contributions:
\begin{equation*}
    (1,1,0,0)\;,\quad(1,0,1,0)\;,\quad (0,1,0,1)\;,\quad(0,0,1,1)\;.
\end{equation*}
From the above power counting follows that the second quadratic order induced by the conserved densities \emph{does not} contain derivatives of the hypersurface normal. To see this consider, for $a,b\ne\mathfrak{S}$:
\begin{equation*}
    \mathsf{D}_\alpha(y)\mathcal{R}^{a,b}_{\mu_{1,2}}(y,x)=\mathfrak{D}_\alpha(y)\mathcal{R}^{a,b}_{\mu_{1,2}}(y,x)\;,\quad \mathsf{D}_\alpha(y)\left[\tau\mathcal{R}^{a,b}_{\mu_{1,2}}(y,x)\right]=\tau\,\mathfrak{D}_\alpha(y)\mathcal{R}^{a,b}_{\mu_{1,2}}(y,x)+\nu_\alpha(y)\mathcal{R}^{a,b}_{\mu_{1,2}}(y,x)\;,
\end{equation*}
which leads to:
\begin{equation*}
    \begin{split}
        \mathsf{D}_\beta(y)\left\{\mathsf{D}_\alpha(y)\left[\tau\mathcal{R}^{a,b}_{\mu_{1,2}}(y,x)\right]\right\}&=\nu_\beta(y)\mathfrak{D}_\alpha(y)\mathcal{R}^{a,b}_{\mu_{1,2}}(y,x)+\tau\, \mathfrak{D}_\beta(y)\left[\mathfrak{D}_\alpha(y)\mathcal{R}^{a,b}_{\mu_{1,2}}(y,x)\right]\\
        &+\nu_\alpha(y)\mathfrak{D}_\beta(y)\mathcal{R}^{a,b}_{\mu_{1,2}}(y,x)+\mathcal{R}^{a,b}_{\mu_{1,2}}(y,x)\mathfrak{D}_\beta(y)\nu_\alpha(y)\;.
    \end{split}
\end{equation*}
On the trivial intersection, being $\tau(x)=0$ and $\mathcal{R}^{a,b}_{\mu_{1,2}}(x,x)=0$ we get: 
\begin{equation*}
    \mathsf{D}_\alpha(y)[\tau\mathcal{R}^{a,b}_{\mu_{1,2}}(y,x)]\big|_{\substack{y=x\\
    \tau=0}}=0\;,
\end{equation*}
which in turn implies:
\begin{equation*}
    \mathsf{D}_\beta(y)\left\{\mathsf{D}_\alpha(y)\left[\tau\mathcal{R}^{a,b}_{\mu_{1,2}}(y,x)\right]\right\}\Big|_{\substack{y=x\\
    \tau=0}}=\nu_\alpha(x)\left[\mathfrak{D}_\beta(y)\mathcal{R}^{a,b}_{\mu_{1,2}}(y,x)\right]\Big|_{\substack{y=x\\
    \tau=0}}+\nu_\beta(x)\left[\mathfrak{D}_\alpha(y)\mathcal{R}^{a,b}_{\mu_{1,2}}(y,x)\right]\Big|_{\substack{y=x\\
    \tau=0}}\;.
\end{equation*}
Thus, quadratic response theory, at order $\mathcal{O}(\partial^2)$ does not receive any contribution from the extrinsic curvature of the decoupling hypersurface like instead does at order $\mathcal{O}(\partial^2)$ in linear response.

We have now everything to compute the explicit second order contribution from the quadratic response theory. As for the linear response, it is convenient to separate the terms in which both sources are conserved densities from those involving the spin potential:
\begin{equation}\label{eq: Delta quad split}
    \Delta^{(2)}_{\rm quad}W^{\mm\nn}_{\lambda\rho}(x,k)=\Delta^{\rm con}_{\rm quad}W^{\mm\nn}_{\lambda\rho}(x,k)+\Delta^{\rm mix}_{\rm quad}W^{\mm\nn}_{\lambda\rho}(x,k)+\Delta^{\mathfrak{S}}_{\rm quad}W^{\mm\nn}_{\lambda\rho}(x,k)\;.
\end{equation}
The first term contains the contributions with $a,b\ne\mathfrak{S}$ and, applying explicitly \eqref{def: D second order} and truncating the series \eqref{eq: QUAD FINAL} at the desired order, reads:
\begin{equation}\label{eq: Delta quad y=x}
    \begin{split}
        &\Delta^{\rm con}_{\rm quad}W^{\mm\nn}_{\lambda\rho}(x,k)=(2\pi)^6\sum_{a,b\ne\mathfrak{S}}\Big\{-\partial^{q_a}_{\alpha_1}\partial^{q_b}_{\beta_1}[\mathsf{G}^{\mu_1\mu_2}_{ab}(k,q_a,q_b)]^{\mm\nn}_{\lambda\rho}\Big|_{\substack{q_a=0\\
        q_b=0}}\,\mathfrak{D}^{\alpha_1}_y\mathcal{R}^a_{\mu_1}(y,x)\Big|_{y=x}\mathfrak{D}^{\beta_1}_y\mathcal{R}^b_{\mu_2}(y,x)\Big|_{y=x}\\
        &-\frac{1}{2}\partial^{q_a}_{\alpha_1}[\mathsf{G}^{\mu_1\mu_2}_{ab}(k,q_a,q_b)^{\mm\nn}_{\lambda\rho}]\Big|_{\substack{q_a=0\\
        q_b=0}}\,\mathfrak{D}^{\gamma}_y\mathcal{R}^b_{\mu_2}(y,x)\Big|_{y=x}\Bigg[\nu^{\alpha_1}(x)\mathfrak{D}^y_{\gamma}\mathcal{R}^a_{\mu_1}(y,x)\Big|_{y=x}+\nu_{\gamma}(x)\mathfrak{D}_y^{\alpha_1}\mathcal{R}^a_{\mu_1}(y,x)\Big|_{y=x}\Bigg]\\
        &+\frac{1}{2}\partial^{q_b}_{\beta_1}[\mathsf{G}^{\mu_1\mu_2}_{ab}(k,q_a,q_b)^{\mm\nn}_{\lambda\rho}]\Big|_{\substack{q_a=0\\
        q_b=0}}\,\mathfrak{D}^\gamma_y\mathcal{R}^a_{\mu_1}(y,x)\Big|_{y=x}\Big[\nu^{\beta_1}(x)\mathfrak{D}^y_\gamma\mathcal{R}^b_{\mu_2}(y,x)\Big|_{y=x}+\nu_\gamma(x)\mathfrak{D}^{\beta_1}_y\mathcal{R}^b_{\mu_2}(y,x)\Big|_{y=x}\Big]\\
        &+\frac{1}{2}\mathsf{G}^{\mu_1\mu_2}_{ab}(k,0,0)^{\mm\nn}_{\lambda\rho}\Big[\nu^2(x)(\mathfrak{D}^\gamma_y\mathcal{R}^a_{\mu_1}(y,x)\mathfrak{D}_\gamma^y\mathcal{R}^b_{\mu_2}(y,x))\Big|_{y=x}+\\
        &+\nu(x)\cdot\mathfrak{D}^y\mathcal{R}^a_{\mu_1}(y,x)\Big|_{y=x}\nu(x)\cdot\mathfrak{D}^y\mathcal{R}^b_{\mu_2}(y,x)\Big|_{y=x}\Big]\Bigg\}+\mathcal{O}(\partial^3)\;.
    \end{split}
\end{equation}
For the terms involving the spin potential the power counting is different, since $\mathcal{R}^{\mathfrak{S}}_\mu(x,x)\ne0$ is already of order $\partial$. From \eqref{Table: B table} and \eqref{def: R e nu maintext}:
\begin{equation}\label{def: R spin}
    \mathcal{R}^{\mathfrak{S}}_{\mu}(x,x)=\frac{1}{2}\nu_\mu(x)\mathfrak{S}_{\sigma\gamma}(x)\;,
\end{equation}
where the indices $\sigma\gamma$ are contracted with those of the spin tensor vertex in the kernel, as in \eqref{eq: W linear spin potential}, and will be understood in the following. A term $(N,M,n,m)$ with $a=\mathfrak{S}$ and $b\ne\mathfrak{S}$ is of order $j_1+j_2+1$, with $j_1=N+m$ and $j_2=M+n$. Since the factor with $\mathcal{R}^b$ requires $j_2\ge1$, at second order we need $j_1=0$ and $j_2=1$, that is $(N,M,n,m)=(0,1,0,0)$ and $(0,0,1,0)$. Similarly, for $a\ne\mathfrak{S}$ and $b=\mathfrak{S}$ the contributing terms are $(1,0,0,0)$ and $(0,0,0,1)$. Using:
\begin{equation*}
    \mathsf{D}_\gamma(y)\left[\tau\,\mathcal{R}^{\mathfrak{S}}_{\mu}(y,x)\right]\Big|_{\substack{y=x\\
    \tau=0}}=\nu_\gamma(x)\,\mathcal{R}^{\mathfrak{S}}_{\mu}(x,x)\;,
\end{equation*}
and the coefficients of \eqref{eq: QUAD FINAL}, the mixed contribution reads:
\begin{equation}\label{eq: Delta quad mix}
    \begin{split}
        \Delta^{\rm mix}_{\rm quad}W^{\mm\nn}_{\lambda\rho}(x,k)&=-\ii(2\pi)^6\sum_{b\ne\mathfrak{S}}\Big\{\partial^{q_b}_{\beta_1}[\mathsf{G}^{\mu_1\mu_2}_{\mathcal{S}b}(k,q_a,q_b)^{\mm\nn}_{\lambda\rho}]\Big|_{\substack{q_a=0\\
        q_b=0}}\,\mathfrak{D}^{\beta_1}_y\mathcal{R}^b_{\mu_2}(y,x)\Big|_{y=x}\\
        &+\frac{1}{2}\mathsf{G}^{\mu_1\mu_2}_{\mathcal{S}b}(k,0,0)^{\mm\nn}_{\lambda\rho}\,\nu(x)\cdot\mathfrak{D}^y\mathcal{R}^b_{\mu_2}(y,x)\Big|_{y=x}\Big\}\,\mathcal{R}^{\mathfrak{S}}_{\mu_1}(x,x)\\
        &-\ii(2\pi)^6\sum_{a\ne\mathfrak{S}}\Big\{\partial^{q_a}_{\alpha_1}[\mathsf{G}^{\mu_1\mu_2}_{a\mathcal{S}}(k,q_a,q_b)^{\mm\nn}_{\lambda\rho}]\Big|_{\substack{q_a=0\\
        q_b=0}}\,\mathfrak{D}^{\alpha_1}_y\mathcal{R}^a_{\mu_1}(y,x)\Big|_{y=x}\\
        &-\frac{1}{2}\mathsf{G}^{\mu_1\mu_2}_{a\mathcal{S}}(k,0,0)^{\mm\nn}_{\lambda\rho}\,\nu(x)\cdot\mathfrak{D}^y\mathcal{R}^a_{\mu_1}(y,x)\Big|_{y=x}\Big\}\,\mathcal{R}^{\mathfrak{S}}_{\mu_2}(x,x)+\mathcal{O}(\partial^3)\;.
    \end{split}
\end{equation}
Finally, the term with two spin potentials is already of second order for $(N,M,n,m)=(0,0,0,0)$:
\begin{equation}\label{eq: Delta quad SS}
    \Delta^{\mathfrak{S}}_{\rm quad}W^{\mm\nn}_{\lambda\rho}(x,k)=(2\pi)^6\,\mathsf{G}^{\mu_1\mu_2}_{\mathcal{S}\mathcal{S}}(k,0,0)^{\mm\nn}_{\lambda\rho}\,\mathcal{R}^{\mathfrak{S}}_{\mu_1}(x,x)\,\mathcal{R}^{\mathfrak{S}}_{\mu_2}(x,x)+\mathcal{O}(\partial^3)\;.
\end{equation}
Also the terms \eqref{eq: Delta quad mix} and \eqref{eq: Delta quad SS} do not contain the extrinsic curvature of the hypersurface. When the spin tensor is absent, as in the Belinfante pseudo-gauge or for the pions of section \ref{SEC: pions}, only the first term of \eqref{eq: Delta quad split} survives and $\Delta^{(2)}_{\rm quad}W=\Delta^{\rm con}_{\rm quad}W$ is given by \eqref{eq: Delta quad y=x}.

%


\section{Application: Dirac field spin polarization vector}\label{SEC: Dirac}
As a first application of the newly derived formula \eqref{eq: QUAD FINAL} we compute the spin polarization vector of a $\Lambda$ hyperon emitted at the decoupling up to the quadratic order in the gradients. A first derivation of the quadratic induced corrections was performed in \cite{Sheng:2024pbw} for the particular case of a flat decoupling hypersurface. It was there shown that the quadratic corrections exactly vanish if the decoupling hypersurface is the flat hyperplane with constant normal vector $\hat{t}_\mu=(1,0,0,0)$ passing through $x$. However for an \emph{average} hypersurface these corrections may be present. We show here that this is not the case and that once the geometry of the hypersurface is retained exactly all the quadratic corrections to the spin polarization vector are bound to vanish.

An hyperon $\Lambda$ has zero isospin hence the associated Dirac field  is $\wpsi_A(x)$, $A=1,2,3,4$ being the spinorial index. 
The covariant Wigner function \eqref{def: Wig op} is a $4\times 4$ complex matrix in the spinorial indices and can be decomposed on the basis of the 16 independent generators of the Clifford algebra:
\begin{equation}\label{eq: DirWig exp}
\begin{split}
    W_{AB}(x,k)=&\frac{1}{4}\bigg[\mathcal{F}(x,k)\mathbb{I}_{AB}+\mathcal{P}(x,k)(\gamma^5)_{AB}+\mathcal{V}^\sigma(x,k)(\gamma_\sigma)_{AB}+\\
    &\mathcal{A}^\sigma(x,k)(\gamma^5\gamma_\sigma)_{AB}+\mathcal{T}^{\sigma\lambda}(x,k)(\sigma_{\sigma\lambda})_{AB}\bigg]\;,
\end{split}
\end{equation}
where $\sigma_{\sigma\lambda}=(\ii/4)[\gamma_\sigma,\gamma_\lambda]$. Each component of the Dirac Wigner function \eqref{eq: DirWig exp} can be obtained by taking opportune traces with the elements of the Clifford algebra. In particular the scalar $\mathcal{F}(x,k)$ and axial part $\mathcal{A}^\sigma(x,k)$ are defined as:
\begin{subequations}\label{def: F e A}
    \begin{align}
        \mathcal{F}(x,k)&=\tr\left[W(x,k)\right]\;,\\
        \mathcal{A}^\sigma(x,k)&=\tr\left[\gamma^\sigma\gamma^5\,W(x,k)\right]\;,
    \end{align}
\end{subequations}
where $\tr[\cdot]$ denotes the trace over the spinorial indexes $A,B$.

The spin polarization vector $S^\mu(k)$ of a Dirac fermion at local equilibrium on the hypersurface $\Sigma_{\rm D}$ can be expressed in terms of the two components of the Wigner function obtained by taking the expectation values of  \eqref{def: F e A} in the corresponding quantum state \cite{Becattini:2020sww}:
\begin{equation}\label{def: S LTE}
    S^\sigma_{\rm LTE}(k)=\frac{1}{2}\frac{\int_{\Sigma_{\rm D}}\di\Sigma(x)\cdot k\,\langle\widehat{\mathcal{A}}^\sigma(x,k)\rangle_{\rm LTE}}{\int_{\Sigma_{\rm D}}\di\Sigma(x)\cdot k\,\langle\widehat{\mathcal{F}}(x,k)\rangle_{\rm LTE}}\equiv\frac{1}{2}\frac{N^\sigma(k)}{D(k)}\;.
\end{equation}
Both the axial and scalar part of the Wigner function at local thermodynamic equilibrium can be expanded in gradients:
\begin{equation}\label{eq: quadratic F e A}
    \begin{split}
    \mathcal{A}^\sigma_{\rm LTE}(x,k)&=\mathcal{A}^\sigma_{\rm GE}(x,k)+\Delta^{(1)}\mathcal{A}^\sigma(x,k)+\Delta^{(2)}\mathcal{A}^\sigma(x,k)\;,\\
        \mathcal{F}_{\rm LTE}(x,k)&=\mathcal{F}_{\rm GE}(x,k)+\Delta^{(1)}\mathcal{F}(x,k)+\Delta^{(2)}\mathcal{F}(x,k)\;,
    \end{split}
\end{equation}
where each term $\Delta^{(1,2)}$ contains exactly 1 and 2, respectively, gradients of the thermodynamic fields and curvature.
In order to compute the quadratic response expression of the spin polarization vector we thus need to compute \eqref{eq: quadratic F e A}.

The global equilibrium Wigner function for a general field has been computed in \eqref{eq: W GE}. The spin polarization of the particles is obtained from the positive frequency part of the Wigner operator $\wW^+=\theta(k^0)\theta(k^2)\wW$, the negative frequency part describing the antiparticles. For an iso-singlet free Dirac field its global equilibrium expectation value reads:
\begin{equation}\label{pettini}
    \langle\wW^+_{AB}(x,k)\rangle_{\rm GE}=\frac{\theta(k^0)\theta(k^2)}{(2\pi)^3}\delta(k^2-m^2)n_{\rm F}(k)(k_\sigma\gamma^\sigma+m\mathbb{I})_{AB}\;.
\end{equation}
From the properties of the Dirac gamma matrices we got:
\begin{equation*}
    \tr(k_\sigma\gamma^\sigma+m\mathbb{I})=4m\;,\qquad  \tr\left[\gamma^\sigma\gamma^5(k_\sigma\gamma^\sigma+m\mathbb{I})\right]=0\;,
\end{equation*}
The computation of the scalar and axial part from \eqref{pettini} is then straightforward and reads:
\begin{equation}\label{eq: F e A GE}
    \mathcal{F}_{\rm GE}(x,k)=\frac{4m}{(2\pi)^3}\theta(k^0)\theta(k^2)\delta(k^2-m^2)n_{\rm F}(k)\;,\qquad\mathcal{A}^\sigma_{\rm GE}(x,k)=0\;.
\end{equation}
By plugging \eqref{eq: quadratic F e A} in \eqref{def: S LTE} and taking into account that, according to \eqref{eq: F e A GE}, $\langle N^\sigma(k)\rangle_{\rm GE}=0$:
\begin{equation*}
    \begin{split}
        S^\sigma(k)&\simeq\frac{1}{2}\frac{ N^\sigma_{(1)}(k)+N^\sigma_{(2)}(k)}{D_{(0)}(k)+ D_{(1)}(k)+D_{(2)}(k)}\\
        &\simeq\frac{1}{2}\frac{N^\sigma_{(1)}(k)}{D_{(0)}(k)}+\frac{1}{2}\frac{N^\sigma_{(2)}(k)}{D_{(0)}(k)}-\frac{1}{2} \frac{N^\sigma_{(1)}(k)}{D_{(0)}(k)}\frac{ D_{(1)}(k)}{D_{(0)}(k)}\;.
    \end{split}
\end{equation*}
Hence the spin polarization vector can be expanded in gradients as follows:
\begin{equation*}
    S^\sigma(k)=S^\sigma_{(1)}(k)+S^\sigma_{(2)}(k)\;,
\end{equation*}
where the linear and quadratic terms are explicitly:
\begin{subequations}\label{eq: S1eS2}
    \begin{align}
        S^\sigma_{(1)}(k)&=\frac{1}{2}\frac{\int_{\Sigma_{\rm D}}\di\Sigma(x)\cdot k\,\tr\left[\gamma^\sigma\gamma^5\,\Delta^{N=1}_{\rm lin}W^+(x,k)\right]}{\int_{\Sigma_{\rm D}}\di\Sigma(x)\cdot k\,\tr\left[\langle\wW^+(x,k)\rangle_{\rm GE}\right]}\;,\label{def: S1}\\
        S^\sigma_{(2)}(k)&=\frac{1}{2}\frac{\int_{\Sigma_{\rm D}}\di\Sigma(x)\cdot k\,\tr\left\{\gamma^\sigma\gamma^5\left[\Delta^{N=2}_{\rm lin}W^+(x,k)+\Delta^{(N,M)=1}_{\rm quad}W^+(x,k)\right]\right\}}{\int_{\Sigma_{\rm D}}\di\Sigma(x)\cdot k\,\tr\left[\langle\wW^+(x,k)\rangle_{\rm GE}\right]}\nonumber\\
        &-S^\sigma_{(1)}(k)\frac{\int_{\Sigma_{\rm D}}\di\Sigma(x)\cdot k\,\tr\left[\,\Delta^{N=1}_{\rm lin}W^+(x,k)\right]}{\int_{\Sigma_{\rm D}}\di\Sigma(x)\cdot k\,\tr\left[\langle\wW^+(x,k)\rangle_{\rm GE}\right]}\label{def: S2}\;.
    \end{align}
\end{subequations}
We will consider the Belinfante and canonical pseudo-gauges separately.

\subsection{Belinfante pseudo-gauge}
In order to compute the linear response of the axial and scalar part of the Wigner function we need to compute the corresponding traces of the functions \eqref{def: G lin} for the case of a Dirac field:
\begin{subequations}\label{eq: Dirac Glin explicit}
    \begin{align}
        G^{\mu\nu}_T(k,q)_{AB}&=\frac{\delta(k^2-m^2+\frac{q^2}{4})}{2(2\pi)^6}\theta(k^0)\theta(k^2)\frac{n_{\rm F}(k_-)-n_{\rm F}(k_+)}{\beta\cdot q}[\Pi(k_-)\mathsf{f}^{\mu\nu}_T(k,q)\Pi(k_+)]_{AB},\\
         G^{\mu}_j(k,q)_{AB}&=\frac{\delta(k^2-m^2+\frac{q^2}{4})}{2(2\pi)^6}\theta(k^0)\theta(k^2)\frac{n_{\rm F}(k_-)-n_{\rm F}(k_+)}{\beta\cdot q}[\Pi(k_-)\mathsf{f}^{\mu}_j(k,q)\Pi(k_+)]_{AB}\;,
    \end{align}
\end{subequations}
where we used $\Lambda_1(\beta\cdot q)n_F(k_+)[1-n_F(k_-)] = [n_F(k_-)-n_F(k_+)]/\beta\cdot q$. The two vertices in the Belinfante pseudo-gauge, evaluated at $p = k_-$ and $p'=k_+$, are explicitly given by:
\begin{equation*}
    \mathsf{f}^{\mu\nu}_T(k,q)=\frac{1}{2}\left(k^\nu\gamma^\mu+k^\mu\gamma^\nu\right)\;,\qquad\mathsf{f}^\mu_j(k,q)=\gamma^\mu\;,
\end{equation*}
and $\Pi(k_\pm)=(\slashed{k}_\pm+m\mathbb{I})$. In order to compute the spin polarization vector we are interested in the axial and scalar part (see appendix \ref{SEC: app dirac}):
\begin{equation}\label{eq: GT e Gj scalar}
   \begin{split}
        \tr[G^{\mu\nu}_T(k,q)]&=\frac{4m\delta(k^2-m^2+\frac{q^2}{4})}{(2\pi)^6}\theta(k^0)\theta(k^2)\frac{n_{\rm F}(k_-)-n_{\rm F}(k_+)}{\beta\cdot q}k^\mu k^\nu\;,\\
        \tr[G^{\mu}_j(k,q)]&=\frac{4m\delta(k^2-m^2+\frac{q^2}{4})}{(2\pi)^6}\theta(k^0)\theta(k^2)\frac{n_{\rm F}(k_-)-n_{\rm F}(k_+)}{\beta\cdot q}k^\mu \;,
   \end{split}
\end{equation}
while for the axial one we obtain:
\begin{equation}\label{eq: GT e Gj axial}
   \begin{split}
        \tr[\gamma^\sigma\gamma^5G^{\mu\nu}_T(k,q)]&=-\ii\frac{\delta(k^2-m^2+\frac{q^2}{4})}{(2\pi)^6}\theta(k^0)\theta(k^2)\frac{n_{\rm F}(k_-)-n_{\rm F}(k_+)}{\beta\cdot q}k_\alpha q_\beta\left(k^\nu\epsilon^{\sigma\mu\alpha\beta}+k^\mu\epsilon^{\sigma\nu\alpha\beta}\right)\;,\\
       \tr[\gamma^\sigma\gamma^5 G^\mu_j(k,q)]&=-2\ii\frac{\delta(k^2-m^2+\frac{q^2}{4})}{(2\pi)^6}\theta(k^0)\theta(k^2)\frac{n_{\rm F}(k_-)-n_{\rm F}(k_+)}{\beta\cdot q}k_\alpha q_\beta\epsilon^{\sigma\mu\alpha\beta}  \;.
   \end{split}
\end{equation}
To derive the linear response expansion up to the second order we are thus interested in computing the Taylor expansion in $q$ of the functions \eqref{eq: GT e Gj scalar} and \eqref{eq: GT e Gj axial}. The common statistical pre-factor can be expanded as follows:
\begin{equation}\label{eq: espansione q delta}
    \begin{split}
        \delta\left(k^2-m^2+\frac{q^2}{4}\right)&=\delta(k^2-m^2)+\frac{q^2}{4}\delta'(k^2-m^2)+\mathcal{O}(q^4)\;,\\
        \frac{n_{\rm F}(k_-)-n_{\rm F}(k_+)}{\beta\cdot q}&=n_{\rm F}(k)[1-n_{\rm F}(k)]\left[1+\frac{1-6n_{\rm F}(k)+6n^2_{\rm F}(k)}{24}(\beta\cdot q)^2\right]+\mathcal{O}(q^4)\;,
    \end{split}
\end{equation}
where $\delta'(k^2-m^2)$ is to be intended as a distribution in $k$ obtained by deriving the $\delta$ with respect to its own argument. Combining this with \eqref{eq: GT e Gj scalar} and \eqref{eq: GT e Gj axial} it follows that the scalar part is \emph{even} in $q$ while the axial part is \emph{odd} in $q$. This allows to obtain that the first order term in the scalar part vanishes altogether.
\begin{equation}\label{Delta F 1=0}
    \begin{split}
        \Delta^{N=1}_{\rm lin}\mathcal{F}(x,k)&=0\;,
    \end{split}
\end{equation}
with a generally non-vanishing second order term $\Delta^{N=2}_{\rm lin}\mathcal{F}(x,k)$, which however does not enter in the second-order spin polarization \eqref{eq: S1eS2}.

On the other hand, for the axial part being odd in $q$, we obtain that the second order is vanishing:
\begin{equation}\label{eq: Delta lin A2=0}
    \Delta^{N=2}_{\rm lin}\mathcal{A}^\sigma=0\;,
\end{equation}
while the first order is non-zero:
\begin{equation*}
    \begin{split}
        \Delta^{N=1}_{\rm lin}\mathcal{A}^\sigma_+(x,k)=\,&\frac{\delta(k^2-m^2)}{(2\pi)^3}\theta(k^0)\theta(k^2)n_{\rm F}(k)[1-n_{\rm F}(k)]k_\beta\,\times\\
        &\times \left[(k^\nu\epsilon^{\sigma\mu\beta}_{\quad\alpha}+k^\mu\epsilon^{\sigma\nu\beta}_{\quad\alpha})\mathfrak{D}^\alpha_y\mathcal{R}^{\Delta\beta}_{\mu\nu}(y,x)\Big|_{y=x}\!\!\!-2\epsilon^{\sigma\mu\beta}_{\quad\alpha}\mathfrak{D}^\alpha_y\mathcal{R}^{\Delta\zeta}_\mu(y,x)\Big|_{y=x}\right]\;.
    \end{split}
\end{equation*}
The axial part can be worked out by using \eqref{eq: DR 1} so that the term in the square brackets becomes:
\begin{equation*}
    \begin{split}
        &k_\beta\left[(k^\nu\epsilon^{\sigma\mu\beta}_{\quad\alpha}+k^\mu\epsilon^{\sigma\nu\beta}_{\quad\alpha})\mathfrak{D}^\alpha_y\mathcal{R}^{\Delta\beta}_{\mu\nu}(y,x)\Big|_{y=x}-2\epsilon^{\sigma\mu\beta}_{\quad\alpha}\mathfrak{D}^\alpha_y\mathcal{R}^{\Delta\zeta}_\mu(y,x)\Big|_{y=x}\right]\\
        &=\epsilon^{\sigma\lambda\beta\alpha}k_\beta\left[+\left(k^\nu g^{\mu}_\lambda+k^\mu g^\nu_\lambda\right)\nu_\mu(x)\mathcal{P}^{\kappa}_{\,\alpha}(x)\partial_\kappa\beta_\nu(x)-2\nu_\mu(x)\mathcal{P}^{\kappa}_{\,\alpha}(x)\partial_\kappa\zeta(x)\right]\\
        &=\epsilon^{\sigma\lambda\beta\alpha}k_\beta\mathcal{P}^{\kappa}_{\,\alpha}\left[\partial_\kappa\beta_\lambda(x)+\nu_\lambda(x)\left(k^\gamma\partial_\kappa\beta_\gamma(x)-2\partial_\kappa\zeta(x)\right)\right]\;,
    \end{split}
\end{equation*}
where in the last step we used that $k^\mu\nu_\mu=1$. Using the definition of the projector \eqref{def: Dy} and taking into account that the contraction $\nu_\lambda\nu_\alpha\epsilon^{\sigma\lambda\beta\alpha}=0$ due to the antisymmetry of the Levi-Civita symbol we obtain:
\begin{equation*}
    \begin{split}
        \Delta^{N=1}_{\rm lin}\mathcal{A}^\sigma_+(x,k)&=\frac{\delta(k^2-m^2)}{(2\pi)^3}\theta(k^0)\theta(k^2)n_{\rm F}(k)[1-n_{\rm F}(k)]\epsilon^{\sigma\lambda\beta\alpha}k_\beta\\
        &\times\left[\partial_\alpha\beta_\lambda(x)-\nu_\alpha(x)k^\gamma\partial_\gamma\beta_\lambda(x)+\nu_\lambda(x)k^\gamma\partial_\alpha\beta_\gamma(x)-2\nu_\lambda(x)\partial_\alpha\zeta(x)\right]\;.
    \end{split}
\end{equation*}
Introducing the \emph{thermal vorticity} $\varpi$ and \emph{thermal shear} $\xi$:
\begin{equation}\label{def: varpi xi}
    \varpi_{\alpha\lambda}(x)\equiv-\frac{1}{2}\left(\partial_\alpha\beta_\lambda(x)-\partial_\lambda\beta_\alpha(x)\right)\;,\qquad \xi_{\lambda\alpha}(x)=\frac{1}{2}\left(\partial_\alpha\beta_\lambda(x)+\partial_\lambda\beta_\alpha(x)\right)\;,
\end{equation}
we have:
\begin{equation*}
    \begin{split}
        \epsilon^{\sigma\lambda\beta\alpha}\left(-\nu_\alpha(x)k^\gamma\partial_\gamma\beta_\lambda(x)+\nu_\lambda(x)k^\gamma\partial_\alpha\beta_\gamma(x)\right)&=2\epsilon^{\sigma\lambda\beta\alpha}\nu_\lambda(x)\xi_{\alpha\gamma}(x)k^\gamma\;,\\
        \epsilon^{\sigma\lambda\beta\alpha}\partial_\alpha\beta_\lambda(x)&=\epsilon^{\sigma\lambda\beta\alpha}\varpi_{\lambda\alpha}(x)\;,
    \end{split}
\end{equation*}
so that the axial part at first order in the gradients is finally given by:
\begin{equation}\label{eq: Delta A lin 1 FINAL}
    \begin{split}
        \Delta^{N=1}_{\rm lin}\mathcal{A}^\sigma_+(x,k)&=-\frac{\delta(k^2-m^2)}{(2\pi)^3}\theta(k^0)\theta(k^2)n_{\rm F}(k)(1-n_{\rm F}(k))\\
        &\times\epsilon^{\sigma\lambda\beta\alpha}k_\lambda\Big[\varpi_{\beta\alpha}(x)+2\nu_\beta(x)\left(\xi_{\alpha\gamma}(x)k^\gamma-\partial_\alpha\zeta(x)\right)\Big]\;,
    \end{split}
\end{equation}
which is exactly the result obtained in \cite{Sheng:2025cjk} for the particular case of trivial intersections $\overline{y}_{i,k}(x)=x$ only.

For the quadratic response the general microscopic kernel reads:
\begin{equation}\label{def: G quad Dirac}
    \left[\mathsf{G}^{\mu\nu}_{ab}(k,q_a,q_b)\right]\equiv\frac{1}{4(2\pi)^9}\,\delta\!\left(k^2-m^2+\frac{q^2}{4}\right)\mathcal{N}(k_\pm,\ell)\left[\Pi(k_-)\,\mathsf{f}^{\mu,\nn\mrr}_a(k_-,\ell)\,\Pi(\ell)\,\mathsf{f}^{\nu,\mrr\mm}_b(\ell,k_+)\,\Pi(k_+)\right]_{\rho\lambda}\;,
\end{equation}
where $\mathcal{N}$ is the statistical factor\footnote{Note that the combination of sign functions reduces to +1 after taking the particle contribution.}:
\begin{equation}\label{def: Sstat Dirac}
    \mathcal{N}(k_+,k_-,\ell)\equiv n_{\rm F}(k_+)\left[1-n_{\rm F}(k_-)\right]\Big[\Lambda_2(z_b,z_a)-n_{\rm F}(\ell)\,\Lambda_1(z_a)\Lambda_1(z_b)\Big]{\rm sgn}(k_-^0)\,{\rm sgn}(\ell^0)\,{\rm sgn}(k_+^0)\;,
\end{equation}
and where $q=q_a+q_b$, $k_\pm=k\pm q/2$, $\ell=k+(q_a-q_b)/2$ and $z_{a,b}$ are given in \eqref{eq: z quad}. As shown in appendix \ref{APP: wick} the statistical factor \eqref{def: Sstat Dirac} possesses the following exchange symmetry:
\begin{equation}\label{eq: Nswap}
    \mathcal{N}(k,q_a,q_b)=\mathcal{N}(k,-q_b,-q_a)\;.
\end{equation}
In order to compute the quadratic correction to the spin polarization vector we only need to compute the axial part of the microscopic kernel \eqref{def: G quad Dirac}, namely:
\begin{equation}\label{def: ST}
    \tr\left[\gamma^\sigma\gamma^5\mathsf{G}^{\mu_1\mu_2}_{ab}(k,q_a,q_b)\right]\equiv\mathcal{G}^{\mu_1\mu_2,\sigma}_{ab}(k,q_a,q_b)\equiv\mathsf{S}(k,q_a,q_b)\mathcal{T}^{\mu_1\mu_2,\sigma}_{ab}(k,q_a,q_b)\;,
\end{equation}
where the statistical factor $\mathsf{S}$ is common between the four-different possible operator combinations and reads:
\begin{equation}\label{def: S}
    \mathsf{S}(k,q_a,q_b)\equiv\frac{1}{4(2\pi)^9}\delta\left[k^2-m^2+\frac{(q_a+q_b)^2}{4}\right]\mathcal{N}(k,q_a,q_b)\;,
\end{equation}
while $\mathcal{T}_{ab}$ represent the generic trace of the double vertices. 
The computation of all the traces necessary to compute the axial part of \eqref{def: ST} are extremely involved and are reported explicitly in appendix \ref{SEC: app dirac}. However it is not necessary to compute them in order to prove that the overall contribution to the axial part of the Wigner function is vanishing. Indeed the result can be obtained from a symmetry argument discussed in detail in appendix \ref{APP: computation G}.

Starting from the general 6 field expectation value and the fundamental axial trace follows:
\begin{equation}\label{eq: Tswap}
    \mathcal{T}^{\mu_1\mu_2,\sigma}_{ab}(q_a,q_b)=-c_ac_b\mathcal{T}^{\mu_2\mu_1,\sigma}_{ba}(-q_b,-q_a)\;,
\end{equation}
where $c_{a,b}=\pm1$ depending on the local vertex. The origin of the above relation is related with the transformation properties of the vertices under the composite transformation $\gamma^5\mathsf{C}$, $\mathsf{C}$ being the charge conjugation, which is equivalent to the inversion of the chain $q_a\to q_b$ to $-q_b\to q_a$. For free fields each vertex has a definite transformation rule (see appendix \ref{APP: computation G} for the proof):
\begin{equation}\label{eq: vertex inversion}
    \widetilde{f}(p,p')\equiv\gamma^5\mathsf{C}\,[\mathsf{f}_a(p',p)]^T\,\mathsf{C}^{-1}\gamma^5\,=-c_a\mathsf{f}_a(p,p')\;,\qquad c_a=\pm1\;.
\end{equation}
Combining \eqref{eq: Tswap} with \eqref{eq: Nswap} implies the following fundamental relation for \eqref{def: ST}:
\begin{equation}\label{eq: Gswap}
    \mathcal{G}^{\mu_1\mu_2,\sigma}_{ab}(q_a,q_b)=-c_ac_b\mathcal{G}^{\mu_2\mu_1,\sigma}_{ba}(-q_b,-q_a)\;.
\end{equation}
The above relation can be derived in $q_a, q_b$ leading to:
\begin{equation*}
    \partial^{q_a}_{\beta_1}\ldots\partial^{q_a}_{\beta_M}\partial^{q_b}_{\alpha_1}\ldots\partial^{q_b}_{\alpha_N}\mathcal{G}^{\mu_2\mu_1,\sigma}_{ba}\Big|_{0}=-c_ac_b(-1)^{N+M}\,\partial^{q_a}_{\alpha_1}\ldots\partial^{q_a}_{\alpha_N}\partial^{q_b}_{\beta_1}\ldots\partial^{q_b}_{\beta_M}\mathcal{G}^{\mu_1\mu_2,\sigma}_{ab}\Big|_{0}\;.
\end{equation*}
In the sum over the ordered pairs we can group the term $(N,M,n,m)$ of the pair $(a,b)$ with the term $(M,N,m,n)$ of the pair $(b,a)$. From the relation above, together with $c_{n,m} = (-1)^{n+m}c_{mn}$ (see equation \eqref{def: coeff c}), the latter is equal to the former multiplied by $-c_ac_b(-1)^{N+M+n+m}$; since the sum runs over all the pairs and all the values of $N,M,n,m$:
\begin{equation}\label{eq: Delta Aquad deriv}
    \Delta_{\rm quad}\mathcal{A}^\sigma=\frac12\sum_{a,b}\sum_{N,M,n,m}\left[1-c_ac_b(-1)^{N+M+n+m}\right]\Big(\text{term }(N,M,n,m)\text{ of }(a,b)\Big)\;.
\end{equation}
Now introducing the following quantities
\begin{subequations}\label{def: III}
    \begin{align}
        I^\sigma_{00}&\equiv\frac12\sum_{ab}\mathcal{G}^{\mu_1\mu_2,\sigma}_{ab}(k,0,0)\Big[\nu^2(x)\,\mathfrak{D}^\gamma\mathcal{R}^a_{\mu_1}(y,x)\Big|_{y=x}\!\!\!\!\!\mathfrak{D}_\gamma\mathcal{R}^b_{\mu_2}(y,x)\Big|_{y=x}\!\!\!\!\!+\\ \nonumber
        &\quad\,\,+\nu_\gamma(x)\mathfrak{D}^\gamma_y\mathcal{R}^a_{\mu_1}(y,x)\Big|_{y=x}\!\!\!\!\!\nu_\sigma(x)\mathfrak{D}^\sigma_y\mathcal{R}^b_{\mu_2}(y,x)\Big|_{y=x}\Big]\;,\\
         I^\sigma_{11}&\equiv-\sum_{ab}\partial^{\rho}_{q_a}\partial^\tau_{q_b}\mathcal{G}^{\mu_1\mu_2,\sigma}_{ab}(k,q_a,q_b)\Big|_{q_a=q_b=0}\mathfrak{D}^y_\rho\mathcal{R}^a_{\mu_1}(y,x)\Big|_{y=x}\mathfrak{D}^y_\tau\mathcal{R}^b_{\mu_2}(y,x)\Big|_{y=x}\;,\\
         I^\sigma_a&\equiv-\frac{1}{2}\sum_{ab}\partial^\rho_{q_a}\mathcal{G}^{\mu_1\mu_2,\sigma}_{ab}(k,q_a,q_b)\Big|_{q_a=q_b=0}\mathfrak{D}^\gamma_y\mathcal{R}^b_{\mu_2}(y,x)\Big|_{y=x}\Big[\nu_\rho(x)\mathfrak{D}_\gamma\mathcal{R}^a_{\mu_1}(y,x)\Big|_{y=x}\!\!\!\!\!\!+\\\nonumber
        &\quad \,\, +\nu_\gamma(x)\mathfrak{D}^y_\rho\mathcal{R}^a_{\mu_1}(y,x)\Big|_{y=x}\Big]\;,\\
        I^\sigma_b&\equiv\frac{1}{2}\sum_{ab}\partial^\tau_{q_b}\mathcal{G}^{\mu_1\mu_2,\sigma}_{ab}(k,q_a,q_b)\Big|_{q_a=q_b=0}\mathfrak{D}^\gamma_y\mathcal{R}^a_{\mu_1}(y,x)\Big|_{y=x}\Big[\nu_\tau(x)\mathfrak{D}^y_\gamma\mathcal{R}^b_{\mu_2}(y,x)\Big|_{y=x}\!\!\!\!\!\!+\\ \nonumber
        &\quad\,\,+\nu_\gamma(x)\mathfrak{D}^y_\tau\mathcal{R}^b_{\mu_2}(y,x)\Big|_{y=x}\Big]\;.
    \end{align}
\end{subequations}
The quadratic response correction of the axial part of the Wigner function can then be written as:
\begin{equation}\label{eq: Delta A III}
    \begin{split}
       \Delta^{(2)}_{\rm quad}\mathcal{A}^\sigma(x,k)=(2\pi)^6\left(I^\sigma_{11}+I^\sigma_a+I^\sigma_b+I^\sigma_{00}\right)\;.
    \end{split}
\end{equation}
Applying \eqref{eq: Delta Aquad deriv} to each term \eqref{def: III} in equation \eqref{eq: Delta A III} and using that:
\begin{equation*}
    c_T=c_j=-1\,,
\end{equation*}
follows that:
\begin{equation}\label{eq: cancellazione I}
    I^\sigma_{11}=-I^\sigma_{11}=0\;,\qquad I^\sigma_a=-I^\sigma_b\;,\qquad I^\sigma_{00}=-I^\sigma_{00}=0\;.
\end{equation}
Plugging the above identities in equation \eqref{eq: Delta A III} we then obtain:
\begin{equation}\label{eq: Delta A2=0}
    \Delta^{(2)}_{\rm quad}\mathcal{A}^\sigma(x,k)=0\;,
\end{equation}
that is the axial contribution of the quadratic response exactly vanishes at second order in the gradients.


\subsection{Canonical pseudo-gauge}\label{SEC: canonical}
We now consider the canonical pseudo-gauge. The currents are the same as in the Belinfante pseudo-gauge, while  the stress-energy tensor and spin tensor are reported in appendix \ref{SEC: app dirac}; the corresponding vertices are:
\begin{equation*}
    \mathsf{f}^{\mu\nu}_{T_{\rm C}}(p,p')=\frac12\gamma^\mu(p+p')^\nu\;,\qquad \mathsf{f}^{\mu,\lambda\rho}_{\mathcal{S}}(p,p')=-\frac12\epsilon^{\mu\lambda\rho\tau}\gamma_\tau\gamma^5\;.
\end{equation*}
The scalar part of the stress-energy tensor vertex coincides with the Belinfante one while the spin tensor vertex has vanishing trace so that it does not contribute to $\mathcal{F}$. We thus only need to compute the contribution to the axial part.
At linear order they are given by:
\begin{subequations}\label{eq: canonical axial f}
    \begin{align}
        \tr[\gamma^\sigma\gamma^5\Pi(k_-)\mathsf{f}^{\mu\nu}_{T_{\rm C}}(k,q)\Pi(k_+)]&=-4\ii\,k^\nu\epsilon^{\sigma\mu\alpha\beta}k_\alpha q_\beta\;,\label{eq: axial fTC}\\
        \tr[\gamma^\sigma\gamma^5\Pi(k_-)\mathsf{f}^{\mu,\lambda\rho}_{\mathcal{S}}(k,q)\Pi(k_+)]&=-2\epsilon^{\mu\lambda\rho\tau}\left[2k^\sigma k_\tau-\frac12q^\sigma q_\tau-\delta^\sigma_\tau\left(k^2+m^2-\frac{q^2}{4}\right)\right]\;.\label{eq: axial FS}
    \end{align}
\end{subequations}
The first-order correction to the axial part has two contributions. The first is the term $N=1$ of the stress-energy tensor and of the current, which is equivalent to the Belinfante case upon the replacement  $\left(k^\nu\epsilon^{\sigma\mu\alpha\beta}+k^\mu\epsilon^{\sigma\nu\alpha\beta}\right)\to2k^\nu\epsilon^{\sigma\mu\alpha\beta}$:
\begin{equation}\label{eq: A canonical T}
    \Delta^{N=1}_{\rm lin}\mathcal{A}^\sigma_{T_{\rm C},j}(x,k)=\frac{2\,\delta(k^2-m^2)}{(2\pi)^3}n_{\rm F}(k)(1-n_{\rm F}(k))\,\epsilon^{\sigma\lambda\beta\alpha}k_\beta\nu_\lambda\left(k^\gamma\partial_\alpha\beta_\gamma-\partial_\alpha\zeta\right)\;.
\end{equation}
 The second term instead originates from the spin tensor contribution \eqref{eq: axial FS}. Expanding in $q$ and using that for $N=1$ the momentum is on-shell: $k^2=m^2$; the linear order contribution to the axial part is:
\begin{equation}\label{eq: A canonical S}
    \Delta^{(1)}_{\rm lin}\mathcal{A}^\sigma_{\mathfrak{S}}(x,k)=-\frac{\delta(k^2-m^2)}{(2\pi)^3}n_{\rm F}(k)(1-n_{\rm F}(k))\,\epsilon^{\mu\lambda\rho\tau}\nu_\mu(x)\left(k^\sigma k_\tau-m^2\delta^\sigma_\tau\right)\mathfrak{S}_{\lambda\rho}(x)\;.
\end{equation}
We can now compare the total axial component in the canonical pseudo-gauge given by the sum of \eqref{eq: A canonical T} and \eqref{eq: A canonical S} with the corresponding term in the  Belinfante pseudo-gauge \eqref{eq: Delta A lin 1 FINAL}.

Using $\partial_\alpha\beta_\lambda-\partial_\lambda\beta_\alpha=-2\varpi_{\alpha\lambda}$, the difference between the stress-tensor canonical contribution \eqref{eq: A canonical T} and the Belinfante expression \eqref{eq: Delta A lin 1 FINAL} is proportional to:
\begin{equation}\label{eq: difference CB}
    \epsilon^{\sigma\lambda\beta\alpha}k_\beta\left[\nu_\lambda k^\gamma\partial_\alpha\beta_\gamma-\partial_\alpha\beta_\lambda+\nu_\alpha k^\gamma\partial_\gamma\beta_\lambda\right]=-\epsilon^{\sigma\lambda\beta\alpha}k_\beta\left[\varpi_{\lambda\alpha}+2\nu_\lambda k^\gamma\varpi_{\alpha\gamma}\right]\;,
\end{equation}
where the thermal shear (and also the gradient of $\zeta$ which is independent of the pseudo-gauge) cancel in the difference. The right hand side can be rewritten by means of the Schouten identity:
\begin{equation}\label{eq: Schouten}
    \delta^\sigma_\kappa\epsilon^{\mu\lambda\rho\tau}=\delta^\mu_\kappa\epsilon^{\sigma\lambda\rho\tau}+\delta^\lambda_\kappa\epsilon^{\mu\sigma\rho\tau}+\delta^\rho_\kappa\epsilon^{\mu\lambda\sigma\tau}+\delta^\tau_\kappa\epsilon^{\mu\lambda\rho\sigma}\;.
\end{equation}
Contracting \eqref{eq: Schouten} with $k^\kappa k_\tau\nu_\mu\varpi_{\lambda\rho}$ and using $k\cdot\nu=1$, $k^2=m^2$ and the antisymmetry of $\varpi$ the contribution from the spin tensor in \eqref{eq: A canonical S} can be written as:
\begin{equation*}
    \epsilon^{\mu\lambda\rho\tau}\nu_\mu\left(k^\sigma k_\tau-m^2\delta^\sigma_\tau\right)\varpi_{\lambda\rho}=\epsilon^{\sigma\lambda\rho\tau}k_\tau\varpi_{\lambda\rho}+2\epsilon^{\mu\sigma\rho\tau}\nu_\mu k_\tau k^\lambda\varpi_{\lambda\rho}\;,
\end{equation*}
which, after renaming the indices, is precisely the right hand side of \eqref{eq: difference CB}. Therefore:
\begin{equation}\label{eq: A canonical}
\begin{split}
    \Delta_{(1)}\mathcal{A}^\sigma_{\rm C}(x,k)=&\Delta_{(1)}\mathcal{A}^\sigma_{\rm B}(x,k)-\frac{\delta(k^2-m^2)}{(2\pi)^3}n_{\rm F}(k)(1-n_{\rm F}(k))\,\times\\
    &\times\epsilon^{\mu\lambda\rho\tau}\nu_\mu(x)\left(k^\sigma k_\tau-m^2\delta^\sigma_\tau\right)\left[\mathfrak{S}_{\lambda\rho}(x)-\varpi_{\lambda\rho}(x)\right]\;,
\end{split}
\end{equation}
where $\Delta_{(1)}\mathcal{A}^\sigma_{\rm B}$ is the Belinfante result \eqref{eq: Delta A lin 1 FINAL}.

For what concerns the quadratic correction, according to equations \eqref{eq: Delta quad mix} and \eqref{eq: Delta quad SS}, one may expect, in principle, contributions to the axial part of the Wigner function like $\partial\mathfrak{S}$ coming from the second order term of the linear response and contributions like  $\mathfrak{S}\varpi$, $\mathfrak{S}\xi$, $\mathfrak{S}\partial\zeta$ and $\mathfrak{S}\mathfrak{S}$ coming from the quadratic response terms. However the symmetry argument can extend to the case of a generic spin tensor and in particular equation \eqref{eq: Delta Aquad deriv} is still valid. 

On the trivial intersection the term $(N,M,n,m)$ is of order $N+M+n+m$ plus the number of spin potentials among the two sources.
Now being:
\begin{equation*}
    c_{T_{\rm C}}=c_j=-1\;,\qquad c_\mathcal{S}=+1\;,
\end{equation*}
each term $c_ac_b$ involving one spin potential $\mathfrak{S}$ is such that $c_ac_b=-1$. Hence only the terms of odd order survive. For the scalar part the sign is the opposite and only the even orders survive. In conclusion, for the free Dirac field coupled to $\beta$, $\zeta$ and $\mathfrak{S}$, in the Belinfante as well as in the canonical pseudo-gauge, within linear and quadratic response:
\begin{equation*}
    \begin{split}
        &\mathcal{A}^\sigma(x,k)\;\;\text{contains only odd orders in the gradients}\,,\\
        &\mathcal{F}(x,k)=\tr \left[W(x,k)\right]\;\;\text{contains only even orders in the gradients}\,,
    \end{split}
\end{equation*}
where $\mathfrak{S}$ and the derivatives of the normal vector count as one gradient. In particular, at second order there is no contribution to the axial part from the mixed terms \eqref{eq: Delta quad mix} $(N,M,n,m)=(0,1,0,0),(0,0,1,0),(1,0,0,0),(0,0,0,1)$ like $\mathfrak{S}\,\partial\beta$ and $\mathfrak{S}\,\partial\zeta$, from the term $(0,0,0,0)$ \eqref{eq: Delta quad SS}, like $\mathfrak{S}\mathfrak{S}$, nor of the type $\partial\mathfrak{S}$ or $\mathfrak{S}\,\partial\nu$ (second term in equation \eqref{eq: W linear spin potential}).
Overall we then conclude:
\begin{equation}\label{eq: vanishing Delta A2 canonical}
    \Delta^{(2)}_{\rm quad}\mathcal{A}^\sigma_{\rm C}(x,k)=0\;.
\end{equation}
Note that the above cancellation only requires that $c_\mathcal{S}=+1$ and thus extends to a much more general class of pseudo-gauges, not only the canonical one.
In particular, from \cite{Speranza:2020ilk, Buzzegoli:2021wlg} we can apply the transformation \eqref{eq: vertex inversion} to a vast class of possible spin tensor vertices. For instance for the two physically relevant pseudo-gauges: Hilgevoord–Wouthuysen (HW) and the de Groot–van Leeuwen–van Weert (GLW), the quadratic term also vanishes. It is possible, however, to build pseudo-gauges for which the condition \eqref{eq: vertex inversion} does not apply. For instance a pseudo-gauge where the spin density tensor is given by:
\begin{equation}\label{counterexample}
    \widehat{\mathcal{S}}^{\lambda,\mu\nu}=\widehat{\mathcal{S}}^{\lambda,\mu\nu}_{\rm C}-a\left(g^{\lambda\mu}\wj^\nu-g^{\lambda\nu}\wj^\mu\right)\;,
\end{equation}
with $\wj^\mu$ vector current would not satisfy the relation \eqref{eq: vertex inversion} given that $\widehat{\mathcal{S}}_{\rm C}$ and $\wj$ have different values of $c$ under the transformation \eqref{eq: vertex inversion}. Hence, in principle, a quadratic correction in $\mathfrak{S}$ may be present.

\subsection{Final result and discussion}\label{SEC: discussion spin pol}

Combining equations \eqref{eq: Delta A lin 1 FINAL}, \eqref{Delta F 1=0}, \eqref{eq: Delta lin A2=0}, \eqref{eq: Delta A2=0} and \eqref{eq: vanishing Delta A2 canonical} with equation \eqref{eq: S1eS2} we obtain that the spin polarization vector has no quadratic contributions in the gradients for a vast class of different pseudo-gauges:
\begin{equation*}
    S^\mu_{(2)}(k)=0\;.
\end{equation*}
Correspondingly we obtain that the spin-polarization vector in the Belinfante pseudo-gauge up to the second order in the gradients is given by:
\begin{equation}\label{eq: S belinfante}
    \boxed{\begin{split}
    S^\sigma_{\rm B}(k)=&-\frac{1}{8mN}\int_{\Sigma_{\rm D}}\!\!\!\!\!\di\Sigma(x)\cdot k\,n_{\rm F}(k)(1-n_{\rm F}(k))\epsilon^{\sigma\lambda\beta\alpha}k_\lambda\times\\
    &\times\Big[\varpi_{\beta\alpha}(x)+2\nu_\beta(x)\left(\xi_{\alpha\gamma}(x)k^\gamma-\partial_{\alpha}\zeta(x)\right)\Big]+\mathcal{O}(\partial^3)\;.
      \end{split}}
\end{equation}
where:
\begin{equation*}
    N=\theta(k^0)\theta(k^2)\int_{\Sigma_{\rm D}}\di\Sigma(x)\cdot k\,n_{\rm F}(k)\;,
\end{equation*}
is proportional to the leading order momentum spectrum of the particles.
As already observed in \cite{Sheng:2025cjk} the combination of thermal vorticity and thermal shear is such that only the components of the derivative along the hypersurface are present leading to a natural implementation of the isothermal prescription in the regime in which $\Sigma_{\rm D}:\, T(x)={\rm constant}$.

Similarly, using equation \eqref{eq: A canonical}, we obtain that the spin polarization vector in the canonical pseudo-gauge up to second order in the gradients reads:
\begin{equation}\label{eq: S canonical}
    \boxed{\begin{split}
    S^\sigma_{\rm C}(k)=&S^\sigma_{\rm B}(k)-\frac{1}{8mN}\int_{\Sigma_{\rm D}}\!\!\!\!\!\di\Sigma(x)\cdot k\,n_{\rm F}(k)(1-n_{\rm F}(k))\,\epsilon^{\mu\lambda\rho\tau}\nu_\mu(x)\times\\
    &\times\left(k^\sigma k_\tau-m^2\delta^\sigma_\tau\right)\left[\mathfrak{S}_{\lambda\rho}(x)-\varpi_{\lambda\rho}(x)\right]+\mathcal{O}(\partial^3)\;.
    \end{split}}
\end{equation}
The above expression coincides with the result obtained in \cite{Buzzegoli:2021wlg} upon the replacement of the constant time vector $\hat{t}_\mu=\delta^0_\mu$ with $\nu_\mu(x)=\hat{n}_\mu(x)/k\cdot\hat{n}(x)$.
At global thermodynamic equilibrium, given that $\mathfrak{S}\to\varpi$ equation \eqref{eq: S belinfante} and equation \eqref{eq: S canonical} coincide as expected.

Our result can be compared with the previous second order computation performed in \cite{Sheng:2024pbw}. There the integration over the decoupling hypersurface was performed assuming an \emph{average} flat hypersurface as first geometrical approximation. It was observed that, in the limiting case in which the aforementioned hypersurface coincided with the flat hyperplane at constant $\hat{t}_\mu=\delta^0_\mu$ the quadratic correction was bound to vanish altogether. It was however observed that for different hypersurfaces in principle a correction may be still present. In our case, where no approximation for the geometry was considered, we obtain that the quadratic correction has to be vanishing, regardless of which normal vector we choose. In this regard we agree with the first conclusion but we disagree with the second one.
Note also that a vanishing quadratic vorticity induced polarization is in agreement with the global equilibrium limit \cite{Palermo:2023cup}. 

The vanishing of the spin polarization at quadratic order is not limited to the canonical pseudo-gauge but extends to a vast class of pseudo-gauges of physical interest like HW and GLW. It is in principle possible to build pseudo-gauges which have a non-vanishing quadratic correction like \eqref{counterexample}. These statements refer to the usual construction of the local equilibrium operator, in which the stress-energy and spin tensors of a given pseudo-gauge are used in \eqref{def: rho}. It has been recently shown \cite{Becattini:2025twu,Hoyos:2026bxb} that it is possible to build a local equilibrium density operator which is invariant under pseudo-gauge transformations coinciding with the Belinfante one. On the other hand the appearance of a finite spin tensor in the local equilibrium state can be motivated by the underlying microscopic theory. For instance in \cite{Buzzegoli:2024mra} it is shown how a system of Dirac fields at local thermodynamic equilibrium in the canonical pseudo-gauge emerges from a precise microscopic interacting theory. 

The obtained formula can also be compared with exact local equilibrium solutions obtained for particular geometries. For longitudinal boost-invariant fluid, for which the decoupling hypersurface is an hyperbola at constant $\tau=(t^2-z^2)^{1/2}$, in the Belinfante pseudo-gauge, no polarization is produced \cite{Palermo:2026mwu}. Indeed the linear order term vanishes. Equation \eqref{eq: S belinfante} agrees with this observation. Introducing the space-time rapidity $\eta_s$, the four-temperature is $\beta^\mu = u^\mu/T(\tau)$ with $u^\mu = (\cosh \eta_s,0,0,\sinh \eta_s)$ and $\zeta = \zeta(\tau)$. Denoting with $z^\mu = (\sinh \eta_s,0,0,\cosh \eta_s)$ the unit vector of the rapidity direction, $\partial_\mu u_\nu = -z_\mu z_\nu/\tau$ we obtain:
\begin{equation*}
    \partial_\mu\beta_\nu = -\dfrac{1}{T\tau}z_\mu z_\nu + u_\mu u_\nu\dfrac{{\rm d}}{{\rm d}\tau}\dfrac{1}{T}\,,\qquad \partial_\mu\zeta = u_\mu\dfrac{{\rm d}\zeta}{{\rm d}\tau}
\end{equation*}
so that $\partial_\mu\beta_\nu$ is symmetric and $\varpi = 0$. On the hypersurface $\tau = \tau_{\rm D}$ the normal vector is $\hat n = u$ and $\nu_\beta = u_\beta/(k\cdot u)$ hence, due to the antisymmetry of the Levi-Civita symbol, equation \eqref{eq: S belinfante} reduces to:
\begin{equation*}
    S_{(1)}^\sigma(k) = \dfrac{1}{4m\,N}\dfrac{1}{T_{\rm D}\tau_{\rm D}}\int_{\Sigma_{\rm D}}\di\Sigma \,n_F(k)(1-n_F(k))(k\cdot z)\epsilon^{\sigma\lambda\beta\alpha}k_\lambda u_\beta z_\alpha \,.
\end{equation*}
The combination $u_\beta z_\alpha - u_\alpha z_\beta$ does not depend on $\eta_s$, whereas, with $k^\mu = (m_{\rm T}\cosh y,\mathbf{k}_{\rm T},m_{\rm T}\sinh y)$, $k\cdot u = m_{\rm T}\cosh(y-\eta_s)$ and $k\cdot z= -m_{\rm T}\sinh(y-\eta_s)$. Since $n_{\rm F}$ depends on $\eta_s$ only through $k\cdot u$ and $\di\Sigma = \tau_{\rm D}\di\eta_s\di^2x_{\rm T}$, the integrand is odd in $\eta_s-y$ and the integral over the space-time rapidity vanishes:
\begin{equation*}
    S_{(1)}^\sigma(k) = 0\;.
\end{equation*}
Note that the integrand does not vanish point by point: the thermal shear term gives no polarization only after the integration over the hypersurface. Together with $S^\sigma_{(2)}=0$,
this shows that the total spin polarization vector agrees with the prediction in \cite{Palermo:2026mwu} at least up to the second order in the gradients. In general the spin density induced polarization in \eqref{eq: S canonical} is non-vanishing.

We conclude by observing that our results have been obtained neglecting the contributions from the non-trivial intersections with the worldline \eqref{eq: worldline} and the time-like branches of the decoupling hypersurface. These contributions are not of immediate interpretation. They are non-local in the gradients and their contribution does not have a general definite sign, being $k\cdot\hat{n}(\bar y_k(x))$ not necessarily positive definite. As such they can also give \emph{negative} contributions to the spectra which is often regarded as a non-physical contribution \cite{Bugaev:1996zq,Anderlik:1998et,Grassi:2004dz, Huovinen:2012is}. It was also observed, at least at linear order \cite{Roselli:2025kbr}, that they are  canceled by the volume term coming from the dissipative term. Their physical origin is motivated by the free-streaming of a particle from $\bar y_k(x)$ to $x$. However in the plasma phase the assumption of free-streaming hadronic particle is not realistic. For all these reasons we assumed these terms to be negligible with respect to the local corrections coming from $\bar y_k(x)=x$.


\section{Application: momentum spectrum of pions}\label{SEC: pions}
As a second application of the general formulae \eqref{eq: QUAD FINAL} we compute the momentum spectrum of pions produced at the decoupling hypersurface up to second order in the gradients. The leading order contribution corresponds to the well known Cooper-Frye formula \cite{Cooper:1974mv}, for the pion with isospin projection $\mm$ and on-shell momentum $k$, $k^0=E_k=(m^2+{\bf k}^2)^{1/2}$:
\begin{equation}\label{eq: CF pions}
    \mathcal{Y}_\mm({\bf k})\equiv E_k\frac{\di N_\mm}{\di^3k}=\frac{1}{(2\pi)^3}\int_{\Sigma_{\rm D}}\di\Sigma(x)\cdot k\, n_{\rm B}(k,\mm)\;.
\end{equation}
The above expression is expected to receive corrections induced by the gradients of the thermodynamic fields and from the curvature of the decoupling hypersurface as well.

For free fields, taking into account that for a particle of the species $\mm$ the corresponding Wigner function is $\theta(k^0)\theta(k^2)W^{\mm\mm}$, it is possible to directly relate the spectra \eqref{eq: CF pions} with the covariant Wigner function:
\begin{equation}\label{eq: CF in terms of W}
    \mathcal{Y}_\mm({\bf k})=2E_k\int_0^{+\infty}\di k^0\int_{\Sigma_{\rm D}}\di\Sigma(x)\cdot k\, \langle\wW^{\mm\mm}(x,k)\rangle\;.
\end{equation}
At local equilibrium the leading order term of the Wigner function is given by \eqref{eq: W GE} which, for pions using \eqref{eq: free spectral} and \eqref{eq: Pi 0 1/2 1}, reduces as expected to the Cooper-Frye formula \eqref{eq: CF pions}. The local equilibrium corrections to the spectra are thus derived from those of the Wigner function. Note that, since $k^\mu W^{\mm\mm}$ is divergence-free for free fields, the integral in \eqref{eq: CF in terms of W} does not depend on the hypersurface for the exact Wigner function.

\subsection{Isospin structure and microscopic kernels}\label{SEC: pion kernels}

The pions are described by an iso-triplet (${\rm I}=1$) of real scalar fields $\wpi_i$, $i=1,2,3$, with Lagrangian density $\mathcal{L}=(1/2)(\partial_\mu\wpi_i\partial^\mu\wpi_i-m^2\wpi_i\wpi_i)$. In the charged basis $\wpi^\mm$, $\mm=+1,0,-1$, introduced in section \ref{SEC: Wigner function}, with $\wpi^{\mm\dagger}=\wpi^{-\mm}$, the stress-energy tensor and the iso-vector current read:
\begin{equation}\label{eq: pion currents}
    \begin{split}
        \wT^{\mu\nu}&=\frac12\sum_{\mm}\Big[\partial^\mu\wpi^{\mm\dagger}\partial^\nu\wpi^\mm+\partial^\nu\wpi^{\mm\dagger}\partial^\mu\wpi^\mm-g^{\mu\nu}\left(\partial_\alpha\wpi^{\mm\dagger}\partial^\alpha\wpi^\mm-m^2\wpi^{\mm\dagger}\wpi^\mm\right)\Big]\;,\\
        \widehat{\mathcal{V}}^\mu_c&=\frac{\ii}{2}\sum_{\mm,\nn}(\mathsf{t}_c)^{\mm\nn}\Big[\wpi^{\mm\dagger}\partial^\mu\wpi^\nn-\partial^\mu\wpi^{\mm\dagger}\wpi^\nn\Big]\;,\qquad c=1,2,3\;,
    \end{split}
\end{equation}
where $\mathsf{t}_c$ are the generators of the representation ${\rm I}=1$ in the charged basis:
\begin{equation}
    \mathsf{t}_1=\frac{1}{\sqrt{2}}\begin{pmatrix}
        0&-1&0\\
        -1&0&1\\
        0&1&0
    \end{pmatrix}\;,\qquad \mathsf{t}_2=\frac{1}{\sqrt{2}}\begin{pmatrix}
        0&\ii&0\\
        -\ii&0&-\ii\\
        0&\ii&0
    \end{pmatrix}\;,\qquad \mathsf{t}_3=\begin{pmatrix}
        1&0&0\\
        0&0&0\\
        0&0&-1
    \end{pmatrix}\;.
\end{equation}
They satisfy $[\mathsf{t}_a,\mathsf{t}_b]=\ii\epsilon_{abc}\mathsf{t}_c$ and $C\mathsf{t}^T_cC=-\mathsf{t}_c$, with $C^{\mm\nn}=\delta^{\mm,-\nn}$ the conjugation matrix introduced in section \ref{SEC: Wigner function}.
Using \eqref{def: FT field} in \eqref{eq: vertex}, the multiplet being self-conjugate, both vertices turn out to be equal to their flip \eqref{eq: flip}, so that the effective vertices to be used in the kernels are twice those obtained from \eqref{eq: pion currents}, that is the vertices of a complex multiplet (see appendix \ref{APP: wick}). With the two momenta on-shell they read:
\begin{equation}\label{eq: pion vertices}
    \mathsf{f}^{\mu\nu,\mm\nn}_T(p,p')=\delta^{\mm\nn}\left[p^\mu p'^\nu+p^\nu p'^\mu+\frac12g^{\mu\nu}(p-p')^2\right]\;,\qquad\mathsf{f}^{\mu,\mm\nn}_{\mathcal{V}_c}(p,p')=(\mathsf{t}_c)^{\mm\nn}(p+p')^\mu\;.
\end{equation}
From the above vertices, approximating ${\rm sgn}(k^0_+){\rm sgn}(k^0_-)=+1$\footnote{Note that the combination ${\rm sgn}(k^0_+){\rm sgn}(k^0_-)$, despite not being globally analytic, is constant and well defined around $q=0$ and equal to 1. We can thus ignore its derivatives.}, it follows that the two microscopic kernels appearing in linear response are given by:
\begin{equation}\label{eq: G lin scalar case}
    \begin{split}
        [G^{\mu\nu}_T(k,q)]^{\mm\nn}&=\frac{\delta^{\mm\nn}}{(2\pi)^6}\delta\left(k^2-m^2+\frac{q^2}{4}\right)L^{\mm\nn}(k,q)\left[k^\mu k^\nu-\frac{1}{4}\left(q^\mu q^\nu-q^2g^{\mu\nu}\right)\right]\;,\\
        [G^{\mu\nu}_{\mathcal{V}_c}(k,q)]^{\mm\nn}&=\frac{({\mathsf{t}}_c)^{\mm\nn}}{(2\pi)^6}\delta\left(k^2-m^2+\frac{q^2}{4}\right)L^{\mm\nn}(k,q)\,k^\mu\;,   
    \end{split}
\end{equation}
where we used that, according to equation \eqref{eq: KMfinal}, $k=(p+p')/2$ and $q=p-p'$ with $p',p=k_\pm$ and we introduced the common statistical pre-factor:
\begin{equation}\label{def: L factor}
    L^{\mm\nn}(k,q)\equiv\Lambda_1[\beta\cdot q-(\mu_\mm-\mu_\nn)]n_{\rm B}(k_+,\mm)[1+n_{\rm B}(k_-,\nn)]=\frac{n_{\rm B}(k_-,\nn)-n_{\rm B}(k_+,\mm)}{\beta\cdot q-(\mu_\mm-\mu_\nn)}\;.
\end{equation}
Note that, for finite isospin, the statistical factor is not symmetric in $q$. Rather:
\begin{equation*}
    L^{\mm\nn}(k,-q)=L^{\nn\mm}(k,q)\;.
\end{equation*}
The computation of the linear and quadratic expansion is reported in detail in appendix \ref{SEC: app pion}. 

We first obtain that the first linear order correction of the Wigner function vanishes for the diagonal components while is generally non-vanishing for the off-diagonal terms:
\begin{equation}
    \Delta^{(1)}_{\rm lin}W^{\mm\mm}(x,k)=0\;,\qquad\Delta^{(1)}_{\rm lin}W^{\mm\nn}(x,k)\;{\rm may\, be\,}\ne0\;,\,{\rm for}\, \mm\ne\nn\;.
\end{equation}
Since the spectra \eqref{eq: CF in terms of W} are given by the diagonal elements of the Wigner function we conclude that the Cooper-Frye formula \eqref{eq: CF pions} does not receive any local equilibrium correction at first order in the gradients. 

The first non-vanishing correction is thus quadratic in the gradients. The first term comes from the second order linear response which requires to compute the second derivative in $q$ of \eqref{eq: G lin scalar case}. The final compact result reads:
\begin{equation}\label{eq: Wscal2 final}
    \begin{split}
        \Delta^{(2)}_{\rm lin}W^{\mm\mm}(x,k)&=\frac{1}{2(2\pi)^3}\Big[\delta(k^2-m^2)L^{\mm\mm}_{[2]}(x,k)\beta_\alpha(x)\beta_\beta(x)\mathcal{Z}^{\alpha\beta}_\mm(x,k)-\frac{\chi^{(1)}_\mm(x,k)}{2}\delta(k^2-m^2)\mathcal{C}(x,k)\\
        &+\frac{\chi^{(1)}_\mm(x,k)}{2}\delta'(k^2-m^2)g_{\alpha\beta}\mathcal{Z}^{\alpha\beta}_\mm(x,k)\Big]\;,
    \end{split}
\end{equation}
where the tensor $\mathcal{Z}^{\alpha\beta}_\mm$ and the scalar $\mathcal{C}$ contain the various gradient terms:
\begin{equation}\label{eq: C e Z final}
    \begin{split}
        \mathcal{Z}^{\alpha\beta}_\mm&=\mathcal{P}^{\rho\alpha}\mathcal{P}^{\sigma\beta}\left(k^\lambda\Xi_{\rho\sigma\lambda}-\mm\,\partial_\rho\partial_\sigma\mathfrak{Z}\right)- K^{\alpha\beta}\left(k^\rho k^\lambda\xi_{\rho\lambda}-\mm\,k^\rho\partial_\rho\mathfrak{Z}\right)\;,\\
        \mathcal{C}&=\nu_\alpha\mathcal{P}^{\rho\alpha}\mathcal{P}^{\sigma\lambda}\Xi_{\rho\sigma\lambda}-\nu^\lambda\left(g^{\rho\sigma}-k^\rho\nu^\sigma-\nu^\rho k^\sigma+\nu^2 k^\rho k^\sigma\right)\Xi_{\rho\sigma\lambda}+ K^\alpha_{\,\alpha}\xi^\gamma_\gamma- K^{\rho\sigma}\xi_{\rho\sigma}\;.
    \end{split}
\end{equation}
while the coefficients $\chi^{(1)}_\mm$ and $L^{\mm\mm}_{[2]}$ are explicitly given in equations \eqref{def: chim} and \eqref{eq: derivatives of L}.
The tensors $K^{\alpha\beta}$ and $\Xi_{\alpha\beta\gamma}$ appearing in the quantities \eqref{eq: C e Z final} and contributing at the second linear order correction \eqref{eq: Wscal2 final} are, respectively, the \emph{extrinsic curvature} tensor:
\begin{equation}\label{eq: Curvature tensor}
    K^{\alpha\beta}\equiv\mathfrak{D}^\alpha\nu^\beta=\frac{1}{k\cdot\hat{n}}\mathcal{P}^{\rho\alpha}\mathcal{P}^{\sigma\beta}\partial_\rho\hat{n}_\sigma\;,
\end{equation}
and:
\begin{equation}\label{def: XI}
    \Xi_{\rho\sigma\lambda}\equiv\partial_\rho\xi_{\sigma\lambda}+\partial_\sigma\xi_{\rho\lambda}-\partial_\lambda\xi_{\rho\sigma}\;.
\end{equation}
That is the second order linear correction \emph{does not} receive any contribution from the thermal vorticity $\varpi$. Neither through its derivative $\partial\varpi$ nor by means of a coupling with the curvature tensor, $K\varpi$.
Overall, in terms of the gradients, the second linear order correction \eqref{eq: Wscal2 final} has the following general structure:
\begin{equation}\label{eq: Wlin2 structure}
    \Delta^{(2)}_{\rm lin}W\sim \Delta_{\partial\xi}W+\Delta_{\partial\partial\mathfrak{Z}}W+\Delta_{K\xi}W+\Delta_{K\partial\mathfrak{Z}}W\;.
\end{equation}
Note that this correction vanishes identically at global equilibrium where $\xi=0$ and $\partial\mathfrak{Z}=0$.

We now come to the computation of the quadratic correction. Again the interested reader may find the detailed computation in appendix \ref{SEC: app pion}. The generic quadratic microscopic kernel entering the quadratic response is, setting as in the linear case the product of the sign functions in \eqref{def: Sstat} to +1 for the particle contribution:
\begin{equation}\label{eq: G general scalar}
    \left[\mathsf{G}^{\mu,\nu}_{ab}(k,q_a,q_b)\right]^{\mm\nn}=\frac{1}{4(2\pi)^9}\delta\left(k^2-m^2+\frac{q^2}{4}\right)\sum_{\rr=-1}^{+1}\mathcal{N}^{\mm\rr\nn}(k,q_a,q_b)[\mathsf{f}^\mu_a(k_-,\ell)]^{\nn\rr}[\mathsf{f}^\nu_b(\ell,k_+)]^{\rr\mm}\;,
\end{equation}
where the statistical factor $\mathcal{N}$ is now given by:
\begin{equation}\label{def: N quad scal}
    \mathcal{N}^{\mm\rr\nn}(k,q_a,q_b)\equiv n_{\rm B}(k_+,\mm)[1+n_{\rm B}(k_-,\nn)]\left[\Lambda_2(z_b,z_a)+n_{\rm B}(\ell,\rr)\Lambda_1(z_a)\Lambda_1(z_b)\right]\;.
\end{equation}
The statistical factor \eqref{def: N quad scal} satisfies an analogous \emph{reverse symmetry} under flip of the chain order even at finite isospin similar to equation \eqref{eq: Nswap} derived for fermions in appendix \ref{APP: computation G}. In particular we find:
\begin{equation}\label{eq: Nswap scalar}
    \mathcal{N}^{\mm\rr\nn}(k,q_a,q_b)=\mathcal{N}^{\nn\rr\mm}(k,-q_b,-q_a)\;.
\end{equation}
From the definition of the vertices \eqref{eq: pion vertices}, using \eqref{eq: Nswap scalar} in \eqref{eq: G general scalar} the following transformations under the reversal of the chain are found:
\begin{equation}\label{eq: scalar Gswap}
   \begin{split}
        \mathsf{G}^{\mu_1\mu_2}_{ab}(k,q_a,q_b)^{\mm\mm}&= \mathsf{G}^{\mu_2\mu_1}_{ba}(k,-q_b,-q_a)^{\mm\mm}\;,\quad a,b=T,\mathcal{V}_3\\
        \mathsf{G}^{\mu_1\mu_2}_{\mathcal{V}_1\mathcal{V}_2}(k,q_a,q_b)^{\mm\mm}&= -\mathsf{G}^{\mu_2\mu_1}_{\mathcal{V}_2\mathcal{V}_1}(k,-q_b,-q_a)^{\mm\mm}\;,
   \end{split}
\end{equation}
where for the second equality we used that $(\mathsf{t}_1)^{\mm\rr}(\mathsf{t}_2)^{\rr\mm}=-(\mathsf{t}_2)^{\mm\rr}(\mathsf{t}_1)^{\rr\mm}$ for each $\rr$,  being $\mathsf{t}_1$ real and symmetric while $\mathsf{t}_2$ being imaginary and antisymmetric. Note that, in principle, the quadratic correction of the diagonal part also depends on the component of the isochemical potential $\mathfrak{Z}_{1,2}$ orthogonal to $\mathfrak{Z}_3$. By taking the derivative of the possible microscopic kernels \eqref{eq: G general scalar} and taking advantage of \eqref{eq: scalar Gswap} we then find
two different kinds of corrections associated with the components of the iso-vector chemical potential parallel and orthogonal to its local direction:
\begin{equation}\label{eq: papo}
    \begin{split}
        \Delta^{(2)}_{\rm quad}W^{\mm\mm}(x,k)=\Delta^{(2)}_{\|}W^{\mm\mm}(x,k)+\Delta^{(2)}_\bot W^{\mm\mm}(x,k)\;,
    \end{split}
\end{equation}
where the explicit expression of the two corrections is given in equations \eqref{eq: Wquad par} and \eqref{eq: Wquad perp}.
From a gradient contribution point of view the quadratic response correction can be written schematically as:
\begin{equation}\label{eq: Wquad structure}
    \Delta^{(2)}_{\rm quad}W\sim\Delta_{\varpi^2}W+\Delta_{\xi^2}W+\Delta_{\varpi\xi}W+\Delta_{\varpi\partial\mathfrak{Z}_3}W+\Delta_{\xi\partial\mathfrak{Z}_3}W+\Delta_{\partial\mathfrak{Z}^2_3}W+\Delta_{\partial\mathfrak{Z}^2_{1}+\partial\mathfrak{Z}^2_{2}}W\;,
\end{equation}
where the last term accounts for the orthogonal isospin correction. 
Note that, unlike the second linear term \eqref{eq: Wscal2 final}, there is no coupling with the curvature $K$ while there are corrections depending on the square of the vorticity. These are terms which, in principle, survive also at global equilibrium where $\varpi={\rm const}$. We further discuss this point at the end of the next subsection.

Overall, combining the second linear order term \eqref{eq: Wscal2 final} and the quadratic correction \eqref{eq: papo}, we find that the total quadratic gradient correction to the diagonal Wigner function is:
\begin{equation}\label{SCALAR W quad FINAL}
\boxed{
    \begin{split}
        \Delta^{(2)}W^{\mm\mm}(x,k)&=\frac{\delta(k^2-m^2)}{4(2\pi)^3}\mathcal{K}^\mm(x,k)-\frac{\delta'(k^2-m^2)}{4(2\pi)^3}g_{\alpha\beta}\mathfrak{D}^\alpha\circ\mathfrak{D}^\beta n_{\rm B}(k,\mm)\\
        &+\Delta^{(2)}_{\rm quad,\bot} W^{\mm\mm}(x,k)
    \end{split}}\;,
\end{equation}
where the \emph{on-shell} coefficient $\mathcal{K}$ is given explicitly by:
\begin{equation}\label{def: on-shell coeff}
    \begin{split}
        \mathcal{K}^\mm&\equiv \frac{\chi^{(2)}_\mm}{2}\left(B^{\;\alpha}_\beta B_{\alpha}^{\;\beta}-2\nu^\alpha U_{\mm\gamma}B^{\;\gamma}_\alpha+(\nu\cdot U_\mm)^2\right)-\chi^{(1)}_\mm\mathcal{C}\\
    &+4L^{\mm\mm}_{[2]}\left(\frac{1}{2}\beta_\alpha\beta_\beta\mathcal{Z}^{\alpha\beta}_\mm+(\beta\cdot\nu)(U_\mm\cdot U_\mm)-\beta_\alpha B_{\beta}^{\;\alpha}U^{\beta}_\mm\right)\;,
    \end{split}
\end{equation}
where $U$ and $B$ contain the following combination of the gradients:
\begin{equation*}
    U^\alpha_\mm\equiv k^\nu B_{\nu}^{\;\alpha}-\mm\, Z^\alpha_3\;,\quad B_\nu^{\;\alpha}\equiv\mathcal{P}^{\rho\alpha}\partial_\rho\beta_\nu\;,\quad Z^\alpha_3\equiv\mathcal{P}^{\rho\alpha}\partial_\rho\ZZ_3.
\end{equation*}
Note that the off-shell contribution in \eqref{SCALAR W quad FINAL} is proportional to a total derivative along the hypersurface, $g_{\alpha\beta}\mathfrak{D}^\alpha\circ\mathfrak{D}^\beta n_{\rm B}=\mathfrak{D}_\beta(\mathfrak{D}^\beta n_{\rm B})$, see \eqref{eq: IBP}. This means that, once integrated over the hypersurface $\Sigma_{\rm D}$ as specified in equation \eqref{eq: CF in terms of W} the term vanishes and thus does not contribute to the spectra.

Besides the spectra equation \eqref{SCALAR W quad FINAL} allows to compute the second order gradient expansion of several local densities such as the stress-energy tensor which have been computed in the global equilibrium case \cite{Buzzegoli:2017cqy} and at local equilibrium \cite{Akkelin:2025lgf} with different methods.

\subsection{Final result and discussion}
In heavy ion collisions the isospin imbalance of the system is fixed by its electric charge, which for hadrons is $Q={\rm I}_3+(B+S)/2$, so that the iso-vector chemical potential has a preferred direction, the third one, associated with the electric charge. We thus consider the particular case where the isovector chemical potential is aligned parallel to the third direction all over the decoupling hypersurface. This choice implies then $\Delta^{(2)}_{\rm quad,\bot}W^{\mm\mm}(x,k)=0$.

Now, since, for a fixed direction of $\bZZ$, the Wigner function is diagonal in the charge basis up to second order, we can collect the three spectra in the diagonal pion spectrum matrix:
\begin{equation}\label{eq: Y matrix}
    \mathsf{Y}({\bf k})\equiv\begin{pmatrix}
        \mathcal{Y}_+({\bf k})&0&0\\
        0&\mathcal{Y}_0({\bf k})&0\\
        0&0&\mathcal{Y}_-({\bf k})
    \end{pmatrix}\;.
\end{equation}
Using that the diagonal components of the linear correction to the Wigner function vanish and the quadratic correction \eqref{SCALAR W quad FINAL}, the spectra of the pions can be written, at quadratic order in the gradients, as:
\begin{equation}\label{eq: pion spectrum final}
    \boxed{\mathcal{Y}_\mm(k)=\frac{1}{(2\pi)^3}\int_{\Sigma_{\rm D}}\di\Sigma(x)\cdot k\left[n^\mm_{\rm B}(k)+\frac{1}{4}\widetilde{\mathcal{K}}^\mm(k)\right]+\mathcal{O}(\partial^3)\;,\quad k^\mu=\left(\sqrt{m^2+|{\bf k}|^2},\,{\bf k}\right)\;,}
\end{equation}
where the off-shell term is eliminated by the hypersurface integration as observed.
The integrand of \eqref{eq: pion spectrum final} is defined up to terms which vanish once integrated over $\Sigma_{\rm D}$. Integrating by parts along the hypersurface (see appendix \ref{SEC: app pion}), the on-shell coefficient $\mathcal{K}$ in the correction to the Wigner function can be replaced by:
\begin{equation}\label{eq: pion K}
    \begin{split}
        \widetilde{\mathcal{K}}^\mm(k)&=\frac{1}{6}n''''_{\rm B}(\beta\cdot U_\mm)^2+\frac{1}{6}n'''_{\rm B}\left(B_\alpha^{\;\alpha}(\beta\cdot U_\mm)+3\beta_\alpha B_\beta^{\;\alpha}U^\beta_\mm-2(U_\mm\cdot U_\mm)\beta\cdot\nu\right)\\
        &+\frac{n''_{\rm B}}{2}\left[B_\beta^{\;\alpha}B_\alpha^{\;\beta}+(\nu\cdot U_\mm)^2-2B_\alpha^{\;\alpha}(\nu\cdot U_\mm)\right]\;,
    \end{split}
\end{equation}
where the ``primes" denote derivatives of the Bose-Einstein distribution with respect to its argument $\chi_\mm=\beta\cdot k-\mm\ZZ$.

After integration then, the form \eqref{eq: pion K} only contains the first derivatives of $\beta$ and $\ZZ$ along the hypersurface and no derivative of the normal vector. Hence the second order correction to the spectra only depends on quadratic combination of first order gradients and \emph{there is no dependence} on the extrinsic curvature and on the second derivatives of the thermodynamic fields. The dependence on the geometry of $\Sigma_{\rm D}$ is only through $\nu$, that is through the orientation of the normal vector with respect to the momentum. This result is non-trivial and we argue that there may be a deeper reason related with the particular dynamics of free fields. Note that, in principle, these corrections are present in the Wigner function \eqref{SCALAR W quad FINAL}, but eventually they disappear from the spectra after integration. 

Using equations \eqref{eq: Wlin2 structure}, \eqref{eq: Wquad structure} together with \eqref{eq: pion spectrum final} we summarize the gradient structures contributing to the Wigner function and to the spectra of the different pions in the table \ref{titta}.
\begin{table}[h!]\label{titta}
    \centering
    \renewcommand{\arraystretch}{1.2}
    \begin{tabular}{lcccc}
        \toprule
        & \multicolumn{2}{c}{Wigner function $W^{\mm\mm}$}
        & \multicolumn{2}{c}{Spectrum $\mathcal{Y}_{\mm}$} \\
        \cmidrule(lr){2-3}
        \cmidrule(l){4-5}
        Gradient structure
        & $\pi^\pm$ & $\pi^0$
        & $\pi^\pm$ & $\pi^0$ \\
        \midrule
        $\xi^2,\ \varpi^2,\ \xi\varpi$
            & $\checkmark$ & $\checkmark$
            & $\checkmark$ & $\checkmark$ \\

        $\xi\,\partial\mathfrak{Z}_3,\
          \varpi\,\partial\mathfrak{Z}_3,\
          (\partial\mathfrak{Z}_3)^2$
            & $\checkmark$ & --
            & $\checkmark$ & -- \\

        \addlinespace[3pt]
        $\partial\xi$
            & $\checkmark$ & $\checkmark$
            & -- & -- \\

        $\partial^2\mathfrak{Z}_3$
            & $\checkmark$ & --
            & -- & -- \\

        \addlinespace[3pt]
        $K\xi$
            & $\checkmark$ & $\checkmark$
            & -- & -- \\

        $K\,\partial\mathfrak{Z}_3$
            & $\checkmark$ & --
            & -- & -- \\
        \bottomrule
    \end{tabular}
    \caption{Second-order gradient structures contributing
    to the diagonal pion Wigner function and momentum
    spectra for a fixed isospin direction.
    Checkmarks indicate potentially nonvanishing
    contributions, while dashes indicate their absence.
    Terms involving second derivatives or explicit
    hypersurface curvature cancel from the spectra
    after hypersurface integration by parts, assuming
    vanishing boundary contributions.}
    \label{tab:pion-gradients}
\end{table}
From the spectra matrix \eqref{eq: Y matrix} we can define the following iso-vector and iso-tensor projections:
\begin{equation}\label{pippo}
   \mathcal{Y}_+({\bf k})- \mathcal{Y}_-({\bf k})=\tr\left(\mathsf{t}_3\mathsf{Y}({\bf k})\right)\;,\qquad \mathcal{Y}_0({\bf k})-\frac{1}{2}\left(\mathcal{Y}_+({\bf k})+\mathcal{Y}_-({\bf k})\right)=-\frac{1}{2}\tr\left[\left(3\mathsf{t}^2_3-2\mathbb{I}\right)\mathsf{Y}({\bf k})\right]\;.
\end{equation}
These are useful because they allow us to define two experimentally relevant quantities sensitive to the local equilibrium corrections; the \emph{charge asymmetry ratio} and the \emph{neutral-to-charged imbalance}:
\begin{equation}\label{def: R ratios}
    {\rm R}_{\pm}({\bf k})\equiv\frac{\mathcal{Y}_+({\bf k})-\mathcal{Y}_-({\bf k})}{\mathcal{Y}_+({\bf k})+\mathcal{Y}_-({\bf k})}\;,\qquad {\rm R}_{0\pm}({\bf k})\equiv\frac{\mathcal{Y}_0({\bf k})-\frac{1}{2}\left(\mathcal{Y}_+({\bf k})+\mathcal{Y}_-({\bf k})\right)}{\mathcal{Y}_0({\bf k})+\mathcal{Y}_+({\bf k})+\mathcal{Y}_-({\bf k})}\;.
\end{equation}
From \eqref{pippo}, ${\rm R}_\pm$ measures a charge-odd iso-vector imbalance, while ${\rm R}_{0\pm}$ measures a charge-even iso-tensor imbalance. Introducing the relative corrections with respect to the Cooper-Frye formula:
\begin{equation}
    \varepsilon_\mm({\bf k})\equiv\frac{\mathcal{Y}^{(2)}_\mm({\bf k})}{\mathcal{Y}^{(0)}_\mm({\bf k})}\;,\qquad\varepsilon_{\rm ch}({\bf k})\equiv\frac{\mathcal{Y}^{(0)}_+({\bf k})\varepsilon_+({\bf k})+\mathcal{Y}^{(0)}_-({\bf k})\varepsilon_-({\bf k})}{\mathcal{Y}^{(0)}_+({\bf k})+\mathcal{Y}^{(0)}_-({\bf k})}\;,
\end{equation}
and expanding \eqref{def: R ratios} to first order in $\varepsilon_\mm$, the second order corrections to the Cooper-Frye ratios ${\rm R}^{\rm CF}$, obtained with $\mathcal{Y}^{(0)}_\mm$, are:
\begin{equation}\label{eq: Delta R}
    \begin{split}
        \Delta{\rm R}_\pm({\bf k})&=\frac{1-({\rm R}^{\rm CF}_\pm({\bf k}))^2}{2}\left(\varepsilon_+({\bf k})-\varepsilon_-({\bf k})\right)\;,\\
        \Delta{\rm R}_{0\pm}({\bf k})&=\frac{3\mathcal{Y}^{(0)}_0({\bf k})\left(\mathcal{Y}^{(0)}_+({\bf k})+\mathcal{Y}^{(0)}_-({\bf k})\right)}{2\left(\mathcal{Y}^{(0)}_0({\bf k})+\mathcal{Y}^{(0)}_+({\bf k})+\mathcal{Y}^{(0)}_-({\bf k})\right)^2}\left(\varepsilon_0({\bf k})-\varepsilon_{\rm ch}({\bf k})\right)\;,
    \end{split}
\end{equation}
with $\varepsilon_\mm$ given by \eqref{eq: pion spectrum final}. Note that, from the property of the pion multiplet:
\begin{equation*}
    \mathcal{Y}_+[\mathfrak{Z}]=\mathcal{Y}_-[-\mathfrak{Z}]\;,\qquad\mathcal{Y}_0[\mathfrak{Z}]=\mathcal{Y}_0[-\mathfrak{Z}]\;,
\end{equation*}
which implies that:
\begin{equation}
    {\rm R}_\pm[-\mathfrak{Z}]=-{\rm R}_\pm[\mathfrak{Z}]\;,\qquad {\rm R}_{0\pm}[-\mathfrak{Z}]={\rm R}_{0\pm}[\mathfrak{Z}]\;,
\end{equation}
that is the charge-imbalance is sensitive to odd powers of the iso-vector chemical potential $\mathfrak{Z}$ while the neutral-to-charged imbalance to even powers. In particular, ${\rm R}_\pm$ starts at first order in $\mathfrak{Z}$ while ${\rm R}_{0\pm}$ starts at second order. In the realistic regime of small $\mathfrak{Z}$ one may thus expect the charge asymmetry to be larger than the neutral-to-charged imbalance; for $\mathfrak{Z}=0$ on the whole hypersurface both ratios, and their corrections, vanish, as required by isospin symmetry. At lower collision energies, where charge asymmetries can be more pronounced, these corrections may provide additional theoretical input for pion emission models. A quantitative assessment however requires to take into account other effects such as, for instance, electromagnetic interactions, hadronic interactions and resonance decays.

We conclude our analysis of the result by computing the global equilibrium limit of the Wigner function correction \eqref{SCALAR W quad FINAL}. In global equilibrium:
\begin{equation*}
    \beta_\lambda=b_\lambda+\varpi_\lambda^{\;\gamma}x_\gamma\;,\qquad \varpi_{\lambda\gamma}={\rm const}\;,\qquad\xi_{\lambda\gamma}=0\;,\qquad\mathfrak{Z}_i={\rm const.}\;,
\end{equation*}
so that\eqref{SCALAR W quad FINAL} reduces to
(see the end of appendix \ref{SEC: app pion} for details):
\begin{equation}\label{eq: pion W2 GE2 final}
    \begin{split}
        \Delta^{(2)}W^{\mm\mm}_{\rm GE}(x,k)&=\frac{\delta(k^2-m^2)}{4(2\pi)^3}n^\mm_{\rm B}(k)(1+n^\mm_{\rm B}(k))\times\\
        &\times\left({\frac{1+2n^\mm_{\rm B}(k)}{2}}\varpi:\varpi+\frac{1+6n^\mm_{\rm B}(k)+6(n^{\mm}_{\rm B}(k))^2}{3T^2}a_\mu k_\nu\varpi^{\mu\nu}\right)\\
        &-\frac{\delta'(k^2-m^2)}{4(2\pi)^3}n^\mm_{\rm B}(k)(1+n^\mm_{\rm B}(k))\left(1+2n_{\rm B}^\mm(k)\right)\varpi_{\sigma\lambda}k^\lambda\varpi^{\sigma}_{\;\gamma}k^\gamma\;,
    \end{split}
\end{equation}
where $a^\mu$ is the four-acceleration while $T$ is the proper temperature.

The on-shell part receives two corrections in the thermal vorticity: $\varpi:\varpi\equiv\varpi^\sigma_{\;\rho}\varpi^\rho_{\;\sigma}$ and $a_\mu k_\nu\varpi^{\mu\nu}$ with $a^\mu$ four-acceleration and $T$ proper temperature. The result is in agreement with the flat space-time limit of an $\hbar$ expansion for the Wigner function in a general curved space-time \cite{paperanna} and with the exact solution for the purely rotational case \cite{Becattini:2020qol}, at least at second order in $\varpi$. Moreover the above solution satisfies the generalized on-shell condition for the Wigner function up to $\mathcal{O}(\varpi^3)$ as expected. 
\section{Conclusions}\label{SEC: conclusions}
In conclusion, we developed a full quadratic response theory expansion for the Wigner function of hadrons of arbitrary spin and isospin and generic pseudo-gauge which allows for exact integration over the decoupling hypersurface retaining the exact geometry.

As a first particular case of analysis we applied the new formula \eqref{eq: QUAD FINAL} in the computation of the spin polarization vector of spin $1/2$ hadrons extending the result obtained in \cite{Sheng:2025cjk} and generalizing the computation performed in \cite{Sheng:2024pbw} to the case of arbitrary geometry. We find that the quadratic corrections of the spin polarization vector are exactly vanishing in a vast class of physically relevant pseudo-gauges, such as Belinfante, canonical, HW and GLW.  

As a second application, we computed the full second order gradient corrections to the spectra of the pion multiplet. We obtained that the full second order term \emph{does not} have any contribution of the vorticity coupled with the curvature of the hypersurface, which only involves the coupling with the thermal shear and the gradient of the iso-vector chemical potential. The first order correction to the spectra vanishes, and the second order one, given by \eqref{eq: pion spectrum final}, after the integration over the decoupling hypersurface only depends on the first derivatives of the four-temperature and of the iso-vector chemical potential along the hypersurface, and not on its extrinsic curvature. We also found that, for a non-Abelian chemical potential whose direction changes along the hypersurface, the gradients of its orthogonal components contribute to the diagonal components of the Wigner function. From the spectra we obtained the second order corrections to the charge asymmetry and to the neutral-to-charged imbalance of the pions. The resulting expressions provide an input for incorporating nondissipative second-order local-equilibrium corrections into particle-emission models and in particular for investigating their influence on isospin-sensitive pion spectra.

Besides the analyzed cases, the formula \eqref{eq: QUAD FINAL} can be applied to derive the local equilibrium correction up to the second order of the derivative of the thermodynamic fields and curvature fields of any other hadronic observables of interest. Among the various possibilities, we point out the spin polarization vector of spin $3/2$ hadrons, extending the computation in \cite{Buzzegoli:2026sge}, the spin density matrix of vector mesons extending the computation in \cite{Yang:2024fkn, Zhang:2024mhs} to include the precise geometry of the decoupling. Moreover the expansion scheme applied in \cite{Sheng:2025cjk} has been extended to the study of non-equilibrium corrections \cite{Becattini:2025rqp} leading to the appearance of several memory corrections which are in principle sensitive to the shape and curvature of the initial hypersurface. Our method allows one to extend the study to second order.


\acknowledgments
D.R acknowledges useful discussions with F. Becattini, A. Palermo, X.L. Sheng and A. Chiarini. A.O. wants to thank C. Hoyos for insightful discussions regarding the role played by pseudo-gauge invariance in the definition of a local equilibrium density operator. The authors acknowledge the use of AI agents for assistance with some technical calculations and
some editing of the draft; they take full responsibility for
the content of the paper. The work of A.O. was partially supported by the In2p3 grant “Extreme Dynamics” and the ANR grant
“XtremeHolo” (ANR project n.284452).



\appendix

\section{Evaluation of the thermal correlators}\label{APP: wick}

In this appendix we derive the thermal expectation values of four and six field insertions in the reference global equilibrium state \eqref{eq: GE} used for the computation  of the three-point Kubo Mori correlator \eqref{def: KuboMori3}.
Being the equilibrium state Gaussian the expectation value of a product of fields is given by the sum over all the complete pairings, each pair being replaced by \eqref{eq: 2pt a} or \eqref{eq: 2pt b} according to the order in which the two fields appear in the product. For fermions each pairing carries the sign $(-1)^{\#}$, where $\#$ is the number of crossings when the pairs are drawn as arcs above the product. In a connected correlator of bilinears only the pairings which connect all the bilinears survive.

Let us introduce the following notation:
\begin{equation*}
    n^\mm_p\equiv n_\eta(p,\mm)\;,\qquad \bar n^\mm_p\equiv1-\eta\,n_\eta(p,\mm)\;,
\end{equation*}
with $\eta=+1$ for fermions and $\eta=-1$ for bosons as usual. With the above the two points correlators \eqref{eq: 2pt a} and \eqref{eq: 2pt b} are written as:
\begin{equation}\label{Cappelli}
    \begin{split}
        \langle\compactphi^\rr_\alpha(p)\wvrphi^\mss_\beta(p')\rangle_{\rm GE}&=(2\pi)^4\delta^{\rr\mss}\delta^4(p-p')n^\rr_p\varrho_{\beta\alpha}(p)\;,\\
        \langle\wvrphi^\mss_\beta(p')\compactphi^\rr_\alpha(p)\rangle_{\rm GE}&=(2\pi)^4\delta^{\rr\mss}\delta^4(p-p')\bar n^\rr_p\varrho_{\beta\alpha}(p)\;.
    \end{split}
\end{equation}
For the linear response theory we need to compute
\begin{equation}\label{de curtis}
    \Lambda_1(z)\e^{\ii(p-p')\cdot y}\mathcal{C}_4(k_+,k_-,p,p')\;,\quad z=-\beta\cdot(p-p')+\mathfrak{Z}(\mrr-\mss)\;,
\end{equation}
with $\Lambda_1(z)$ given in \eqref{eq: Lambda1} and:
\begin{equation*}
    \mathcal{C}_4(k_+,k_-,p,p')\equiv\langle\underbrace{\compactphi^\mm_\lambda(k_+)}_{A_1}\underbrace{\wvrphi^\nn_\rho(k_-)}_{A_2}\,,\underbrace{\compactphi^\mrr_\alpha(p)}_{B_1}\mathsf{f}^{\mrr\mss}_a(p,p')\underbrace{\wvrphi^\mss_\beta(p')}_{B2}\rangle_{c,\rm GE}\;,
\end{equation*}
is the connected expectation value for 4 field insertions with one vertex.

Ordering the fields from left to right the only connected contribution between two sets $A$ and $B$ is $(A_1B_2)(A_2B_1)$. The corresponding ring does not have any intersection hence the overall sign is $+1$ for both fermions and bosons. Using \eqref{Cappelli} we get:
\begin{equation*}
    \begin{split}
        \mathcal{C}_4(k_+,k_-,p,p')&=\langle \compactphi^\mm_\lambda(k_+)\wvrphi^\mss_\beta(p'))\rangle_{\rm GE}\mathsf{f}^{\mrr\mss}_a(p,p')\langle\wvrphi^\nn_\rho(k_-)\compactphi^\mrr_\alpha(p)\rangle_{\rm GE}\\
    &=(2\pi)^8\delta^{\nn\mrr}\delta^{\mm\mss}\delta^4(p-k_-)\delta^4(p'-k_+)\bar n^\nn_{k_-}n^\mm_{k_+}\left[\varrho(k_-)\mathsf{f}^{\nn\mm}_a(k_-,k_+)\varrho(k_+)\right]_{\rho\lambda}\;,
    \end{split}
\end{equation*}
where we used that the two $\delta$ force $p=k_-$ and $p'=k_+$ so that $q = p'-p$ and $z = \beta\cdot q - \mathfrak{Z}(\mm-\nn)$. Then equation \eqref{de curtis} becomes:
\begin{equation}\label{eq: RISULTATO C4}
   \Lambda_1(z)\e^{-\ii q\cdot y}(2\pi)^8\delta^{\nn\mrr}\delta^{\mm\mss}\delta^4(p-k_-)\delta^4(p'-k_+)\bar n^\nn_{k_-}n^\mm_{k_+}\left[\varrho(k_-)\mathsf{f}^{\nn\mm}_a(k_-,k_+)\varrho(k_+)\right]_{\rho\lambda}\;,
\end{equation}
For the quadratic response theory we need to compute:
\begin{equation}\label{de curtis 2}
    \Lambda_2(z_a,z_b)\e^{\ii(p_1-p'_1)\cdot y_1}\e^{\ii(p_2-p'_2)\cdot y_2}\mathcal{C}_6(k_\pm,p_{1,2},p'_{1,2})
\end{equation}
where $\Lambda_2$ is defined in equation \eqref{A: Lambda2} and the six field expectation value reads:
\begin{equation}\label{eq: 6phi}
    \mathcal{C}_6(k_\pm,p_{1,2},p'_{1,2})\equiv\Big\langle\underbrace{\compactphi^\mm_\lambda(k_+)}_{A_1}\underbrace{\wvrphi^\nn_\rho(k_-)}_{A_2}\,,\,\underbrace{\compactphi^{\mrr_2}_{\alpha_2}(p_2)}_{B_1}\mathsf{f}^{\mrr_2\mss_2}_b\underbrace{\wvrphi^{\mss_2}_{\beta_2}(p_2')}_{B_2}\,,\,\underbrace{\compactphi^{\mrr_1}_{\alpha_1}(p_1)}_{C_1}\mathsf{f}^{\mrr_1\mss_1}_a\underbrace{\wvrphi^{\mss_1}_{\beta_1}(p_1')}_{C_2}\Big\rangle_c\;.
\end{equation}
The terms $B$ come from $\widehat{\Gamma}_b$ at $(y_2,\lambda_2)$ while the terms $C$ come from $\widehat{\Gamma}_a$ at $(y_1,\lambda_1)$. In a connected pairing each $\wvrphi$ must be paired with the $\compactphi$ of a different bilinear. Hence the possible pairing, i.e. the two cyclic orderings  of the three bilinear. We refer to these cyclic ordering as \emph{rings} and for the expectation value \eqref{eq: 6phi} they are:
\begin{itemize}
    \item {\rm{Ring I}}: $(A_2C_1)(C_2B_1)(B_2A_1)$, i.e.\ $\wvrphi(k_-)\to\widehat{\Gamma}_a\to\widehat{\Gamma}_b\to\compactphi(k_+)$. Using the same numeration of the fields the arcs are now $(1,4),(2,5),(3,6)$ for a total of three crossings and on overall sign of the ring $(-\eta)^3=-\eta$.
    \item {\rm{Ring II}}: $(A_2B_1)(B_2C_1)(C_2A_1)$, i.e.\ $\wvrphi(k_-)\to\widehat{\Gamma}_b\to\widehat{\Gamma}_a\to\compactphi(k_+)$. Numbering the fields from left to right as 1 to 6, the arcs originating from the pairings are $(1,6),(2,3),(4,5)$ and they involve no crossing so the overall sign of the ring is $(-\eta)^0=+1$.
\end{itemize}
Each ring contains \emph{three} contractions each one bringing a spectral functions for a total of three on-shell lines. The two external ones, $k_+$ and $k_-$, from the Wigner operator insertions and an internal one, from the resulting contraction which we denote with the overall momentum $\ell$, $\ell^2=m^2$\;.

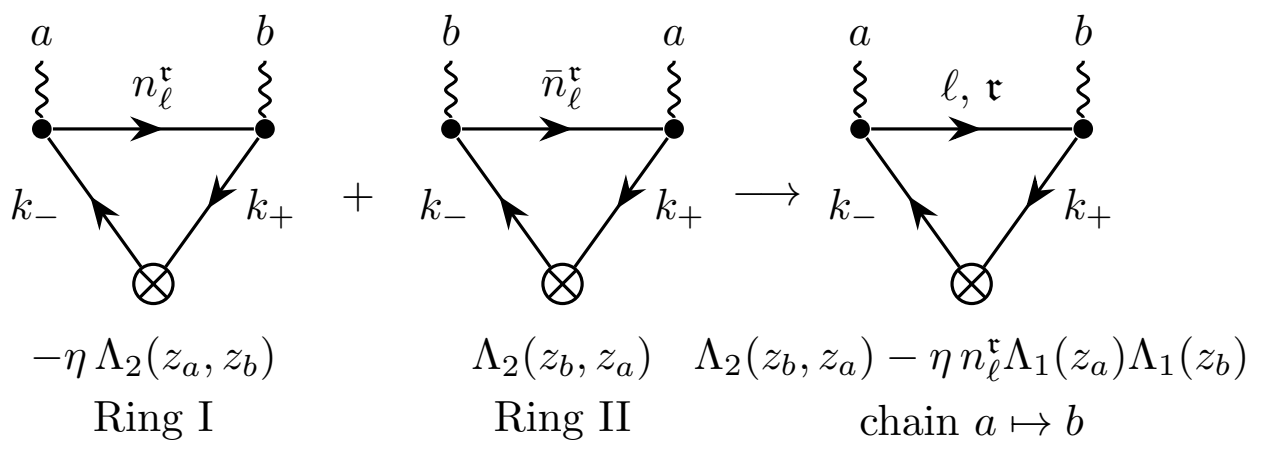
\begin{figure}
\centering
\begin{tikzpicture}[>=Stealth,line width=0.6pt,scale=0.82,every node/.style={scale=0.82},
    fline/.style={postaction={decorate},decoration={markings,mark=at position .55 with {\arrow{Stealth[length=2.2mm]}}}},
    src/.style={decorate,decoration={snake,amplitude=1pt,segment length=4pt}},
    wig/.style={circle,draw,fill=white,inner sep=0pt,minimum size=9pt,path picture={\draw (path picture bounding box.south west)--(path picture bounding box.north east) (path picture bounding box.north west)--(path picture bounding box.south east);}},
    vtx/.style={circle,fill,inner sep=0pt,minimum size=4.5pt}]
    \newcommand{\tri}[4]{
        \begin{scope}[shift={(#1,0)}]
            \node[wig] (W) at (0,0) {};
            \node[vtx] (L) at (-0.9,1.25) {};
            \node[vtx] (R) at (0.9,1.25) {};
            \draw[fline] (W)--(L) node[pos=0.45,left=5pt]{$k_-$};
            \draw[fline] (L)--(R) node[midway,above=1pt]{$#4$};
            \draw[fline] (R)--(W) node[pos=0.55,right=5pt]{$k_+$};
            \draw[src] (L)--(-0.9,1.8); \node[anchor=south] at (-0.9,1.8) {$#2$};
            \draw[src] (R)--(0.9,1.8); \node[anchor=south] at (0.9,1.8) {$#3$};
        \end{scope}}
    \tri{0}{a}{b}{n^\mrr_\ell}
    \node at (0,-0.6) {$-\eta\,\Lambda_2(z_a,z_b)$};
    \node at (0,-1.1) {Ring I};
    \node at (1.65,0.7) {$+$};
    \tri{3.3}{b}{a}{\bar n^\mrr_\ell}
    \node at (3.3,-0.6) {$\Lambda_2(z_b,z_a)$};
    \node at (3.3,-1.1) {Ring II};
    \node at (4.95,0.7) {$\longrightarrow$};
    \tri{6.6}{a}{b}{\ell,\,\mrr}
    \node at (6.6,-0.6) {$\Lambda_2(z_b,z_a)-\eta\,n^\mrr_\ell\Lambda_1(z_a)\Lambda_1(z_b)$};
    \node at (6.6,-1.1) {chain $a\mapsto b$};
\end{tikzpicture}
\caption{\label{fig: diagrams} Diagrammatic representation of the connected six-field function \eqref{eq: 6phi}. The crossed vertex is the Wigner operator at $x$, from which the line $k_-$ leaves and into which the line $k_+$ enters; the dots are the bilinears $\wGamma_{a,b}$ at the points $y_{1,2}$ of $\Sigma_{\rm D}$, attached to their sources $\vartheta_{a,b}$ (wavy lines). Each line is a contraction \eqref{Cappelli}, $(2\pi)^4\delta^4\,\delta^{\mrr\mss}\varrho(p)$, and the arrow follows the order $\wvrphi(k_-)\mapsto\wGamma\mapsto\wGamma\mapsto\compactphi(k_+)$, so that at each vertex the transferred momentum is the outgoing minus the incoming one. Below each ring we report the internal factor, the Wick sign and the imaginary time weight of table \eqref{Table: RINGS}; the external lines always carry $\bar n^\nn_{k_-}n^\mm_{k_+}$. Exchanging $a\leftrightarrow b$ in ring II, which is allowed under the sum over $a,b$, both rings become the chain $a\mapsto b$, whose total weight is \eqref{eq: Lambda combination}.}
\end{figure}

Let us start from the first ring. 
Using \eqref{Cappelli} the contractions between the field insertions are:
\begin{align*}
    (A_2C_1)&:\quad(2\pi)^4\delta^{\nn\mrr_1}\delta^4(p_1-k_-)\bar n^\nn_{k_-}\varrho_{\rho\alpha_1}(k_-)\;,\\
    (B_1C_2)&:\quad(2\pi)^4\delta^{\mrr_2\mss_1}\delta^4(p_2-p'_1)\,n^{\mrr_2}_{p_2}\,\varrho_{\beta_1\alpha_2}(p_2)\;,\\
    (B_2A_1)&:\quad(2\pi)^4\delta^{\mm\mss_2}\delta^4(p'_2-k_+)\,n^\mm_{k_+}\,\varrho_{\beta_2\lambda}(k_+)\;,
\end{align*}
where the contraction $(C_2B_1)$ is the internal line. Overall the ring sets:
\begin{equation*}
    {\rm Ring\, I}\,:\qquad\begin{cases}
        p_1=k_-\;,\quad p'_1=\ell=p_2\,,\quad p'_2=k_+\;,\\
        \mrr_1=\nn\,,\quad\mss_1=\mrr=\mrr_2\,\quad \mss_2=\mm\;,
    \end{cases}
\end{equation*}
leading to:
\begin{equation}\label{eq: Ring I}
    {\rm Ring\, I}=-\eta\,\bar n^\nn_{k_-}n^\rr_\ell n^\mm_{k_+}\left[\varrho(k_-)\mathsf{f}^{\nn\rr}_a(k_-,\ell)\varrho(\ell)\mathsf{f}^{\mrr\mm}_b(\ell,k_+)\varrho(k_+)\right]_{\rho\lambda}\;.
\end{equation}
For the ring II we proceed in the same way. The contractions are:
\begin{equation*}
    \begin{split}
        (A_2 B_1)&=(2\pi)^4\delta^{\nn\mrr_2}\delta^4(p_2-k_-)\bar n^\nn_{k_-}\varrho_{\rho\alpha_2}(k_-)\;,\\
        (B_2C_1)&=(2\pi)^4\delta^{\mss_2\rr_1}\delta^4(p'_2-p_1)\bar n^{\mrr_1}_{p_1}\varrho_{\beta_2\alpha_1}(p_1)\;,\\
        (C_2A_1)&=(2\pi)^4\delta^{\mm\mss_1}\delta^4(p'_1-k_+)n^\mm_{k_+}\varrho_{\beta_1\lambda}(k_+)\;.
    \end{split}
\end{equation*}
Overall, taking into account the constraints of the $\delta$ and naming the internal on-shell momentum $\ell\equiv p'_2\equiv p_1$, the internal line reduces to:
\begin{equation}\label{eq: Ring II}
    {\rm Ring\, II}=\bar n^\nn_{k_-}\bar n^\rr_\ell n^\mm_{k_+}\left[\varrho(k_-)\mathsf{f}^{\nn\rr}_b(k_-,\ell)\varrho(\ell)\mathsf{f}^{\rr\mm}_a(\ell,k_+)\varrho(k_+)\right]_{\rho\lambda}\;.
\end{equation}
Now consider the ordered chain $\wvrphi^\nn(k_-)\mapsto\wGamma_a\mapsto\wGamma_b\mapsto\compactphi^\mm(k_+)$. In terms of the two rings \eqref{eq: Ring I} and \eqref{eq: Ring II} we have:
\begin{equation}\label{Table: RINGS}
    \begin{array}{c|cccc}
        {\rm Ring} & {\rm internal\, factor} & {\rm Wick\, sign} & {\rm Imaginary\, time\, weight} \\ \hline
        {\rm I} & n^\rr_\ell & -\eta& \Lambda_2(z_a,z_b) \\ \hline
        {\rm II} & \bar n^\rr_\ell & +1& \Lambda_2(z_b,z_a)
    \end{array}
\end{equation}
The double integral over $\lambda_1,\lambda_2$, $\Lambda_2$ satisfies:
\begin{equation*}
    \Lambda_2(z_a,z_b)+\Lambda_2(z_b,z_a)=\Lambda_1(z_a)\Lambda_1(z_b)\;,
\end{equation*}
therefore the combination of the two rings \eqref{Table: RINGS} gives:
\begin{equation}\label{eq: Lambda combination}
    \Lambda_2(z_b,z_a)\bar n^\mrr_\ell-\eta\Lambda_2(z_a,z_b)n^\mrr_\ell=\Lambda_2(z_b,z_a)-\eta n^\rr_\ell\Lambda_1(z_a)\Lambda_1(z_b)\;.
\end{equation}
For what concern the complex phases in \eqref{de curtis 2}, for the chain $a\mapsto b$ we have:
\begin{equation*}
   \begin{matrix}
        q_a=\ell-k_-\;,\quad &q_b= k_+-\ell\;,\\
        q=q_a+q_b\;,\quad &\ell=k+\frac{q_a-q_b}{2}
   \end{matrix}\,\,\implies\e^{\ii q\cdot x}\e^{-\ii q_a\cdot y_1}\e^{-\ii q_{b}\cdot y_2}=\e^{\ii q_a\cdot(x-y_1)}\e^{\ii q_b\cdot(x-y_2)}\;,
\end{equation*}
which sets $z_a+z_b=\beta\cdot q-\mathfrak{Z}(\mm-\nn)$ and leads to:
\begin{equation}\label{eq: goku}
    \e^{\ii q_a\cdot(x-y_1)}\e^{\ii q_b\cdot(x-y_2)}\left[\Lambda_2(z_b,z_a)-\eta n^\rr_\ell\Lambda_1(z_a)\Lambda_1(z_b)\right]\bar n^\nn_{k_-}n^\mm_{k_+}\left[\varrho(k_-)\mathsf{f}^{\nn\rr}_a(k_-,\ell)\varrho(\ell)\mathsf{f}^{\rr\mm}_b(\ell,k_+)\varrho(k_+)\right]_{\rho\lambda}\;.
\end{equation}
The relation between the two rings and the chain is reported diagrammatically in figure \ref{fig: diagrams}.

We conclude by observing that the statistical factor coming from the contraction of the two rings posses a clear exchange symmetry for $q_a,q_b\mapsto-q_b,-q_a$.
For the sake of simplicity we consider the case of a Dirac field discussed in Section \ref{SEC: Dirac}.

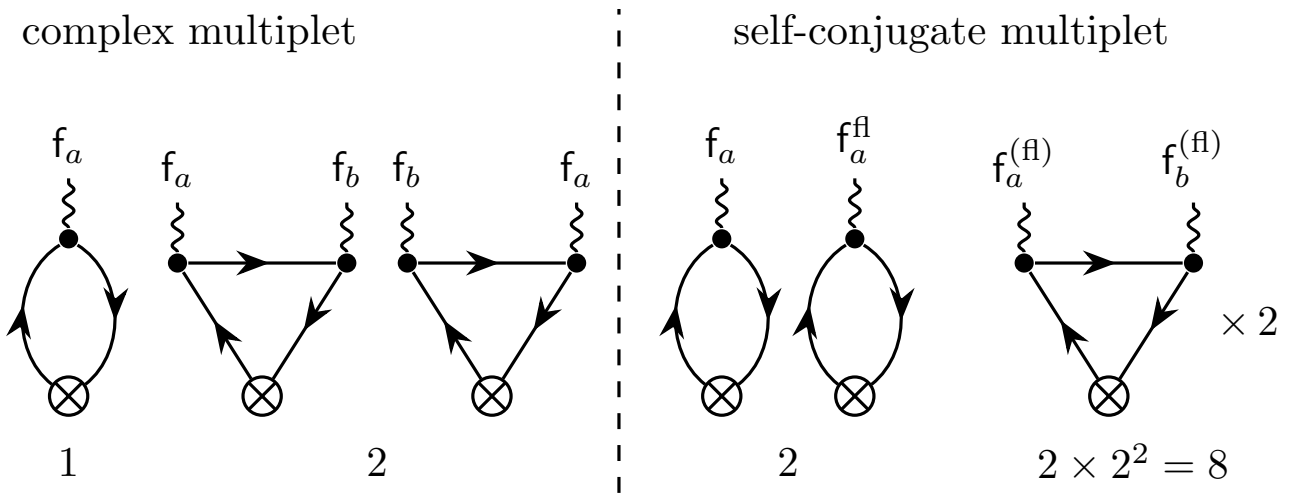
\begin{figure}
\centering
\begin{tikzpicture}[>=Stealth,line width=0.6pt,scale=0.8,every node/.style={scale=0.8},
    fline/.style={postaction={decorate},decoration={markings,mark=at position .55 with {\arrow{Stealth[length=2.2mm]}}}},
    src/.style={decorate,decoration={snake,amplitude=1pt,segment length=4pt}},
    wig/.style={circle,draw,fill=white,inner sep=0pt,minimum size=9pt,path picture={\draw (path picture bounding box.south west)--(path picture bounding box.north east) (path picture bounding box.north west)--(path picture bounding box.south east);}},
    vtx/.style={circle,fill,inner sep=0pt,minimum size=4.5pt}]
    \newcommand{\loopone}[2]{
        \begin{scope}[shift={(#1,0)}]
            \node[wig] (W) at (0,0) {};
            \node[vtx] (A) at (0,1.3) {};
            \draw[fline] (W) to[bend left=55] (A);
            \draw[fline] (A) to[bend left=55] (W);
            \draw[src] (A)--(0,1.8); \node[anchor=south] at (0,1.8) {$#2$};
        \end{scope}}
    \newcommand{\tri}[3]{
        \begin{scope}[shift={(#1,0)}]
            \node[wig] (W) at (0,0) {};
            \node[vtx] (L) at (-0.7,1.1) {};
            \node[vtx] (R) at (0.7,1.1) {};
            \draw[fline] (W)--(L); \draw[fline] (L)--(R); \draw[fline] (R)--(W);
            \draw[src] (L)--(-0.7,1.6); \node[anchor=south] at (-0.7,1.6) {$#2$};
            \draw[src] (R)--(0.7,1.6); \node[anchor=south] at (0.7,1.6) {$#3$};
        \end{scope}}
    \node at (1.0,3.0) {complex multiplet};
    \loopone{0}{\mathsf{f}_a}
    \node at (0,-0.55) {$1$};
    \tri{1.6}{\mathsf{f}_a}{\mathsf{f}_b}
    \tri{3.5}{\mathsf{f}_b}{\mathsf{f}_a}
    \node at (2.55,-0.55) {$2$};
    \draw[dashed] (4.55,-0.8)--(4.55,3.2);
    \node at (7.3,3.0) {self-conjugate multiplet};
    \loopone{5.4}{\mathsf{f}_a}
    \loopone{6.5}{\mathsf{f}^{\rm fl}_a}
    \node at (5.95,-0.55) {$2$};
    \tri{8.6}{\mathsf{f}^{(\rm fl)}_a}{\mathsf{f}^{(\rm fl)}_b}
    \node at (9.75,0.6) {$\times\,2$};
    \node at (8.8,-0.55) {$2\times2^2=8$};
\end{tikzpicture}
\caption{\label{fig: diagrams2} Connected contractions of the Wigner operator with one (four-field function) and two (six-field function) bilinears, with the conventions of figure~\ref{fig: diagrams}. Left: for a complex multiplet each $\wvrphi$ can only be contracted with a $\compactphi$, and there is one diagram with one bilinear and two rings with two bilinears. Right: for a self-conjugate multiplet also $\langle\wvrphi\wvrphi\rangle_{\rm GE}$ and $\langle\compactphi\compactphi\rangle_{\rm GE}$ are non-vanishing and each bilinear can be crossed by the line in both directions. Crossing $\wGamma[\mathsf{f}]$ in the opposite direction is equivalent to crossing $\wGamma[\mathsf{f}^{\rm fl}]$ in the original one, see \eqref{eq: flip}; hence each vertex appears both as $\mathsf{f}$ and as $\mathsf{f}^{\rm fl}$, for a total of 2 diagrams with one bilinear and $2\times2^2=8$ with two bilinears (the factor $\times2$ stands for the two rings).}
\end{figure}

Let us introduce explicitly the statistical factor coming from \eqref{eq: goku}:
\begin{equation*}
    \mathcal{N}(k_+,k_-,\ell)\equiv n_{\rm F}(k_+)\left[1-n_{\rm F}(k_-)\right]\Big[\Lambda_2(z_b,z_a)-n_{\rm F}(\ell)\,\Lambda_1(z_a)\Lambda_1(z_b)\Big]{\rm sgn}(k_-^0)\,{\rm sgn}(\ell^0)\,{\rm sgn}(k_+^0)\;,
\end{equation*}
with $z_a=\beta\cdot q_a$ and $z_b=\beta\cdot q_b$ for an iso-singlet. The above term satisfies the following fundamental relation:
\begin{equation}\label{eq: symm N}
    \mathcal{N}(k,q_a,q_b)=\mathcal{N}(k,-q_b,-q_a)\;.
\end{equation}
The above relation is direct consequence of the inversion of the chain $a\to b$: $k_-\to\ell\to k_+$  where the exchanged momenta are $q_a=\ell-k_{-}$ and $q_b=k_+-\ell$ in the chain $b\to a$: $k_+\to \ell\to k_- $ where the exchange momenta are now $q'_a=\ell-k_+$ and $q'_b=k_--\ell$. Being $q'_a=-q_b$ and $q'_b=-q_a$ the exchange in \eqref{eq: symm N} follows. To identity \eqref{eq: symm N} can be proved explicitly by considering the following trivial properties of the Dirac factor:
\begin{equation*}
    n_{\rm F}(k_+)[1-n_{\rm F}(k_-)]=\e^{-\beta\cdot q}n_{\rm F}(k_-)[1-n_{\rm F}(k_+)]\;,
\end{equation*}
and of the imaginary time weights:
\begin{equation*}
    \Lambda_1(-z)=\e^{-z}\Lambda_1(z)\;,\qquad \Lambda_2(-z_a,-z_b)=\e^{-(z_a+z_b)}\Lambda_2(z_b,z_a)\;.
\end{equation*}
Now the above derivation can be extended to the case of self-conjugate fields.
 For a self-conjugate multiplet, such as the pion, the modes are related by \eqref{eq: mode self-adjoint}, and also $\langle\wvrphi\wvrphi\rangle_{\rm GE}$ and $\langle\compactphi\compactphi\rangle_{\rm GE}$ are different from zero. The number of connected pairings increases accordingly: a ring through $n$ bilinears can be closed in $(n-1)!\,2^{n-1}$ ways instead of $(n-1)!$, because each bilinear other than the first one can be crossed in two directions. Thus there are 2 pairings instead of 1 in the four-field function and 8 instead of 2 in the six-field function (see figure \ref{fig: diagrams2}).

Consider a generic bilinear with vertex $\mathsf{f}$, 
\begin{equation*}
    \wGamma[\mathsf{f}](y)=\frac{1}{(2\pi)^8}\int\di^4p\,\di^4p'\,\e^{\ii(p-p')\cdot y}\,\compactphi^\mrr_\alpha(p)\,\mathsf{f}^{\mrr\mss}_{\alpha\beta}(p,p')\,\wvrphi^\mss_\beta(p')\;.
\end{equation*}
Using \eqref{eq: mode self-adjoint} for both fields, $\compactphi^\mrr_\alpha(p)=C^{\mrr\mrr'}\wvrphi^{\mrr'}_\alpha(-p)$ and $\wvrphi^\mss_\beta(p')=C^{\mss\mss'}\compactphi^{\mss'}_\beta(-p')$, commuting the two (bosonic) fields and renaming $P=-p'$, $P'=-p$, one finds that, up to a c-number which does not contribute to the connected correlators,
\begin{equation}\label{eq: flip}
    \wGamma[\mathsf{f}]=\wGamma[\mathsf{f}^{\rm fl}]\;,\qquad \left[\mathsf{f}^{\rm fl}(P,P')\right]^{\mss\mrr}_{\beta\alpha}\equiv C^{\mss\mss'}\left[\mathsf{f}(-P',-P)\right]^{\mrr'\mss'}_{\alpha\beta}C^{\mrr'\mrr}\;,
\end{equation}
that is, only the part of the vertex which is symmetric under the ``flip'' \eqref{eq: flip}, $\mathsf{f}^{\rm sym}\equiv(\mathsf{f}+\mathsf{f}^{\rm fl})/2$, is physical. The pairing in which a bilinear is crossed in the opposite direction is obtained from the original one precisely by the exchange of the roles of its two fields, i.e. it is equal to the original pairing with $\mathsf{f}\to\mathsf{f}^{\rm fl}$. Therefore the sum over the two directions of each vertex amounts to the replacement:
\begin{equation*}
    \mathsf{f}_a\;\longrightarrow\;\mathsf{f}_a+\mathsf{f}^{\rm fl}_a=2\,\mathsf{f}^{\rm sym}_a\;,
\end{equation*}
in the linear kernel \eqref{def: G lin} and, for both vertices, in the quadratic kernel \eqref{def: G quad}; nothing else changes, in particular the structure of the two rings, the statistical factor \eqref{def: Sstat} and the hypersurface integration. 


\section{Integration over the decoupling hypersurface}\label{APP: integration over Sigma}
In the linear response expansion, the $N-$th order correction depends on the following double integral:
\begin{equation}\label{eq: I lin}
    I^{\alpha_1\ldots\alpha_N}_\mu[\vartheta](x)\equiv\int_{\Sigma_{\rm D}}\di\Sigma_\mu(y)\,\vartheta(y,x)\int\di^4q\,\delta(k\cdot q)\,q^{\alpha_1}\cdots q^{\alpha_N}\,\e^{\ii q\cdot(x-y)}\;.
\end{equation}
The point $x$ appears in \eqref{eq: I lin} in two places: in the phase, where it is the starting point of the worldline, and in $\vartheta(y,x)$, where it is the reference point of the gradient expansion. It is convenient to keep them distinct: we write the phase as $\ii q\cdot(x'-y)$ and set $x'=x$ at the end. Then all the $q$'s can be traded for derivatives with respect to $x'$, of this phase and these derivatives can be taken out of the integrals, in formulae:
\begin{equation}\label{eq: I as derivative}
\begin{split}
    I^{\alpha_1\ldots\alpha_N}_\mu(x)&\equiv(-\ii)^N\,\partial^{\alpha_1}_{x'}\cdots\partial^{\alpha_N}_{x'}\,L_\mu(x,x')\Big|_{x'=x}\;,\\
    L_\mu(x,x')&\equiv\int_{\Sigma_{\rm D}}\di\Sigma_\mu(y)\,\vartheta_a(y,x)\int\di^4q\,\delta(k\cdot q)\,\e^{\ii q\cdot(x'-y)}\;,
\end{split}
\end{equation}
where is understood that $x'$ is set equal to $x$ \emph{only after} all the derivatives have acted upon $L_\mu$.

Now the innermost integral in $\di^4q$ of $L_\mu$ can be performed by taking advantage of the integral representation of $\delta$:
\begin{equation*}
    \delta(k\cdot q)=\frac{1}{2\pi}\int\di\tau\,\e^{-\ii\tau k\cdot q}\;,\quad\tau\in\mathbb{R}\;,
\end{equation*}
so that:
\begin{equation*}
    \int\di^4q\,\delta(k\cdot q)\,\e^{\ii q\cdot(x'-y)}=(2\pi)^3\int_{-\infty}^{+\infty}\di\tau\,\delta^4(x'-y-\tau k)\;.
\end{equation*}
The decoupling hypersurface $\Sigma_{\rm D}$ can be implicitly defined by the condition $\Phi(y)=0$, with $\di\Sigma_\mu(y)=\di^4y\,\delta(\Phi(y))\,\partial_\mu\Phi(y)$ (the sign of $\Phi$ fixes the orientation). The integration over $y$ sets $y=x'-\tau k$, and:
\begin{equation*}
    \delta\big[\Phi(x'-\tau k)\big]=\sum_i\frac{\delta(\tau-\tau_i(x'))}{|k\cdot\partial\Phi(\overline{y}_{i,k}(x))|}\;,\qquad\overline{y}_{i,k}(x')\equiv x'-\tau_i(x')\,k\;,\qquad\Phi(\overline{y}_{i,k}(x'))=0\;,
\end{equation*}
where the sum runs over all the intersections of the line $x'-\tau k$ with $\Sigma_{\rm D}$ (if the hypersurface has time-like parts there can be more than one). Since $\partial_\mu\Phi/|k\cdot\partial\Phi|=\hat{n}_\mu/|k\cdot \hat{n}|$, with $\hat{n}_\mu$ the unit normal with the orientation of $\di\Sigma_\mu$ we finally obtain:
\begin{equation}\label{eq: L result}
    L_\mu(x,x')\equiv(2\pi)^3\sum_i[\mathcal{R}^a_\mu(y,x)]_{y=\overline{y}_{i,k}(x')}=(2\pi)^3\sum_i\left[\frac{\hat{n}_\mu(y)}{|k\cdot \hat{n}(y)|}\vartheta_a(y,x)\right]_{y=\overline{y}_{i,k}(x')}\;.
\end{equation}
Now in order to compute \eqref{eq: I as derivative} from \eqref{eq: L result} we need the derivatives with respect to $x'$ of a functions of the form $H(\overline{y}_k(x'),x)$, with $H$ representing a generic $N-$th derivative of $L_\mu(x,x')$ with respect to $x'$. 

To compute the action of such derivative we start differentiating $\Phi(x'-\tau(x')k)=0$ with respect to $x'$:
\begin{equation*}
    \partial^{x'}_\alpha\Phi(x')-(k^\sigma\partial^{x'}_\sigma\Phi(x'))\frac{\partial\tau(x')}{\partial x'^\alpha}=0\quad\Longrightarrow\quad\frac{\partial\tau(x')}{\partial x'^\alpha}=\frac{\hat{n}_\alpha(\overline{y}_k(x))}{k\cdot \hat{n}(\overline{y}_k(x))}\;,
\end{equation*}
that is:
\begin{equation}\label{eq: dtau}
    \frac{\partial\overline{y}^\beta_k(x')}{\partial x'^\alpha}=\delta^\beta_\alpha-\frac{k^\beta \hat{n}_\alpha(\overline{y}_k(x'))}{k\cdot \hat n(\overline{y}_k(x'))}\equiv\mathcal{P}^\beta{}_\alpha(\overline{y}_k(x'))\;.
\end{equation}
Hence the derivative with respect to $x'$ becomes:
\begin{equation*}
    \frac{\partial H(\overline{y}(x'),x)}{\partial x'^\alpha}=\mathcal{P}^\beta_\alpha(\overline{y}_k(x'))\frac{\partial H(\overline{y}(x'),x)}{\partial\overline{y}^\beta_k(x')}\;.
\end{equation*}
Introducing:
\begin{equation}\label{eq: Dsig}
    \mathfrak{D}^y_\alpha(y)\equiv\mathcal{P}^\beta_{\,\alpha}(y)\partial^y_\beta\;,\qquad \frac{\partial H(y(x'),x)}{\partial x'^\alpha}=\mathfrak{D}^y_\alpha(y) H(y,x)\;,
\end{equation}
the $N-$th derivative with respect to $x'$ is given by the repeated combination of $N$ operator $\mathfrak{D}^{\alpha_i}_y(y)$:
\begin{equation}\label{eq: Dsig iterated}
    \partial^{x'}_{\alpha_1}\cdots\partial^{x'}_{\alpha_N}H(x,\overline{y}(x'))=\left[\mathfrak{D}_{\alpha_1}^y(y)\circ\mathfrak{D}_{\alpha_2}^y(y)\circ\ldots\circ\mathfrak{D}_{\alpha_N}^y(y)H(y,x)\right]\Big|_{y=\overline{y}(x')}\;,
\end{equation}
where each $\mathfrak{D}$ acts on everything on its right, including the normal vectors contained in the operators $\mathfrak{D}$ on its right. In particular, for $N=2$, since $\partial_{\beta_1}\mathcal{P}^{\beta_2}_{\alpha_2} = - k^{\beta_2}\partial_{\beta_1}\nu_{\alpha_2}$ with $\nu_\alpha = \hat n_\alpha/k\cdot \hat n$:
\begin{equation}\label{eq: D2 explicit}
    \mathfrak{D}_{\alpha_1}\circ\mathfrak{D}_{\alpha_2}H = \mathcal{P}^{\beta_1}\,_{\alpha_1}\mathcal{P}^{\beta_2}\,_{\alpha_2}\partial_{\beta_1}\partial_{\beta_2}H - \mathcal{P}^{\beta_1}\,_{\alpha_1}(\partial_{\beta_1}\nu_{\alpha_2})k^{\beta_2}\partial_{\beta_2}H.
\end{equation}
The second term is of the same order of the first one when $H$ is the product $\nu_\mu\vartheta_\alpha$, compare with \eqref{eq: DR 2}, and it is necessary in order for the result to depend only on the values of $H$ on $\Sigma_{\rm D}$: the combination $\mathfrak{D}_{\alpha_1}\circ\mathfrak{D}_{\alpha_2}H$ vanishes for $H = \Phi\,h$, with $h$ arbitrary, at $\Phi=0$, whereas the first term alone does not. 

Plugging the equation \eqref{eq: Dsig iterated} in \eqref{eq: I as derivative} and using that $H(x,y(x'))$ is given by \eqref{eq: L result} we thus obtain that the linear response expansion of the Wigner function is:
\begin{equation}\label{A: LIN FINAL}
    \begin{split}
        \Delta_{\rm lin}W(x,k)&=(2\pi)^3\sum_a\sum_{N=0}^\infty\frac{(-\ii)^N}{N!}\,\partial^q_{\alpha_1}\cdots\partial^q_{\alpha_N}[G^\mu_a(k,q)]^{\mm\nn}_{\lambda\rho}\Big|_{q=0}\\
        &\times\sum_i\left[\mathfrak{D}^{\alpha_1}_y(y)\circ\ldots\circ\mathfrak{D}^{\alpha_N}_y(y)\left(\frac{\hat{n}_\mu(y)}{|k\cdot \hat{n}(y)|}\,\vartheta_a(y,x)\right)\right]\Big|_{y=\overline{y}_{i,k}(x)}\;.
    \end{split}
\end{equation}
Two properties of the mixed projector $\mathcal{P}^\beta_{\,\alpha}$ are worth to emphasize. First:
\begin{equation*}
    \hat{n}_\beta(y)\mathcal{P}^\beta_{\,\alpha}(y)=0\;,
\end{equation*}
which implies that $\mathfrak{D}^y_\alpha(y)$ differentiates only inside $\Sigma_{\rm D}$ and not in the directions orthogonal to it. This means that $\mathfrak{D}^\alpha_y(y)H(x,y)$, and in general the result of the $N-$th composition with $\mathfrak{D}_y(y)$, only depends on the values of the function $H$ on the hypersurface and is itself a function on $\Sigma_{\rm D}$ for any $N$. Second:
\begin{equation*}
    k^\alpha\mathcal{P}^\beta_{\,\alpha}(y)=0\;,
\end{equation*}
which implies that $k_\alpha\mathfrak{D}^\alpha_y(y)=0$ consistently with the on-shell condition $k\cdot q=0$.

To make notation more compact we introduce (for the trivial intersection, which lies on the space-like part of $\Sigma_{\rm D}$ where $k\cdot\hat n>0$, one has $\mathfrak{s} =1$ and the definition with \eqref{def: R e nu maintext}):
\begin{equation}\label{def : R e nu}
    \mathcal{R}^a_{\mu}(y,x)\equiv \mathfrak{s}(y,x)\nu_\mu(y)\vartheta_a(y,x)\;,\quad \nu_\mu(y)\equiv\frac{\hat{n}_\mu(y)}{k\cdot\hat{n}(y)}\;,\qquad\mathfrak{s}(y,x)\equiv{\rm sgn}(k\cdot\hat{n}(\bar y_k(x)))\;,
\end{equation}
so that the derivative operator \eqref{eq: Dsig} reads:
\begin{equation*}
    \mathfrak{D}^\alpha(y)=\partial^\alpha_y-\nu^\alpha(y)k\cdot\partial^y\;,
\end{equation*}
while, considering only the trivial intersection $\overline{y}_{i,k}=x$, the linear response expansion of the Wigner function \eqref{A: LIN FINAL} reads:
\begin{equation*}
    \begin{split}
        \Delta_{\rm lin}W(x,k)&=(2\pi)^3\sum_a\!\sum_{N=0}^\infty\!\!\frac{(-\ii)^N}{N!}\partial^q_{\alpha_1}\cdots\partial^q_{\alpha_N}[G^\mu_a(k,q)]^{\mm\nn}_{\lambda\rho}\Big|_{q=0}\left[\mathfrak{D}^{\alpha_1}_y(y)\circ\ldots\circ\mathfrak{D}^{\alpha_N}_y(y)\mathcal{R}^a_\mu(y,x)\right]\Big|_{y=x}\;.
    \end{split}
\end{equation*}
We now come to the quadratic response term. 
The double hypersurface integral has a similar structure to the linear one \eqref{eq: I lin}:
\begin{equation}\label{eq: master}
    \begin{split}
        \mathcal{I}^{\alpha_1\ldots\alpha_N\beta_1\ldots\beta_M}_{\mu_1\mu_2}(x)&\equiv\int\di\Sigma_{\mu_1}(y_1)\vartheta_a(y_1,x)\int\di\Sigma_{\mu_2}(y_2)\vartheta_b(y_2,x)\int\di^4q_a\di^4q_b\\
        &\times\,\delta(k\cdot q_a-s)\,\delta(k\cdot q_b+s)\,q^{\alpha_1}_a\ldots q^{\alpha_N}_aq^{\beta_1}_b\ldots q^{\beta_M}_b\,\e^{\ii q_a\cdot(x-y_1)+\ii q_b\cdot(x-y_2)}\;,
    \end{split}
\end{equation}
with  $s\equiv q_a\cdot q_b/2$. Note that the two $\delta$'s arguments are equivalent to $q_a\cdot (k - q_b/2) = k\cdot q_a-s$ and $q_b\cdot(k+q_a/2)=k\cdot q_b +s$.

As in the linear case we separate the roles of $x$ in the integral. We denote with $x_{1,2}$ the starting points of the two worldlines that appear in the phases, while $x$ is the reference point in $\vartheta(y,x)$. In this way the factors $q_a$ and $q_b$ are brought-down by repeated action of $\partial_{x_{1,2}}$ which can be therefore extracted from the integral and \eqref{eq: master} can be written as:
\begin{equation*}
    \begin{split}
        \mathcal{I}^{\alpha_1\ldots\alpha_N\beta_1\ldots\beta_M}_{\mu_1\mu_2}(x)&\equiv(-\ii)^{N+M}\partial^{\alpha_1}_{x_1}\ldots\partial^{\alpha_N}_{x_1}\partial^{\beta_1}_{x_2}\ldots\partial^{\beta_M}_{x_2}\mathcal{L}_{\mu_1\mu_2}(x,x_1,x_2)\Big|_{x_1=x_2=x}\;,\\
        \mathcal{L}_{\mu_1\mu_2}(x,x_1,x_2)&\equiv \int\di\Sigma_{\mu_1}(y_1)\vartheta_a(y_1,x)\int\di\Sigma_{\mu_2}(y_2)\vartheta_b(y_2,x)\\
        &\times\int\di^4q_a\di^4q_b\,\delta(k\cdot q_a-s)\,\delta(k\cdot q_b+s)\,\e^{\ii q_a\cdot(x_1-y_1)+\ii q_b\cdot(x_2-y_2)}\;,
    \end{split}
\end{equation*}
where $s\equiv q_a\cdot q_b/2$.
After the integral is computed and all the derivatives acted on $\mathcal{L}$ we set $x_{1}=x_2=x$. 

In order to perform the inner integral $\mathcal{L}$ we represent both $\delta$'s in Fourier transform:
\begin{equation*}
    \delta(k\cdot q_a-s)\,\delta(k\cdot q_b+s)=\frac{1}{(2\pi)^2}\int\di\tau_1\di\tau_2\,\e^{-\ii\tau_1k\cdot q_a-\ii\tau_2k\cdot q_b}\,\e^{\ii(\tau_1-\tau_2)s}\;.
\end{equation*}
The two deltas are coupled through the mixed exponential $\exp(\ii s(\tau_1-\tau_2))$. It can be expanded in powers of $s$ as follows:
\begin{equation}\label{eq: exp etau}
    \e^{\ii(\tau_1-\tau_2)s}=\sum_{j\ge0}\frac{\ii^j(\tau_1-\tau_2)^j}{j!}s^j=\sum_{n,m\ge0}\frac{\ii^{n+m}(-1)^m}{n!\,m!2^{n+m}}\,\tau_1^n\tau_2^m\,(q_a\cdot q_b)^{n+m}\;,
\end{equation}
which is an expansion in powers of $q_a\cdot q_b$. Now using that each factor $q$ can be written in terms of a derivative with respect to $x$ of the exponential,
 every factor $(q_a\cdot q_b)^{n+m}$ in \eqref{eq: exp etau} can be replaced by $(-1)^{n+m}(\partial_{x_1}\cdot\partial_{x_2})^{n+m}$ once acting on the exponent and we can thus write:
\begin{equation}\label{eq: matcal L 2}
    \begin{split}
        \mathcal{L}_{\mu_1\mu_2}(x,x_1,x_2)&\equiv \frac{1}{(2\pi)^2}\int\di\Sigma_{\mu_1}(y_1)\vartheta_a(y_1,x)\int\di\Sigma_{\mu_2}(y_2)\vartheta_b(y_2,x)\int\di\tau_1\di\tau_2\\
        &\times\sum_{n,m\ge 0}c_{nm}(\partial_{x_1}\cdot\partial_{x_2})^{n+m}\tau^n_1\tau^m_2\int\di^4q_a\di^4q_b\e^{\ii q_a\cdot(x_1-y_1-\tau_1k)}\e^{\ii q_b\cdot(x_2-y_2-\tau_2 k)}\;,
    \end{split}
\end{equation}
where the coefficient $c_{nm}$ is defined as:
\begin{equation}\label{def: coeff c}
    c_{nm}\equiv\frac{\ii^{n+m}(-1)^n}{n!m!\,2^{n+m}}\;,\quad c_{00}=1,\; c_{01}=c^*_{10}=\frac{\ii}{2},\;c_{11}=\frac{1}{4}\;.
\end{equation}
In \eqref{eq: matcal L 2} the two integrals in $q_{a,b}$ are now factorized in two identical copies:
\begin{equation*}
    L^{(\tau^n_i)}_{\mu_i}(x_i,x)\equiv\frac{1}{2\pi}\int{\di\tau}_i\tau^n_i\int_{\Sigma_{\rm D}}\di\Sigma_{\mu_i}(y_i)\vartheta_{a,b}(y_i,x)\int\di^4 q_{a,b}\e^{\ii q_{a,b}\cdot(x_i-y_i-\tau_ik)}\;.
\end{equation*}
The integral above is formally equivalent to \eqref{eq: I as derivative} with the only difference of the external $\tau^n_i$ factor. The computation then proceeds analogously to the passages from \eqref{eq: I as derivative} to \eqref{eq: L result} leading to:
\begin{equation}\label{eq: tau L result}
    L^{(\tau^n_j)}_{\mu_i}(x_i,x)
    =(2\pi)^3\sum_{\overline{y}_{i,k}}
    [\tau^n_j\,\mathcal{R}^{a,b}_{\mu_i}(y,x)]
    \bigg|_{\substack{
        y=\overline{y}_{i,k}(x_i)\\
        \tau_i=\tau(x_i)
    }}\;,
\end{equation}
where $\mathcal{R}^a$ has been defined in \eqref{def : R e nu}. Now plugging back \eqref{eq: tau L result} in \eqref{eq: matcal L 2} the double integral \eqref{eq: master} turns out to be:
\begin{equation}\label{eq: master2}
    \begin{split}
        \mathcal{I}^{\alpha_1\ldots\alpha_N\beta_1\ldots\beta_M}_{\mu_1\mu_2}(x)&=(2\pi)^6(-\ii)^{N+M}\sum_{\overline{y}_{i,k}(x_1)}\sum_{\overline{y}_{j,k}(x_2)}\partial^{\alpha_1}_{x_1}\ldots\partial^{\alpha_N}_{x_1}\partial^{\beta_1}_{x_2}\ldots\partial^{\beta_M}_{x_2}\\
        &\times\Bigg\{\sum_{n,m\ge0}c_{nm}(\partial_{x_1}\cdot\partial_{x_2})^{n+m}\Big[(\tau^n_1\mathcal{R}_a(y_1,x))\bigg|_{\substack{
        y=\overline{y}_{i,k}(x_1)\\
        \tau_1=\tau(x_1)
    }}(\tau^m_{2}\mathcal{R}_b(y_2,x))\bigg|_{\substack{
        y=\overline{y}_{j,k}(x_2)\\
        \tau_2=\tau(x_2)
    }}\Big]\Bigg\}\;.
    \end{split}
\end{equation}
Now the derivative with respect to $x_{1,2}$ acts on a function $H(\overline{y}(x_{1,2}),\tau(x_{1,2}),x)$. We thus have:
\begin{equation*}
    \begin{split}
        \partial_\gamma^{x_{1,2}}H(\overline{y}(x_{1,2}),\tau(x_{1,2}))&=\frac{\partial\tau(x_{1,2})}{\partial x^{\gamma}_{1,2}}\frac{\partial H(\overline{y},x_{1,2})}{\partial\tau}+\frac{\partial\overline{y}^\sigma(x_{1,2})}{\partial x^\gamma_{1,2}}\frac{\partial H(y,x_{1,2})}{\partial y^\sigma}\\
    &=\left[\nu_\gamma(x_{1,2})\frac{\partial}{\partial\tau}+\mathcal{P}^\sigma_{\,\gamma}(x_{1,2})\frac{\partial}{\partial y^\sigma}\right]H(y,\tau)\Big|_{y=\overline{y}(x_{1,2})}\;.
    \end{split}
\end{equation*}
Thus the derivative with respect to $x_{1,2}$ can be expressed in terms of the differential operator:
\begin{equation}
    \mathsf{D}^\alpha(y)\equiv\mathfrak{D}^\alpha(y)+\nu^\alpha(y)\partial_\tau=\mathcal{P}^\beta_{\,\alpha}(y)\partial^y_{\beta}+\nu^\alpha(y)\partial_\tau\;,
\end{equation}
defined in equation \eqref{def: D second order} in the main text,
where we used the definition \eqref{eq: Dsig} of the linear order differential operator $\mathfrak{D}(y)$.
Now, taking into account that:
\begin{equation*}
    (\partial_{x_1}\cdot\partial_{x_2})^r=\partial^{\gamma_1}_{x_1}\ldots\partial^{\gamma_r}_{x_1}\partial_{\gamma_1}^{x_2}\ldots\partial_{\gamma_r}^{x_2}\;,
\end{equation*}
the integral \eqref{eq: master2} can be finally written as:
\begin{equation*}
    \begin{split}
        \mathcal{I}^{\alpha_1\ldots\alpha_N\beta_1\ldots\beta_M}_{\mu_1\mu_2}(x)=(2\pi)^6(-\ii)^{N+M}\sum_{i,j}&\sum_{n,m\ge0}c_{nm}\left[\mathsf{D}^{\alpha_1}_y\circ\ldots\circ\mathsf{D}_y^{\alpha_N}\circ\mathsf{D}^y_{\gamma_1}\circ\ldots\mathsf{D}^y_{\gamma_{n+m}}\left(\tau^n\mathcal{R}^a_{\mu_1}(y,x)\right)\right]\Big|_{i}\\
        &\times \left[\mathsf{D}^{\beta_1}_y\circ\ldots\circ\mathsf{D}_y^{\beta_M}\circ\mathsf{D}_y^{\gamma_1}\circ\ldots\mathsf{D}_y^{\gamma_{n+m}}\left(\tau^m\mathcal{R}^b_{\mu_2}(y,x)\right)\right]\Big|_{j}\;,
    \end{split}
\end{equation*}
with $i,j$ is a shorthand for $y=\overline{y}_{i,j}(x,k)$ and $\tau=\tau_{i,j}(x)$.
With this we finally obtain:
\begin{equation*}
    \begin{split}
        \Delta_{\rm quad}W(x,k)&=(2\pi)^6\sum_{a,b}\sum_{i,j}\sum_{N,M}\frac{(-\ii)^{N+M}}{N!M!}\partial^{q_a}_{\alpha_1}\ldots\partial^{q_a}_{\alpha_N}\partial^{q_b}_{\beta_1}\ldots\partial^{q_b}_{\beta_M}\,[\mathsf{G}^{\mu_1\mu_2}_{ab}(k,q_a,q_b)^{\mm\nn}_{\lambda\rho}]\Big|_{q_a=q_b=0}\\
        &\times\sum_{n,m\ge0}c_{nm}\Bigg\{\left[\mathsf{D}^{\alpha_1}_y\circ\ldots\circ\mathsf{D}_y^{\alpha_N}\circ\mathsf{D}^y_{\gamma_1}\circ\ldots\mathsf{D}^y_{\gamma_{n+m}}\left(\tau^n\mathcal{R}^a_{\mu_1}(y,x)\right)\right]\Big|_{i}\\
        &\times \left[\mathsf{D}^{\beta_1}_y\circ\ldots\circ\mathsf{D}_y^{\beta_M}\circ\mathsf{D}_y^{\gamma_1}\circ\ldots\mathsf{D}_y^{\gamma_{n+m}}\left(\tau^m\mathcal{R}^b_{\mu_2}(y,x)\right)\right]\Big|_{j}\Bigg\}\;,
    \end{split}
\end{equation*}
which is equation \eqref{eq: QUAD FINAL} in the main text.

\section{Parity-gradient selection rules}\label{APP: computation G}
In this appendix we prove the fundamental identities \eqref{eq: Tswap} and \eqref{eq: Gswap} used in the computation of the quadratic response correction to the Wigner function of a Dirac field.

Let $\mathsf{C}$ be the charge-conjugation matrix. We have:
\begin{equation*}
    \mathsf{C}(\gamma^{\mu})^T\mathsf{C}^{-1}=-\gamma^\mu\;,\qquad \mathsf{C}(\gamma^{5})^T\mathsf{C}^{-1}=\gamma^5\;.
\end{equation*}
Each microscopic kernel $G_{ab}$ and $\mathcal{G}_{ab}$ are in general matrices in the spinor space. Let denote the most general of these matrices with $X$. We define the transformed of $X$ under the matrix $\gamma^5\mathsf{C}$ as:
\begin{equation}\label{def: tilde}
    \widetilde{X}\equiv\gamma^5\,\mathsf{C}X^T\mathsf{C}^{-1}\gamma^5\;.
\end{equation}
This operation reverses the order of the products, $\widetilde{XY}=\widetilde{Y}\widetilde{X}$, and leaves the trace invariant, $\tr\widetilde{X}=\tr X$. 

We are now interested in its action on the general matrices appearing in the microscopic kernels  For the projector in the spectral function we have:
\begin{equation*}
    \mathsf{C}(\slashed{p}+m)^T\mathsf{C}^{-1}=-\slashed{p}+m\implies \widetilde{\Pi}(p)=\Pi(p)\;,
\end{equation*}
while for the axial insertion we have:
\begin{equation*}
     \widetilde{\gamma^\sigma\gamma^5}=-\gamma^\sigma\gamma^5\;.
\end{equation*}
The overall transformation of the given microscopic kernel then is determined from those of the vertices inserted:
\begin{equation}\label{eq: transfo f}
    \widetilde{f}(p,p')\equiv\gamma^5\mathsf{C}\,[\mathsf{f}_a(p',p)]^T\,\mathsf{C}\gamma^5\,,
\end{equation}
where $p,p'\to p',p$ is necessary to preserve the reversing of the vertex.
All the vertices of this work, under \eqref{eq: transfo f} are such that:
\begin{equation*}
    \widetilde{\mathsf{f}}_a(p,p')=-c_a\,\mathsf{f}_a(p,p')\;,\qquad c_T=c_{T_{\rm C}}=c_j=-1\;,\quad c_{\mathcal{S}}=+1\;,
\end{equation*}
which is exactly the relation \eqref{eq: vertex inversion} in the main text.

\begin{figure}
\centering
\begin{tikzpicture}[>=Stealth,line width=0.6pt,scale=0.82,every node/.style={scale=0.82},
    fline/.style={postaction={decorate},decoration={markings,mark=at position .55 with {\arrow{Stealth[length=2.2mm]}}}},
    src/.style={decorate,decoration={snake,amplitude=1pt,segment length=4pt}},
    vtx/.style={circle,fill,inner sep=0pt,minimum size=4.5pt},
    ax/.style={rectangle,draw,fill=white,inner sep=2pt}]
    \begin{scope}[shift={(0,0)}]
        \node[ax] (G) at (0,0) {$\gamma^\sigma\gamma^5$};
        \node[vtx] (A) at (-1.0,1.4) {};
        \node[vtx] (B) at (1.0,1.4) {};
        \draw[fline] (G)--(A) node[pos=0.45,left=5pt]{$k_-$};
        \draw[fline] (A)--(B) node[midway,above=1pt]{$\ell$};
        \draw[fline] (B)--(G) node[pos=0.55,right=5pt]{$k_+$};
        \draw[src] (A)--(-1.0,1.95); \node[anchor=south] at (-1.0,1.95) {$a$};
        \draw[src] (B)--(1.0,1.95); \node[anchor=south] at (1.0,1.95) {$b$};
        \node at (0,-0.65) {$\mathcal{T}^{\mu_1\mu_2,\sigma}_{ab}(q_a,q_b)$};
    \end{scope}
    \node at (2.35,0.8) {$=\;-c_ac_b$};
    \begin{scope}[shift={(4.7,0)}]
        \node[ax] (G) at (0,0) {$\gamma^\sigma\gamma^5$};
        \node[vtx] (A) at (1.0,1.4) {};
        \node[vtx] (B) at (-1.0,1.4) {};
        \draw[fline] (G)--(B) node[pos=0.45,left=5pt]{$k_+$};
        \draw[fline] (B)--(A) node[midway,above=1pt]{$\ell$};
        \draw[fline] (A)--(G) node[pos=0.55,right=5pt]{$k_-$};
        \draw[src] (A)--(1.0,1.95); \node[anchor=south] at (1.0,1.95) {$a$};
        \draw[src] (B)--(-1.0,1.95); \node[anchor=south] at (-1.0,1.95) {$b$};
        \node at (0,-0.65) {$\mathcal{T}^{\mu_2\mu_1,\sigma}_{ba}(-q_b,-q_a)$};
    \end{scope}
    \begin{scope}[shift={(-1.3,-1.9)}]
        \draw[fline] (0,0)--(0.8,0) node[midway,above]{$p$};
        \node at (1.15,0) {$\mapsto$};
        \draw[fline] (2.3,0)--(1.5,0) node[midway,above]{$p$};
        \node[anchor=west] at (2.35,0) {$\Pi(p)$};
        \draw (3.6,0)--(4.4,0); \node[vtx] at (4.0,0) {}; \draw[src] (4.0,0)--(4.0,0.4); \node[anchor=south] at (4.0,0.4) {$a$};
        \node at (4.75,0) {$\mapsto$};
        \node[anchor=west] at (5.0,0) {$-c_a\,\mathsf{f}_a$};
        \node[ax] at (6.95,0) {$\gamma^\sigma\gamma^5$};
        \node at (7.85,0) {$\mapsto$};
        \node[anchor=west] at (8.05,0) {$-\gamma^\sigma\gamma^5$};
    \end{scope}
\end{tikzpicture}
\caption{\label{fig: flip} Graphical meaning of the operation \eqref{def: tilde} applied to the axial trace of a chain with two vertices, with the conventions of figure~\ref{fig: diagrams}: the direction of the chain is reversed. Bottom: action on the single elements; the projector is unchanged, the vertex $\mathsf{f}_a$ gets the sign $-c_a$ and the axial insertion changes sign. Since the trace is invariant, the two chains are related by \eqref{eq: Tswap general}. Without the axial insertion (scalar part) the overall sign is $+c_ac_b$.}
\end{figure}
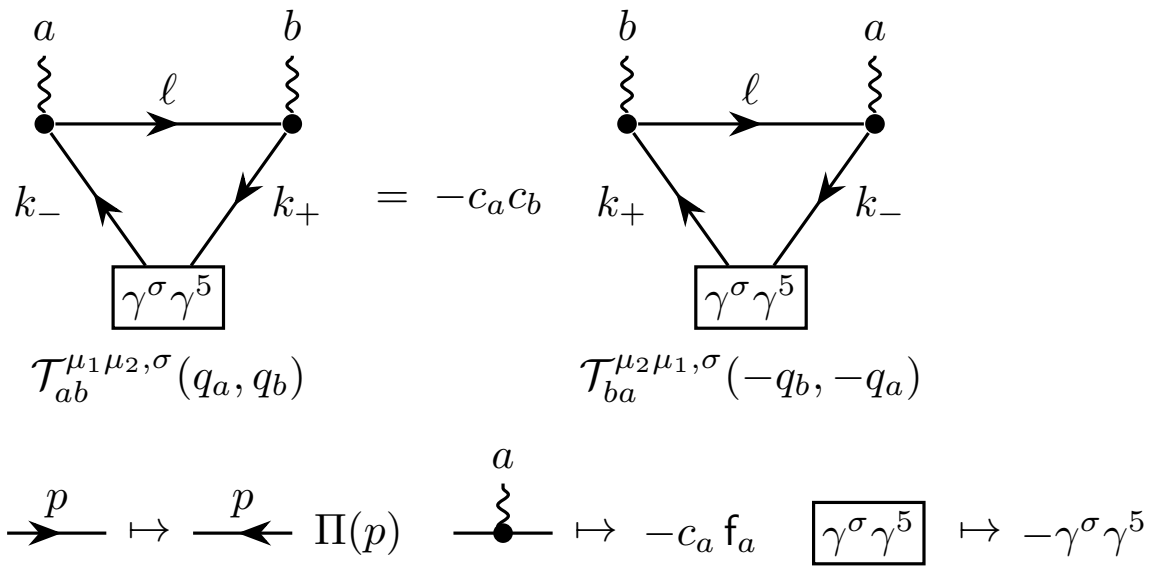

The physical meaning of the operation \eqref{def: tilde} is clear once it is applied to a chain, see figure~\ref{fig: flip}: it reverses the direction of the chain. The projectors are left unchanged, each vertex gets the sign $-c_a$, the axial insertion changes sign and the order of the factors is reversed. Since the vertices only depend on $p+p'$, the reversed chain is again a chain of the same type, with the vertices in the opposite order and $k_+\leftrightarrow k_-$, i.e.\ $(q_a,q_b)\to(-q_b,-q_a)$, while the internal momentum $\ell$ is unchanged.

At linear response order we are interested in:
\begin{equation*}
    \mathcal{T}^{\mu,\sigma}_a(q)\equiv\tr[\gamma^\sigma\gamma^5\Pi(k_-)\mathsf{f}^\mu_a(k_-,k_+)\Pi(k_+)]\;.
\end{equation*}
Applying \eqref{def: tilde} to the above trace  the whole product transform as:
\begin{equation*}
    \mathcal{T}^{\mu,\sigma}_a(q)=\tr\left[\Pi(k_+)\left(-c_a\mathsf{f}^\mu_a\right)\Pi(k_-)\left(-\gamma^\sigma\gamma^5\right)\right]=c_a\tr\left[\gamma^\sigma\gamma^5\Pi(k_+)\mathsf{f}^\mu_a(k_+,k_-)\Pi(k_-)\right]=c_a\,\mathcal{T}^{\mu,\sigma}_a(-q)\;,
\end{equation*}
where we used that $q\to-q$ for $k_+\to k_-$. The remaining statistical factor of the kernel \eqref{eq: Dirac Glin explicit} is even in $q$. Hence the axial current is \emph{always odd} in $q$ which was clear from its general expression \eqref{eq: vertex inversion}.
Proceeding analogously for the scalar trace we obtain that for $a=T,j$ it is even in $q$ while  is odd for $a=\mathcal{S}$, in agreement with \eqref{eq: LINEAR scalar f} and \eqref{eq: LINEAR axial f}.

For the quadratic response the symmetry is less obvious given that the expressions are much more involved. The general axial trace is:
\begin{equation*}
    \begin{split}
        \mathcal{T}^{\mu_1\mu_2,\sigma}_{ab}(q_a,q_b)&=\tr\left[\gamma^\sigma\gamma^5\,\Pi(k_-)\,\mathsf{f}^{\mu_1}_a(k_-,\ell)\,\Pi(\ell)\,\mathsf{f}^{\mu_2}_b(\ell,k_+)\,\Pi(k_+)\right]\\
        &=\tr\left[\Pi(k_+)\left(-c_b\mathsf{f}^{\mu_2}_b\right)\Pi(\ell)\left(-c_a\mathsf{f}^{\mu_1}_a\right)\Pi(k_-)\left(-\gamma^\sigma\gamma^5\right)\right]\\
        &=-c_ac_b\,\tr\left[\gamma^\sigma\gamma^5\,\Pi(k_+)\,\mathsf{f}^{\mu_2}_b(k_+,\ell)\,\Pi(\ell)\,\mathsf{f}^{\mu_1}_a(\ell,k_-)\,\Pi(k_-)\right]\;.
    \end{split}
\end{equation*}
The exchange $(q_a,q_b)\to(-q_b,-q_a)$ sends $q\to-q$, i.e. $k_+\leftrightarrow k_-$, and leaves $\ell=k+(q_a-q_b)/2$ unchanged; therefore:
\begin{equation}\label{eq: Tswap general}
    \mathcal{T}^{\mu_1\mu_2,\sigma}_{ab}(q_a,q_b)=-c_ac_b\,\mathcal{T}^{\mu_2\mu_1,\sigma}_{ba}(-q_b,-q_a)\;.
\end{equation}
Given that the statistical factor is invariant for $q_a,q_b\to-q_b,-q_a$ the identity \eqref{eq: Gswap} follows.
For the scalar part the insertion $\gamma^\sigma\gamma^5$ is absent and the overall sign is then $+c_ac_b$.

Note that the argument only uses the charge-conjugation properties of the vertices and their symmetry in the two momenta, so that it applies to any other pseudo-gauge whose vertices have a definite sign $c_a$. Also it can be extended to a chain with an arbitrary number of vertices allowing for a similar analysis for higher order response corrections.

For the particular case of the quadratic response we can use these relations to determine which terms of the gradient expansion survive. In the linear response \eqref{A: LIN FINAL} the term with $N$ derivatives of the axial kernel at $q=0$ is non-vanishing only if $(-1)^N=c_a$, that is for $N$ odd if $a=T,j$ and for $N$ even if $a=\mathcal{S}$: in both cases the term is of odd order in the gradients. 

In the quadratic response \eqref{eq: QUAD FINAL} consider the term $(N,M,n,m)$ of the pair $(a,b)$ and the term $(M,N,m,n)$ of the pair $(b,a)$. They contain the same product of derivatives of $\mathcal{R}^a$ and $\mathcal{R}^b$; their numerical coefficients are related by $c_{mn}=(-1)^{n+m}c_{nm}$, see \eqref{def: coeff c}, and, differentiating \eqref{eq: Gswap},
\begin{equation*}
    \partial^{q_a}_{\beta_1}\ldots\partial^{q_a}_{\beta_M}\partial^{q_b}_{\alpha_1}\ldots\partial^{q_b}_{\alpha_N}\mathcal{G}^{\mu_2\mu_1,\sigma}_{ba}\Big|_{0}=-c_ac_b(-1)^{N+M}\,\partial^{q_a}_{\alpha_1}\ldots\partial^{q_a}_{\alpha_N}\partial^{q_b}_{\beta_1}\ldots\partial^{q_b}_{\beta_M}\mathcal{G}^{\mu_1\mu_2,\sigma}_{ab}\Big|_{0}\;.
\end{equation*}
Therefore the second term is equal to the first one multiplied by $-c_ac_b(-1)^{N+M+n+m}$, and since the sum runs over all the pairs and all the values of $N,M,n,m$:
\begin{equation*}
    \Delta_{\rm quad}\mathcal{A}^\sigma=\frac12\sum_{a,b}\sum_{N,M,n,m}\left[1-c_ac_b(-1)^{N+M+n+m}\right]\Big(\text{term }(N,M,n,m)\text{ of }(a,b)\Big)\;.
\end{equation*}
On the trivial intersection the term $(N,M,n,m)$ is of order $N+M+n+m$ plus the number of spin potentials among the two sources, and $c_ac_b=-1$ precisely when there is one spin potential. Hence only the terms of odd order survive. For the scalar part the sign is the opposite and only the even orders survive. In conclusion, for the free Dirac field coupled to $\beta$, $\zeta$ and $\mathfrak{S}$, in the Belinfante as well as in the canonical pseudo-gauge, within linear and quadratic response:
\begin{equation*}
    \begin{split}
        &\mathcal{A}^\sigma(x,k)\;\;\text{contains only odd orders in the gradients}\,,\\
        &\mathcal{F}(x,k)=\tr \left[W(x,k)\right]\;\;\text{contains only even orders in the gradients}\,,
    \end{split}
\end{equation*}
as explicitly stated in the main text.

\section{Second-order corrections: Dirac field}\label{SEC: app dirac}
In  this appendix we report for completeness the explicit computation of the traces appearing at linear and quadratic order. The relations derived in appendix \ref{APP: computation G} and used in Section \ref{SEC: Dirac} in the main text can thus be tested with the exact expressions.

For the $\Lambda$ hyperon the two conserved densities are the stress-energy tensor $\wT^{\mu\nu}$ and the Abelian current. In the Belinfante pseudo-gauge these read:
\begin{equation}\label{eq: Dirac T e j}
    \begin{split}
        \wT^{\mu\nu}_{\rm B}(y)&=\frac{\ii}{4}\compactpsi(y)\gamma^\mu\partial^\nu\wpsi(y)-\frac{\ii}{4}\partial^\nu\compactpsi(y)\gamma^\mu\wpsi(y)+\left(\mu\leftrightarrow\nu\right)\;,\\
        \wj^{\mu}(y)&=\compactpsi(y)\gamma^\mu\wpsi(y)\;,\\
    \end{split}
\end{equation}
where the contraction with the spinorial indexes is understood. In the canonical pseudo-gauge we have instead:
\begin{equation}\label{eq: densities canonical}
    \begin{split}
        \wT^{\mu\nu}_{\rm C}(y)&=\frac{\ii}{2}\compactpsi(y)\gamma^\mu\partial^\nu\wpsi(y)-\frac{\ii}{2}\partial^\nu\compactpsi(y)\gamma^\mu\wpsi(y)\;,\\
        \widehat{\mathcal{S}}^{\mu,\lambda\rho}_{\rm C}(y)&=-\frac{1}{2}\epsilon^{\mu\lambda\rho\tau}\compactpsi(y)\gamma_\tau\gamma^5\wpsi(y)\;.
    \end{split}
\end{equation}
Note that $\wT^{\mu\nu}_{\rm B}=(1/2)(\wT^{\mu\nu}_{\rm C}+\wT^{\nu\mu}_{\rm C})$.
The densities \eqref{eq: Dirac T e j} and \eqref{eq: densities canonical} are naturally expressed in Fourier transform by using equations \eqref{def: FT field}. For instance, for the derivative term we obtain:
\begin{equation*}
    \begin{split}
        \compactpsi(y)\gamma^\mu\partial^\nu\wpsi(y)&=\frac{1}{(2\pi)^8}\int\di^4p\di^4p'\,(\gamma^\mu)_{AB}\left(\e^{\ii p\cdot y}\compactpsi_B(p)\right)\partial^\nu\left(\e^{-\ii p'\cdot y}\,\wpsi_A(p')\right)\\
        &=\frac{(-\ii)}{(2\pi)^8}\int\di^4p\di^4p'\,\e^{\ii(p-p')\cdot y}\,p'^\nu(\gamma^\mu)_{AB}\compactpsi_B(p)\wpsi_A(p')\;,
    \end{split}
\end{equation*}
which in turn implies:
\begin{equation}\label{eq: T Dir as integral}
    \begin{split}
        \wT^{\mu\nu}(y)&=\frac{1}{4(2\pi)^8}\int\di^4p\di^4p'\,\e^{\ii(p-p')\cdot y}\left[(p+p')^\nu\gamma^\mu+(p+p')^\mu\gamma^\nu\right]_{AB}\compactpsi_B(p)\wpsi_A(p')\;.
    \end{split}
\end{equation}
Proceeding in the same way for the Abelian four-current we get:
\begin{equation}\label{eq: j Dir as integral}
    \begin{split}
        \wj^{\mu}(y)&=\frac{1}{(2\pi)^8}\int\di^4p\di^4p'\,\e^{\ii(p-p')\cdot y}(\gamma^\mu)_{AB}\compactpsi_B(p)\wpsi_A(p')\;.
    \end{split}
\end{equation}
Then, comparing equations \eqref{eq: T Dir as integral} and \eqref{eq: j Dir as integral} with the definition \eqref{eq: Dirac T e j} we obtain the momentum representation of the two vertices:
\begin{equation*}
    \begin{split}
        \mathsf{f}^{\mu\nu}_{T_{\rm B}}(p,p')&=\frac{1}{4}\left[(p+p')^\nu\gamma^\mu+(p+p')^\mu\gamma^\nu\right]\;,\qquad\mathsf{f}^\mu_j(p,p')=\gamma^\mu\;,\\
        \mathsf{f}^{\mu\nu}_{T_{\rm C}}(p,p')&=\frac{1}{2}(p+p')^\nu\gamma^\mu\;,\qquad\mathsf{f}^{\mu\lambda\rho}_\mathcal{S}(p,p')=-\frac{1}{2}\epsilon^{\mu\lambda\rho\tau}\gamma_\tau\gamma^5\;.
    \end{split}
\end{equation*}
The microscopic kernel in linear response are thus, for the Belinfante case:
\begin{equation}\label{eq: GT lin}
    \begin{split}
        G^{\mu\nu}_{T_{\rm B}}(k,q)_{AB}&=\frac{1}{4(2\pi)^6}\delta\left(k^2-m^2+\frac{q^2}{4}\right)\Lambda_1(\beta\cdot q\,n_{\rm F}(k_+)[1-n_{\rm F}(k_-)]\\
        &\times\left[(\slashed{k}_-+m\mathbb{I})\left(k^\nu\gamma^\mu+k^\mu\gamma^\nu\right)(\slashed{k}_++m\mathbb{I})\right]_{AB}\;,
    \end{split}
\end{equation}
with a similar expression for the canonical stress tensor insertion.
For the Abelian four-current:
\begin{equation}\label{eq: Gj lin}
    \begin{split}
        G^{\mu}_j(k,q)_{AB}&=\frac{1}{2(2\pi)^6}\delta\left(k^2-m^2+\frac{q^2}{4}\right)\Lambda_1(\beta\cdot q)\,n_{\rm F}(k_+)[1-n_{\rm F}(k_-)]\\
        &\times\left[(\slashed{k}_-+m\mathbb{I})\gamma^\mu(\slashed{k}_++m\mathbb{I})\right]_{AB}\;.
    \end{split}
\end{equation}
and finally for the spin tensor:
\begin{equation*}
    \begin{split}
        G^{\mu\lambda\rho}_\mathcal{S}(k,q)_{AB}&=-\frac{1}{4(2\pi)^6}\delta\left(k^2-m^2+\frac{q^2}{4}\right)\Lambda_1(\beta\cdot q)\,n_{\rm F}(k_+)[1-n_{\rm F}(k_-)]\\
        &\times\epsilon^{\mu\lambda\rho\tau}\left[(\slashed{k}_-+m\mathbb{I})\gamma_\tau\gamma^5(\slashed{k}_++m\mathbb{I})\right]_{AB}\;.
    \end{split}
\end{equation*}
We are interested in computing the axial and scalar part of both the contributions. Using the following properties of the Dirac $\gamma$ matrices:
\begin{equation}\label{eq: gamma prop}
    \begin{split}
        &\tr(\gamma^\mu\gamma^{\nu_1}\ldots\gamma^{\nu_{2N+1}})=0\;,\quad \tr(\gamma^\alpha\gamma^\mu)=4g^{\alpha\mu}\;,\\
        &\tr(\gamma^\alpha\gamma^\mu\gamma^\beta\gamma^\nu)=4(g^{\alpha\mu}g^{\beta\nu}-g^{\alpha\beta}g^{\mu\nu}+g^{\alpha\nu}g^{\mu\beta}) \\
    &\tr(\gamma^\sigma\gamma^5\gamma^\alpha\gamma^\mu)=\tr(\gamma^\sigma\gamma^5\gamma^\mu)=\tr(\gamma^5\gamma^\mu)=\tr(\gamma^5)=0\;,\\
    &\tr(\gamma^\sigma\gamma^5\gamma^\alpha\gamma^\mu\gamma^\beta)=-4\ii\epsilon^{\sigma\mu\alpha\beta}\;,
    \end{split}
\end{equation}
after some simple algebra we obtain for the scalar part the following traces:
\begin{equation}\label{eq: LINEAR scalar f}
    \begin{split}
        \tr[\Pi(k_-)\mathsf{f}^{\mu\nu}_T(k,q)\Pi(k_+)]&=8mk^\mu k^\nu\;,
        \\ \tr[\Pi(k_-)\mathsf{f}^{\mu}_j(k,q)\Pi(k_+)]&=8mk^\mu \;,\\
        \tr[\Pi(k_-)\mathsf{f}^{\mu\lambda\rho}_\mathcal{S}(k,q)\Pi(k_+)]&=0\;,
    \end{split}
\end{equation}
while for the axial part the following ones:
\begin{equation}\label{eq: LINEAR axial f}
    \begin{split}
        \tr[\gamma^\sigma\gamma^5\Pi(k_-)\mathsf{f}^{\mu\nu}_T(k,q)\Pi(k_+)]&=-2\ii k_\alpha q_\beta\left(k^\nu\epsilon^{\sigma\mu\alpha\beta}+k^\mu\epsilon^{\sigma\nu\alpha\beta}\right)\;,\\
        \tr[\gamma^\sigma\gamma^5\Pi(k_-)\mathsf{f}^{\mu}_j(k,q)\Pi(k_+)]&=-4\ii k_\alpha q_\beta\epsilon^{\sigma\mu\alpha\beta} \;,\\
        \tr[\gamma^\sigma\gamma^5\Pi(k_-)\mathsf{f}^{\mu\lambda\rho}_\mathcal{S}(k,q)\Pi(k_+)]&=-2\epsilon^{\mu\lambda\rho\tau}\left[2k^\sigma k_\tau-\frac12q^\sigma q_\tau-\delta^\sigma_\tau\left(k^2+m^2-\frac{q^2}{4}\right)\right]\;,
    \end{split}
\end{equation}
where we used and that $k_\pm=k\pm q/2$ to express the final result in terms of $k$ and the momentum $q$.
Combining the results \eqref{eq: LINEAR scalar f} and \eqref{eq: LINEAR axial f} with \eqref{eq: GT lin} and \eqref{eq: Gj lin} we obtain the expression \eqref{eq: GT e Gj scalar} and \eqref{eq: GT e Gj axial} in the main text.

The computation of the functions $\mathsf{G}^{\mu_1\mu_2}_{ab}(k,q_a,q_b)$ appeareing in the quadratic response theory \eqref{def: G quad} can be computed in a similar manner
For the specific case of a iso-singlet Dirac field the general structure is:
\begin{equation*}
    \left[\Pi(k_-)\,\mathsf{f}^{\mu,\nn\mrr}_a(k_-,\ell)\,\Pi(\ell)\,\mathsf{f}^{\nu,\mrr\mm}_b(\ell,k_+)\,\Pi(k_+)\right]_{\rho\lambda}\;,
\end{equation*}
that is, for the Belinfante pseudo-gauge:
\begin{equation*}
    \begin{split}
        \mathcal{T}_{TT}&\equiv\frac{1}{16} \left\{[\slashed{k}_-+m][(\ell+k_-)^{\nu_1}\gamma^{\mu_1}+(\ell+k_-)^{\mu_1}\gamma^{\nu_1}][\slashed{\ell}+m][(\ell+k_+)^{\nu_2}\gamma^{\mu_2}+(\ell+k_+)^{\mu_2}\gamma^{\nu_2}][\slashed{k}_++m]\right\}\\
        \mathcal{T}_{Tj}&\equiv\frac{1}{4}\left\{[\slashed{k}_-+m][(\ell+k_-)^{\nu_1}\gamma^{\mu_1}+(\ell+k_-)^{\mu_1}\gamma^{\nu_1}][\slashed{\ell}+m]\gamma^{\mu_2}[\slashed{k}_++m]\right\}\\
         \mathcal{T}_{jT}&\equiv\frac{1}{4}\left\{[\slashed{k}_-+m]\gamma^\mu[\slashed{\ell}+m][(\ell+k_+)^{\nu_2}\gamma^{\mu_2}+(\ell+k_+)^{\mu_2}\gamma^{\nu_2}][\slashed{k}_++m]\right\}\\
        \mathcal{T}_{jj}&\equiv\left[\slashed{k}_-+m\right]\gamma^{\mu_1}[\slashed{\ell}+m]\gamma^{\mu_2}[\slashed{k}_++m]\;.
    \end{split}
\end{equation*}
For the spin polarization vector we are interested in computing only the axial part of each of the above combinations.

Let us define the fundamental trace:
\begin{equation*}
    \mathcal{H}^{\alpha\beta\sigma}\equiv\tr\left\{\left[\left(\slashed{k}-\frac{\slashed{q}}{2}\right)+m\right]\gamma^\alpha(\slashed{\ell}+m)\gamma^\beta\left[\left(\slashed{k}+\frac{\slashed{q}}{2}\right)+m\right]\gamma^\sigma\gamma^5\right\}\;.
\end{equation*}
We can compute it using the properties of the $\gamma$ matrices \eqref{eq: gamma prop} and we find:
\begin{equation}\label{eq: trace H}
    \mathcal{H}^{\alpha\beta\sigma}=-4\ii\left(\mathcal{H}^{\alpha\beta\sigma}_{(0)}+\mathcal{H}^{\alpha\beta\sigma}_{(1)}+\mathcal{H}^{\alpha\beta_\sigma}_{(2)}\right)\;,
\end{equation}
where each $\mathcal{H}_{(i)}$ contains, respectively, $i$ powers of $q$:
\begin{equation*}
    \begin{split}
        \mathcal{H}^{\alpha\beta\sigma}_{(0)}&=(k^2+m^2)\epsilon^{\alpha\beta\sigma\lambda}\ell_\lambda-2m^2\epsilon^{\alpha\beta\sigma\lambda}k_\lambda-2k^\sigma\epsilon^{\alpha\beta\lambda\rho}k_\lambda\ell_\rho\;,\\
        \mathcal{H}^{\alpha\beta\sigma}_{(1)}&=-\frac{1}{2}(\ell\cdot q)\epsilon^{\alpha\beta\sigma\lambda}k_\lambda+\frac{1}{2}(k\cdot\ell)\epsilon^{\alpha\beta\sigma\lambda}q_\lambda-\frac{1}{2}\left(q^\alpha\epsilon^{\beta\sigma\lambda\rho}k_\lambda\ell_\rho+q^\beta\epsilon^{\alpha\sigma\lambda\rho}k_\lambda\ell_\rho+q^\sigma\epsilon^{\alpha\beta\lambda\rho}k_\lambda\ell_{\rho}\right)\\
        &-\frac{1}{2}\left(k^\alpha\epsilon^{\beta\sigma\lambda\rho}\ell_\lambda q_\rho+k^\beta\epsilon^{\alpha\sigma\lambda\rho}\ell_\lambda q_\rho+k^\sigma\epsilon^{\alpha\beta\lambda\rho}\ell_\lambda q_\rho\right)-\ell^\alpha\epsilon^{\beta\sigma\lambda\rho}k_\lambda q_\rho+g^{\beta\sigma}\epsilon^{\alpha\lambda\rho\gamma}k_\lambda\ell_\rho q_\gamma\;,\\
        \mathcal{H}^{\alpha\beta\sigma}_{(2)}&=-\frac{q^2}{4}\epsilon^{\alpha\beta\sigma\lambda}\ell_\lambda-\frac{q^\sigma}{2}\epsilon^{\alpha\beta\lambda\rho}\ell_\lambda q_\rho\;.
    \end{split}
\end{equation*}
Note that the trace \eqref{eq: trace H} fulfills $\mathcal{H}^{\alpha\beta\sigma}(k,\ell,q)=-\mathcal{H}^{\beta\alpha\sigma}(k,\ell,-q)$, where $q\to-q$ is equivalent to $q_a,q_b\to-q_b,-q_a$ and $\ell\to\ell$, which is just the realization of \eqref{eq: Tswap}.

Now, introducing:
\begin{equation*}
    P^\mu\equiv \ell^\mu+k^\mu\;,\quad P^\mu_\pm=P^\mu\pm\frac{q^\mu}{2}\;,
\end{equation*}
we obtain that the fundamental traces are:
\begin{equation*}
    \begin{split}
        \mathcal{T}_{TT}\equiv\tr\left[\mathsf{G}^{\mu_1\nu_1\mu_2\nu_2}_{TT}\,\gamma^\sigma\gamma^5\right]&=\frac{1}{16}\left[P^{\nu_1}_-P^{\nu_2}_+\mathcal{H}^{\mu_1\mu_2\sigma}+P^{\nu_1}_-P^{\mu_2}_+\mathcal{H}^{\mu_1\nu_2\sigma}+P^{\mu_1}_-P^{\nu_2}_+\mathcal{H}^{\nu_1\mu_2\sigma}+P^{\mu_1}_-P^{\mu_2}_+\mathcal{H}^{\nu_1\nu_2\sigma}\right]\;,\\
        \mathcal{T}_{Tj}\equiv\tr\left[\mathsf{G}^{\mu_1\nu_1\mu_2}_{Tj}\,\gamma^\sigma\gamma^5\right]&=\frac{1}{4}\left[\left(P^{\nu_1}-\frac{q^{\nu_1}}{2}\right)\mathcal{H}^{\mu_1\mu_2\sigma}+\left(P^{\mu_1}-\frac{q^{\mu_1}}{2}\right)\mathcal{H}^{\nu_1\mu_2}\sigma\right]\;,\\
        \mathcal{T}_{jT}\equiv\tr\left[\mathsf{G}^{\mu_1\nu_1\mu_2}_{jT}\,\gamma^\sigma\gamma^5\right]&=\frac{1}{4}\left[\left(P^{\nu_2}+\frac{q^{\nu_2}}{2}\right)\mathcal{H}^{\mu_1\mu_2\sigma}+\left(P^{\mu_2}+\frac{q^{\mu_2}}{2}\right)\mathcal{H}^{\mu_1\nu_2\sigma}\right]\;,\\
        \mathcal{T}_{jj}\equiv\tr\left[\mathsf{G}^{\mu_1\nu_1\mu_2}_{jj}\,\gamma^\sigma\gamma^5\right]&=\mathcal{H}^{\mu_1\mu_2\sigma}\;,
    \end{split}
\end{equation*}
so that the microscopic kernel \eqref{def: G quad} can be written in the following way:
\begin{equation*}
    \tr\left[\mathsf{G}_{ab}(k,q_a,q_b)\gamma^\sigma\gamma^5\right]=\frac{1}{4(2\pi)^9}\delta\left[k^2-m^2+\frac{(q_a+q_b)^2}{4}\right]\mathcal{N}(k,q_a,q_b)\,\mathcal{T}_{ab}\;.
\end{equation*}
The functions appearing in the quadratic response are intended to be expanded in $q_{a,b}$ rather than $q$ and $\ell$. Using:
\begin{equation*}
    \ell= k+\frac{q_a-q_b}{2}\;,\qquad q=q_a+q_b\;,\qquad k_\pm=k\pm\frac{q_a+q_b}{2}\;,
\end{equation*}
and introducing the following compact notation:
\begin{equation}\label{def: X}
    \mathcal{S}_{1,2}\equiv 1+\left(\mu_{1,2}\leftrightarrow\nu_{1,2}\right)\;,\qquad \mathcal{T}_{ab}\equiv\tr\left[\mathsf{G}_{ab}\gamma^\sigma\gamma^5\right]\;,\qquad \kappa^{\nu_{1,2}}_{1,2}\equiv 2k^{\nu_{1,2}}\mp q^{\nu_{1,2}}_{b,a}\;,
\end{equation}
the four quadratic traces read:
\begin{align*}
        \mathcal{T}^\sigma_{TT}&=\frac{1}{16}\mathcal{S}_1\mathcal{S}_2 \kappa^{\nu_1}_1\kappa^{\nu_2}_2\mathcal{H}^{\mu_1\mu_2\sigma}\;,\\
        \mathcal{T}^\sigma_{Tj}&=\frac{1}{4}\mathcal{S}_1\kappa^{\nu_1}_1\mathcal{H}^{\mu_1\mu_2\sigma}\;,\\
        \mathcal{T}^\sigma_{jT}&=\frac{1}{4}\mathcal{S}_2\kappa^{\nu_2}_2\mathcal{H}^{\mu_1\mu_2\sigma}\;,\\
        \mathcal{T}^\sigma_{jj}&=\mathcal{H}^{\mu_1\mu_2\sigma}\;.
\end{align*}

The needed derivatives of the various traces are thus obtained by differentiating the above expressions. For the tensor-tensor term:
\begin{align*}
        \partial^\rho_{q_a}\mathcal{T}_{TT}&=\frac{1}{16}\mathcal{S}_1\mathcal{S}_2\left(\kappa^{\nu_1}_1\kappa^{\nu_2}_2\mathcal{H}^\rho_a+\kappa^{\nu_1}_1g^{\nu_2\rho}\mathcal{H}\right)\;,\\
        \partial^\tau_{q_b}\mathcal{T}_{TT}&=\frac{1}{16}\mathcal{S}_1\mathcal{S}_2\left(\kappa^{\nu_1}_1\kappa^{\nu_2}_2\mathcal{H}^\tau_b-\kappa^{\nu_2}_2g^{\nu_1\tau}\mathcal{H}\right)\;,\\
        \partial^{\rho}_{q_a}\partial^\tau_{q_b}\mathcal{T}_{TT}&=\frac{1}{16}\mathcal{S}_1\mathcal{S}_2\left(\kappa^{\nu_1}_1\kappa^{\nu_2}_2\mathcal{H}^{\rho\tau}_{ab}+\kappa^{\nu_1}_1g^{\nu_2\rho}\mathcal{H}^\tau_b-\kappa^{\nu_2}_2g^{\nu_1\tau}\mathcal{H}^{\rho}_a-g^{\nu_1\tau}g^{\nu_2\rho}\mathcal{H}\right)\;,
\end{align*}
for the tensor-current term
\begin{align*}
        \partial^\rho_{q_a}\mathcal{T}_{Tj}&=\frac{1}{4}\mathcal{S}_1\kappa^{\nu_1}_1\mathcal{H}^\rho_a\;,\\
        \partial^\tau_{q_b}\mathcal{T}_{Tj}&=\frac{1}{4}\mathcal{S}_1\left(\kappa^{\nu_1}_1\mathcal{H}^\tau_b-g^{\nu_1\tau}\mathcal{H}\right)\;,\\
        \partial^{\rho}_{q_a}\partial^\tau_{q_b}\mathcal{T}_{Tj}&=\frac{1}{4}\mathcal{S}_1\left(\kappa^{\nu_1}_1\mathcal{H}^{\rho\tau}_{ab}-g^{\nu_1\tau}\mathcal{H}^\rho_a\right)\;,
\end{align*}
for the current-tensor term:
\begin{align*}
        \partial^\rho_{q_a}\mathcal{T}_{jT}&=\frac{1}{4}\mathcal{S}_2\left(\kappa^{\nu_2}_2\mathcal{H}^\rho_a+g^{\nu_2\rho}\mathcal{H}\right)\;,\\
        \partial^\tau_{q_b}\mathcal{T}_{jT}&=\frac{1}{4}\mathcal{S}_2\kappa^{\nu_2}_2\mathcal{H}^\tau_b\;,\\
        \partial^{\rho}_{q_a}\partial^\tau_{q_b}\mathcal{T}_{jT}&=\frac{1}{4}\mathcal{S}_2\left(\kappa^{\nu_2}_2\mathcal{H}^{\rho\tau}_{ab}+g^{\nu_2\rho}\mathcal{H}^\tau_b\right)\;,
\end{align*}
and finally for the current-current term:
\begin{align*}
        \partial^\rho_{q_a}\mathcal{T}_{jj}&=\mathcal{H}^\rho_a\;,\\
        \partial^\tau_{q_b}\mathcal{T}_{jj}&=\mathcal{H}^\tau_b\;,\\
        \partial^\rho_{q_a}\partial^\tau_{q_b}\mathcal{T}_{jj}&=\mathcal{H}^{\rho\tau}_{ab}\;,
\end{align*}
where we denoted with $\mathcal{H}\equiv\mathcal{H}^{\mu_1\mu_2\sigma}$, $\mathcal{H}^\rho_a\equiv\partial^\rho_{q_a}\mathcal{H}$ and $\mathcal{H}^{\rho\tau}_{ab}\equiv\partial^{\rho}_{q_a}\partial^\tau_{q_b}\mathcal{H}$

In order to compute the quadratic response we need the mixed derivatives of the traces \eqref{def: ST}. Introducing the following quantities:
\begin{equation}\label{def: C e E fermi}
    C\equiv n_{\rm F}(k)\left(1-n_{\rm F}(k)\right)\left(\frac{1}{2}-n_{\rm F}(k)\right)\;,\quad E\equiv\frac{n_{\rm F}(k)\left(1-n_{\rm F}(k)\right)}{12}\left[1-6n_{\rm F}(k)+6n^2_{\rm F}(k)\right]\;,
\end{equation}
the corresponding derivatives for the statistical factor \eqref{def: S} are:
\begin{equation*}
    \begin{split}
        \mathsf{S}(k,0,0)&=\frac{1}{4(2\pi)^9}\delta(k^2-m^2)C\;,\\
        \partial^\rho_{q_a}\mathsf{S}(k,0,0)&=-\frac{1}{4(2\pi)^9}\delta(k^2-m^2)E\beta^\rho\;,\\
       \partial^\rho_{q_b}\mathsf{S}(k,0,0)&=\frac{1}{4(2\pi)^9}\delta(k^2-m^2)E\beta^\rho\;,\\
        \partial^{\rho}_{q_a}\partial_{q_b}^\lambda\mathsf{S}(k,0,0)&=\frac{1}{4(2\pi)^9}\frac{C}{2}g^{\rho\lambda}\delta'(k^2-m^2)\,+\,\frac{1}{4(2\pi)^9}\delta(k^2-m^2)\,\partial^\rho_{q_a}\partial^\lambda_{q_b}\mathcal{N}\Big|_{0}\;.
    \end{split}
\end{equation*}
The above can be combined with the derivatives of the traces computed earlier to obtain the corresponding expansion in $q_a,q_b$ of \eqref{def: ST}. At leading order we find:
\begin{equation}\label{eq: Gab 00}
    \mathcal{G}_{ab}(k,0,0)=0\;.
\end{equation}
For the first derivatives we get instead:
\begin{equation}\label{eq: partialGab 00}
    \begin{split}
        \partial^\rho_{q_a}\mathcal{G}_{ab}(k,q_a,q_b)\Big|_{q_a=q_b=0}&=\frac{1}{4(2\pi)^9}C\delta(k^2-m^2)\partial^\rho_{q_a}\mathcal{T}_{ab}(k,q_a,q_b)\Big|_{q_a=q_b=0}\;,\\
        \partial^\tau_{q_b}\mathcal{G}_{ab}(k,q_a,q_b)\Big|_{q_a=q_b=0}&=\frac{1}{4(2\pi)^9}C\delta(k^2-m^2)\partial^\tau_{q_b}\mathcal{T}_{ab}(k,q_a,q_b)\Big|_{q_a=q_b=0}\;.
    \end{split}
\end{equation}
Finally for the second mixed derivative we obtain:
\begin{equation}\label{eq: partial2Gab 00}
    \begin{split}
         &\partial^\rho_{q_a}\partial^\tau_{q_b}\mathsf{G}_{ab}(k,q_a,q_b)\Big|_{q_a=q_b=0}=\frac{1}{4(2\pi)^9}\Bigg\{\frac{C}{2}g^{\rho\tau}\delta'(k^2-m^2)\mathcal{T}_{ab}(k,0,0)+\delta(k^2-m^2)\times\\
         &\times\Big[C\partial^\rho_{q_a}\partial^\tau_{q_b}\mathcal{T}_{ab}(k,q_a,q_b)\Big|_{q_a=q_b=0}+E\beta^\tau\partial^\rho_{q_a}\mathcal{T}_{ab}(k,q_a,q_b)\Big|_{q_a=q_b=0}-E\beta^\rho\partial^\tau_{q_b}\mathcal{T}_{ab}(k,q_a,q_b)\Big|_{q_a=q_b=0}\Big]\Bigg\}\;.
    \end{split}
\end{equation}
By plugging the explicit expressions for the traces in equations \eqref{eq: Gab 00}, \eqref{eq: partialGab 00} and \eqref{eq: partial2Gab 00} it is possible to verify the relations \eqref{eq: cancellazione I} and the consequential vanishing of the quadratic response correction to the axial part of the Wigner function.


\section{Second-order corrections: pions}\label{SEC: app pion}
In this appendix, we collect the computations for the pions, the microscopic kernels and their statistical factors, the quadratic response of the Wigner function, the full second-order Wigner function and the momentum spectrum.

We start from the linear response term which microscopic kernel is given by:
\begin{equation*}
    \begin{split}
        [G^{\mu\nu}_T(k,q)]^{\mm\nn}&=\frac{\delta^{\mm\nn}}{(2\pi)^6}\delta\left(k^2-m^2+\frac{q^2}{4}\right)L^{\mm\nn}(k,q)\left[k^\mu k^\nu-\frac{1}{4}\left(q^\mu q^\nu-q^2g^{\mu\nu}\right)\right]\;,\\
        [G^{\mu\nu}_{\mathcal{V}_c}(k,q)]^{\mm\nn}&=\frac{({\mathsf{t}}_c)^{\mm\nn}}{(2\pi)^6}\delta\left(k^2-m^2+\frac{q^2}{4}\right)L^{\mm\nn}(k,q)k^\mu\;.    
    \end{split}
\end{equation*}
The statistical factor $L^{\mm\nn}$ defined in equation \eqref{def: L factor} can be expanded in Taylor series around $q=0$:
\begin{equation*}
    L^{\mm\nn}(k,q)=\sum_{r=0}^{+\infty}\frac{(\beta\cdot q)^r}{r!}L_{[r]}^{\mm\nn}(k)\;,\qquad L^{\mm\nn}_{[r]}(k)\equiv\frac{\di^r L^{\mm\nn}(k,q)}{\di(\beta\cdot q)^r}\Big|_{q=0}\;,
\end{equation*}
the various coefficients satisfy:
\begin{equation*}
    L^{\mm\nn}_{[r]}(k)=(-1)^rL^{\nn\mm}_{[r]}(k)\;.
\end{equation*}
Now introducing:
\begin{equation}\label{def: chim}
    \chi^{(1)}_\mm(k)\equiv n_{\rm B}(k,\mm)[1+n_{\rm B}(k,\mm)]\;,\qquad \chi^{(2)}_\mm(k)\equiv\chi^{(1)}_\mm(k)[1+2n_{\rm B}(k,\mm)]\;,\qquad \Delta_{\mm\nn}\equiv \mu_\mm-\mu_\nn\;, 
\end{equation}
the first three coefficients of the Taylor expansion of \eqref{def: L factor} are:
\begin{equation}\label{eq: derivatives of L}
    \begin{split}
        L^{\mm\nn}_{[0]}(k)&=\frac{n_{\rm B}(k,\mm)-n_{\rm B}(k,\nn)}{\Delta_{\mm\nn}}\;,\\
        L^{\mm\nn}_{[1]}(k)&=\frac{L^{\mm\nn}_{[0]}(k)-\frac{1}{2}\left(\chi^{(1)}_\mm(k)+\chi^{(1)}_\nn(k)\right)}{\Delta_{\mm\nn}}\;,\\
        L^{\mm\nn}_{[2]}(k)&=\frac{2L^{\mm\nn}_{[1]}(k)+\frac{1}{4}\left(\chi^{(2)}_\mm(k)-\chi^{(2)}_\nn(k)\right)}{\Delta_{\mm\nn}}\;,
    \end{split}
\end{equation}
Now, being the vertices \eqref{eq: pion vertices} and the overall $\delta$ function  even in $q$, from \eqref{eq: derivatives of L} it follows that in principle a linear correction is possible for the off-diagonal terms of the Wigner function. However, setting $\mm=\nn$ in \eqref{eq: derivatives of L}, gives:
\begin{equation}\label{eq: derivatives of L diagonal}
   \begin{split}
       L^{\mm\mm}_{[0]}(k)&=n_{\rm B}(k,\mm)[1+n_{\rm B}(k,\mm)]\;,\\
       L^{\mm\mm}_{[1]}(k)&=0\;,\\
       L^{\mm\mm}_{[2]}(k)&=\frac{L^{\mm\mm}_{[0]}(k)}{12}\left[1+6n_{\rm B}(k,\mm)+6n^2_{\rm B}(k,\mm)\right]\;.
   \end{split}
\end{equation}
It then follows:
\begin{equation*}
    \Delta^{(1)}_{\rm lin}W^{\mm\mm}(x,k)=0\;,\qquad\Delta^{(1)}_{\rm lin}W^{\mm\nn}(x,k)\ne0\;,\,{\rm for}\, \mm\ne\nn\;.
\end{equation*}
Since the spectra \eqref{eq: CF in terms of W} are given by the diagonal elements of the Wigner function we conclude that the Cooper-Frye formula does not receive any local equilibrium correction at first order in the gradients. 

The first non-vanishing correction is thus quadratic in the gradients. The first term comes from the second order linear response which requires to compute the second derivative in $q$ of \eqref{eq: G lin scalar case}. Expanding the $\delta$ function in analogy with equation \eqref{eq: espansione q delta} and using \eqref{eq: derivatives of L diagonal}:
\begin{equation*}
    \begin{split}
        &\delta\left(k^2-m^2+\frac{q^2}{4}\right)L^{\mm\mm}(k,q)=\delta(k^2-m^2)L^{\mm\mm}_{[0]}(k)\\
        &+\left[\frac{1}{2}\delta(k^2-m^2)L^{\mm\mm}_{[2]}(k)\,\beta_{\lambda_1}\beta_{\lambda_2}+\frac{1}{4}\delta'(k^2-m^2)L^{\mm\mm}_{[0]}(k)g_{\lambda_1\lambda_2}\right]q^{\lambda_1}q^{\lambda_2}+\mathcal{O}(q^4)\;,
    \end{split}
\end{equation*}
plugging the expansion in the second line of \eqref{eq: W linear N2} we obtain:
\begin{equation}\label{eq: Wscal1}
    \begin{split}
        \Delta^{(2)}_{\rm lin}W^{\mm\mm}(x,k)&=\frac{1}{4(2\pi)^3}\delta(k^2-m^2)L^{\mm\mm}_{[0]}(k)\left(\delta^\mu_{\alpha_1}\delta^\nu_{\alpha_2}-g_{\alpha_1\alpha_2}g^{\mu\nu}\right)\left[\mathfrak{D}^{\alpha_1}_y(\mathfrak{D}^{\alpha_2}_y\mathcal{R}^T_{\mu\nu})\right]\Big|_{y=x}\\
        &-\frac{1}{2(2\pi)^3}\left[\delta(k^2-m^2)L^{\mm\mm}_{[2]}(k)\,\beta_{\alpha_1}\beta_{\alpha_2}+\frac{1}{2}\delta'(k^2-m^2)L^{\mm\mm}_{[0]}(k)g_{\alpha_1\alpha_2}\right]\\
        &\times \left[k^\mu k^\nu\mathfrak{D}^{\alpha_1}_y[\mathfrak{D}^{\alpha_2}_y\mathcal{R}^T_{\mu\nu}]+k^\mu(\mathsf{t}_3)^{\mm\mm}\mathfrak{D}^{\alpha_1}_y[\mathfrak{D}^{\alpha_2}_y\mathcal{R}^\mathcal{V}_{\mu}]\right]\Big|_{y=x}\;.
    \end{split}
\end{equation}
Let us recall the geometrical quantities of interest:
\begin{equation*}
    \nu^\mu(y)=\frac{\hat{n}^\mu(y)}{k\cdot\hat{ n}(y)}\;,\qquad \mathcal{P}^{\alpha\beta}(y)=g^{\alpha\beta}-k^\alpha\nu^\beta(y)\;,\qquad \mathfrak{D}^\alpha_y=\mathcal{P}^{\rho\alpha}(y)\partial^y_\rho\;.
\end{equation*}
We define the \emph{extrinsic curvature} tensor $K^{\alpha\beta}$ as:
\begin{equation*}
    K^{\alpha\beta}\equiv\mathfrak{D}^\alpha\nu^\beta=\frac{1}{k\cdot\hat{n}}\mathcal{P}^{\rho\alpha}\mathcal{P}^{\sigma\beta}\partial_\rho\hat{n}_\sigma\;,
\end{equation*}
where the dependence on $y$ is understood. From the properties \eqref{eq: ortho P maintext} of the projector $\mathcal{P}$ follows:
\begin{equation}\label{eq: prop B}
    K^{\alpha\beta}=K^{\beta\alpha}\;,\quad k_\alpha K^{\alpha\beta}=k_\beta K^{\alpha\beta}=0\;,\qquad \mathfrak{D}^\alpha\mathcal{P}^{\rho\beta}=-k^\rho K^{\alpha\beta}\;.
\end{equation}
The second derivative present in equation \eqref{eq: Wscal1} has been computed in \eqref{eq: DR 2}. Taking into account that $\mathcal{R}^a_{\mu}=\nu_\mu\vartheta_a$, with $\vartheta_T=-\Delta\beta$ and $\vartheta_{\mathcal{V}_3}=\Delta\ZZ$, and the properties \eqref{eq: prop B} we can arrange the two derivatives $\mathfrak{D}^\alpha\mathfrak{D}^\beta$ in the second term of \eqref{eq: Wscal1} as follows:
\begin{equation}\label{EEE def: matZ}
    \begin{split}
        \mathcal{Z}^{\alpha\beta}_\mm&\equiv-\left(k^\mu k^\nu\mathfrak{D}^{\alpha}\mathfrak{D}^\beta \mathcal{R}^T_{\mu\nu}+\mm\, k^\mu\mathfrak{D}^{\alpha}\mathfrak{D}^\beta \mathcal{R}^\mathcal{V}_{\mu}\right)=\mathfrak{D}^\alpha\circ\mathfrak{D}^\beta\left[k\cdot\beta (y)-\mm\mathfrak{Z}(y)\right]\\
        &=\mathcal{P}^{\rho\alpha}\mathcal{P}^{\sigma\beta}\partial_\rho\partial_\sigma\left[k\cdot\beta (y)-\mm\mathfrak{Z}(y)\right]- K^{\alpha\beta}k\cdot\partial\left[k\cdot\beta (y)-\mm\mathfrak{Z}(y)\right]\;.
    \end{split}
\end{equation}
For the other term let us define:
\begin{equation*}
    \mathcal{C}\equiv(g_{\mu\alpha_1}g_{\nu\alpha_2}-g_{\mu\nu}g_{\alpha_1\alpha_2})\mathfrak{D}^{\alpha_1}\mathfrak{D}^{\alpha_2}[\nu^\mu\Delta\beta^\nu]\;.
\end{equation*}
Using the definition of $\mathfrak{D}$ and \eqref{eq: prop B} the above expression can be equivalently written as:
\begin{equation}\label{EEE def: matC}
    \mathcal{C}=\nu_\alpha\mathcal{P}^{\rho\alpha}\mathcal{P}^{\sigma\lambda}\partial_\rho\partial_\sigma\beta_\lambda-\nu^\lambda\mathrm{Z}^{\rho\sigma}\partial_\rho\partial_\sigma\beta_\lambda+ K^\alpha_{\,\alpha}\partial_\gamma\beta^\gamma- K^{\rho\sigma}\partial_\rho\beta_\sigma\;,
\end{equation}
where we defined:
\begin{equation}\label{EEE def: Zrm}
    \mathrm{Z}^{\rho\sigma}\equiv g_{\alpha_1\alpha_2}\mathcal{P}^{\rho\alpha_1}\mathcal{P}^{\sigma\alpha_2}=g^{\rho\sigma}-k^\rho\nu^\sigma-\nu^\rho k^\sigma+\nu^2 k^\rho k^\sigma\;.
\end{equation}
In terms of these quantities the second order linear correction \eqref{eq: Wscal1} can be written as:
\begin{equation}\label{eq: Wscal2}
    \begin{split}
        &\Delta^{(2)}_{\rm lin}W^{\mm\mm}(x,k)=-\frac{1}{4(2\pi)^3}\delta(k^2-m^2)L^{\mm\mm}_{[0]}(k)\,\mathcal{C}(k,x)\\
        &+\frac{1}{2(2\pi)^3}\left[\delta(k^2-m^2)L^{\mm\mm}_{[2]}(k)\,\beta_{\alpha_1}\beta_{\alpha_2}+\frac{1}{2}\delta'(k^2-m^2)L^{\mm\mm}_{[0]}(k)g_{\alpha_1\alpha_2}\right]\mathcal{Z}^{\alpha_1\alpha_2}_\mm(k,x)\;.
    \end{split}
\end{equation}
Now let us split the four-temperature gradient in thermal vorticity and thermal shear \eqref{def: varpi xi}:
\begin{equation*}
    \partial_\mu\beta_\nu=\xi_{\mu\nu}-\varpi_{\mu\nu}\;.
\end{equation*}
Let us also introduce the following derivative shear tensor:
\begin{equation}\label{EEE def: XI}
    \Xi_{\rho\sigma\lambda}\equiv\partial_\rho\xi_{\sigma\lambda}+\partial_\sigma\xi_{\rho\lambda}-\partial_\lambda\xi_{\rho\sigma}\;.
\end{equation}
Then, being the thermal shear symmetric and the thermal vorticity antisymmetric, one has:
\begin{equation}\label{EEE eq: prop XI}
    \partial_\rho\partial_\sigma\beta_\lambda=\Xi_{\rho\sigma\lambda}\;,\qquad  K^{\rho\lambda}\partial_\rho\beta_\lambda= K^{\rho\lambda}\xi_{\rho\lambda}\;.
\end{equation}
Then using equation \eqref{EEE def: XI} and using the properties \eqref{eq: prop B} and \eqref{EEE eq: prop XI} the two factors \eqref{EEE def: matZ} and \eqref{EEE def: matC} can be written as:
\begin{equation}\label{EEE eq: C e Z final}
    \begin{split}
        \mathcal{Z}^{\alpha\beta}_\mm&=\mathcal{P}^{\rho\alpha}\mathcal{P}^{\sigma\beta}\left(k^\lambda\Xi_{\rho\sigma\lambda}-\mm\,\partial_\rho\partial_\sigma\mathfrak{Z}\right)- K^{\alpha\beta}\left(k^\rho k^\lambda\xi_{\rho\lambda}-\mm\,k^\rho\partial_\rho\mathfrak{Z}\right)\;,\\
        \mathcal{C}&=\nu_\alpha\mathcal{P}^{\rho\alpha}\mathcal{P}^{\sigma\lambda}\Xi_{\rho\sigma\lambda}-\nu^\lambda\mathrm{Z}^{\rho\sigma}\Xi_{\rho\sigma\lambda}+ K^\alpha_{\,\alpha}\xi^\gamma_\gamma- K^{\rho\sigma}\xi_{\rho\sigma}\;.
    \end{split}
\end{equation}
Overall, plugging the above expressions into \eqref{eq: Wscal2} and using \eqref{eq: derivatives of L diagonal} we finally obtain that the second order term of the linear response correction is:
\begin{equation}\label{EEE eq: Wscal2 final}
    \begin{split}
        \Delta^{(2)}_{\rm lin}W^{\mm\mm}(x,k)&=\frac{1}{2(2\pi)^3}\Big[\delta(k^2-m^2)L^{\mm\mm}_{[2]}(x,k)\beta_\alpha(x)\beta_\beta(x)\mathcal{Z}^{\alpha\beta}_\mm(x,k)-\frac{\chi^{(1)}_\mm(x,k)}{2}\delta(k^2-m^2)\mathcal{C}(x,k)\\
        &+\frac{\chi^{(1)}_\mm(x,k)}{2}\delta'(k^2-m^2)g_{\alpha\beta}\mathcal{Z}^{\alpha\beta}_\mm(x,k)\Big]\;,
    \end{split}
\end{equation}
with $\mathcal{Z}$ and $\mathcal{C}$ given in \eqref{EEE eq: C e Z final} and $L^{\mm\mm}_{[2]}$ and $\chi^{(1)}_\mm=L^{\mm\mm}_{[0]}$ given in \eqref{eq: derivatives of L diagonal}.

We now pass at the computation of the quadratic correction.
The general corresponding microscopic kernel is:
\begin{equation*}
    \left[\mathsf{G}^{\mu,\nu}_{ab}(k,q_a,q_b)\right]^{\mm\nn}=\frac{1}{4(2\pi)^9}\delta\left(k^2-m^2+\frac{q^2}{4}\right)\sum_{\rr=-1}^{+1}\mathcal{N}^{\mm\rr\nn}(k,q_a,q_b)[\mathsf{f}^\mu_a(k_-,\ell)]^{\nn\rr}[\mathsf{f}^\nu_b(\ell,k_+)]^{\rr\mm}\;,
\end{equation*}
with the statistical factor given in equation \eqref{def: N quad scal}.
Using the relation
\begin{equation*}
    \mathcal{N}^{\mm\rr\nn}(k,q_a,q_b)=\mathcal{N}^{\nn\rr\mm}(k,-q_b,-q_a)\;,
\end{equation*}
and defining:
\begin{equation*}
    \begin{split}
        \mathcal{N}^{\mm\rr}_0&\equiv\mathcal{N}^{\mm\rr\mm}(k,0,0),\\
        \mathcal{N}^{\mm\rr}_a&\equiv\partial_{\beta\cdot q_a}\mathcal{N}^{\mm\rr\mm}(k,q_a,q_b)\Big|_{q_a=q_b=0}\;,\\
        \mathcal{N}^{\mm\rr}_b&\equiv\partial_{\beta\cdot q_b}\mathcal{N}^{\mm\rr\mm}(k,q_a,q_b)\Big|_{q_a=q_b=0}\;,\\
        \mathcal{N}^{\mm\rr}_{ab}&\equiv\partial_{\beta\cdot q_a}\partial_{\beta\cdot q_b}\mathcal{N}^{\mm\rr\mm}(k,q_a,q_b)\Big|_{q_a=q_b=0}\;,
    \end{split}
\end{equation*}
we have:
\begin{equation}\label{EEE: der N}
    \begin{split}
        &\mathcal{N}^{\mm\mm}_0=\frac{\chi^{(2)}_\mm}{2}\;,\qquad \mathcal{N}_a^{\mm\mm}=-\mathcal{N}_b^{\mm\mm}=-L^{\mm\mm}_{[2]}\;,\qquad \mathcal{N}_{ab}^{\mm\mm}=0\;,\\
    &\mathcal{N}_0^{\mm\rr}=\dfrac{\chi^{(1)}_\mm-L^{\mm\rr}_{[0]}}{\Delta_{\mm\rr}}\;,\qquad\mathcal{N}^{\mm\rr}_a=-\mathcal{N}^{\mm\rr}_b=\frac{L^{\mm\rr}_{[1]}}{\Delta_{\mm\rr}}\;.
    \end{split}
\end{equation}
where $\Delta_{\mm\rr} =\mu_\mm-\mu_\rr$. Introducing the tensors:
\begin{equation*}
    t^{\mu\nu}_{a}\equiv \frac{1}{2}\left(\kappa^\mu_1\kappa^\nu_1-q^\mu_aq^\nu_a+q^2_ag^{\mu\nu}\right)\;,\quad t^{\mu\nu}_{b}\equiv \frac{1}{2}\left(\kappa^\mu_2\kappa^\nu_2-q^\mu_bq^\nu_b+q^2_bg^{\mu\nu}\right)\;,\quad\;,
\end{equation*}
where $\kappa^\mu_{1,2}$ are given as in \eqref{def: X} the four possible diagonal microscopic kernels are:
\begin{subequations}\label{eq: kernel list scalar}
    \begin{align}
        \left[\mathsf{G}^{\mu\nu,\rho\sigma}_{TT}(k,q_a,q_b)\right]^{\mm\mm}&\equiv \frac{1}{4(2\pi)^9}\delta\left(k^2-m^2+\frac{q^2}{4}\right)\mathcal{N}^{\mm\mm\mm}(k,q_a,q_b)t^{\mu\nu}_{a}(k,q_a,q_b)t^{\rho\sigma}_{b}(k,q_a,q_b)\;,\\
        \left[\mathsf{G}^{\mu\nu,\rho}_{T\mathcal{V}_i}(k,q_a,q_b)\right]^{\mm\mm}&\equiv \frac{1}{4(2\pi)^9}\delta\left(k^2-m^2+\frac{q^2}{4}\right)(\mathsf{t}_i)^{\mm\mm}\mathcal{N}^{\mm\mm\mm}(k,q_a,q_b)t^{\mu\nu}_{a}(k,q_a,q_b)\kappa^\rho_2\;,\\
        \left[\mathsf{G}^{\mu,\rho\sigma}_{\mathcal{V}_iT}(k,q_a,q_b)\right]^{\mm\mm}&\equiv \frac{1}{4(2\pi)^9}\delta\left(k^2-m^2+\frac{q^2}{4}\right)(\mathsf{t}_i)^{\mm\mm}\mathcal{N}^{\mm\mm\mm}(k,q_a,q_b)t^{\rho\sigma}_{b}(k,q_a,q_b)\kappa^\mu_1\;,\\
        \left[\mathsf{G}^{\mu\nu}_{\mathcal{V}_i\mathcal{V}_j}(k,q_a,q_b)\right]^{\mm\mm}&\equiv \frac{1}{4(2\pi)^9}\delta\left(k^2-m^2+\frac{q^2}{4}\right)\sum_{\rr=-1}^{+1}(\mathsf{t}_i)^{\mm\rr}(\mathsf{t}_j)^{\rr\mm}\mathcal{N}^{\mm\rr\mm}(k,q_a,q_b)\kappa^\mu_1 \kappa^\nu_2\;.
    \end{align}
\end{subequations}
We only kept the diagonal terms $\mm\mm$ given that we are interested in the spectra. Now being $({\mathsf{t}}_1)^{\mm\mm}=({\mathsf{t}}_2)^{\mm\mm}=0$ and $({\mathsf{t}}_3)^{\mm\mm}=\mm$ we obtain that not all the kernels \eqref{eq: kernel list scalar} contribute to the corrections to the Cooper-Frye formula. In particular:
\begin{equation*}
    \begin{split}
        \mathsf{G}^{\mm\mm}_{T\mathcal{V}_{1,2}}=\mathsf{G}^{\mm\mm}_{\mathcal{V}_{1,2}T}=\mathsf{G}^{\mm\mm}_{\mathcal{V}_3\mathcal{V}_{1,2}}=\mathsf{G}^{\mm\mm}_{\mathcal{V}_{1,2}\mathcal{V}_3}=0\,.
    \end{split}
\end{equation*}
Note that, however, the purely transverse isospin kernels are generally non-vanishing being:
\begin{equation*}
    \begin{split}
        \sum_{\rr=-1}^{+1}(\mathsf{t}_1)^{\mm\rr}(\mathsf{t}_1)^{\rr\mm}\mathcal{N}^{\mm\rr\mm}=\sum_{\rr=-1}^{+1}(\mathsf{t}_2)^{\mm\rr}(\mathsf{t}_2)^{\rr\mm}\mathcal{N}^{\mm\rr\mm}\equiv\mathcal{M}^{\mm}=\begin{cases}
            \mathcal{M}^+=\frac{1}{2}\mathcal{N}^{+0+}\;,\\
            \mathcal{M}^0=\frac{1}{2}\left(\mathcal{N}^{0+0}+\mathcal{N}^{0-0}\right)\;,\\
            \mathcal{M}^-=\frac{1}{2}\mathcal{N}^{-0-}\;.
        \end{cases}
    \end{split}
\end{equation*}
Let $a\in\{T,\mathcal{V}_i\}$.
Combining \eqref{eq: Nswap scalar} and \eqref{eq: kernel list scalar}, we recover \eqref{eq: scalar Gswap}:
\begin{equation*}
   \begin{split}
        \mathsf{G}^{\mu_1\mu_2}_{ab}(k,q_a,q_b)^{\mm\mm}&= \mathsf{G}^{\mu_2\mu_1}_{ab}(k,-q_b,-q_a)^{\mm\mm}\;,\quad a,b=T,\mathcal{V}_3\\
        \mathsf{G}^{\mu_1\mu_2}_{\mathcal{V}_1\mathcal{V}_2}(k,q_a,q_b)^{\mm\mm}&= -\mathsf{G}^{\mu_2\mu_1}_{\mathcal{V}_2\mathcal{V}_1}(k,-q_b,-q_a)^{\mm\mm}\;,
   \end{split}
\end{equation*}
We can now compute the various derivatives of the microscopic kernels that appear in the quadratic correction. Let us define:
\begin{equation*}
    h^{\mu\nu}_{T}\equiv 2k^\mu k^\nu\;,\qquad h^\mu_\mathcal{V}\equiv2k^\mu\;,\qquad b_{T,a}\equiv\delta^\mu_\alpha k^\nu+\delta^\nu_\alpha k^\mu\;,\qquad b_{\mathcal{V},a}\equiv\delta^\mu_\alpha\;,
\end{equation*}
then the various derivatives of the vertices are:
\begin{equation*}
    t_a\big|_0=t_b\big|_0=h_T\;,\qquad \partial_{q^\alpha_a}t_a\big|_0=\partial_{q^\alpha _b}t_b\big|_0=0\;, \qquad \partial_{q^\alpha_b}t_a\big|_0=-\partial_{q^\alpha _a}t_b\big|_0=-b_{T,a}\;.
\end{equation*}
We then introduce the following tensor in the isospin space:
\begin{equation*}
    \mathsf{t}_a=\left(\mathsf{t}_T\equiv\mathbb{I},\;\mathsf{t}_{\mathcal{V}_i}\equiv\mathsf{t}_i\right)\;,\qquad \mathsf{T}^{\mm\rr}_{ab}\equiv (\mathsf{t}_a)^{\mm\rr}(\mathsf{t}_b)^{\rr\mm}\;,
\end{equation*}
so that:
\begin{equation*}
    \begin{split}
        \mathsf{G}_{ab}(k,0,0)&=\frac{1}{4(2\pi)^9}\delta(k^2-m^2)\sum_{\rr}\mathsf{T}^{\mm\rr}_{ab}\mathcal{N}_0^{\mm\rr}h_ah_b\;,\\
        \partial_{q^\alpha_a}\mathsf{G}_{ab}(k,0,0)&=\frac{1}{4(2\pi)^9}\delta(k^2-m^2)\sum_{\rr}\mathsf{T}^{\mm\rr}_{ab}\left[\mathcal{N}^{\mm\rr}_a\beta_\alpha h_a h_b+\mathcal{N}^{\mm\rr}_0 h_a b_{b,\alpha}\right]\;,\\
        \partial_{q^\beta_b}\mathsf{G}_{ab}(k,0,0)&=\frac{1}{4(2\pi)^9}\delta(k^2-m^2)\sum_{\rr}\mathsf{T}^{\mm\rr}_{ab}\left[\mathcal{N}^{\mm\rr}_b\beta_\beta h_a h_b-\mathcal{N}^{\mm\rr}_0 b_{a,\beta} h_{b}\right]\;,\\
        \partial_{q^\alpha_a}\partial_{q^\beta_b}\mathsf{G}_{ab}(k,0,0)&=\frac{1}{4(2\pi)^9}\sum_{\rr}\mathsf{T}^{\mm\rr}_{ab}\Big\{\delta(k^2-m^2)\Big[\mathcal{N}^{\mm\rr}_{ab}\beta_\alpha\beta_\beta h_a h_b-\mathcal{N}^{\mm\rr}_a\beta_\alpha b_{b,\beta}h_b\\
        &\qquad\qquad\qquad+\mathcal{N}^{\mm\rr}_b\beta_\beta h_a b_{b,\alpha}-\mathcal{N}^{\mm\rr}_0b_{a,\beta}b_{b,\alpha}\Big]+\frac{1}{2}\delta'(k^2-m^2)g_{\alpha\beta}\mathcal{N}^{\mm\rr}_0 h_a h_b\Big\}\;.
    \end{split}
\end{equation*}
We can now define, in analogy with \eqref{def: III}, the various contraction $I_a$, $I_b$ and $I_{00}$ contributing to total quadratic correction.
Then using \eqref{eq: scalar Gswap} it follows:
\begin{equation*}
    I^{\mathcal{V}_1\mathcal{V}_2}_a+I^{\mathcal{V}_2\mathcal{V}_1}_b=0\;,
\end{equation*}
which in turn implies that the mixed $\mathcal{V}_1\mathcal{V}_2$ kernels do not contribute at the quadratic correction of the diagonal term:
\begin{equation*}
    \Delta_{\rm quad}W^{\mm\mm}(x,k)\Big|_{\mathcal{V}_1\mathcal{V}_2+\mathcal{V}_2\mathcal{V}_1}=0\;.
\end{equation*}
Now let us introduce the two following quantities:
\begin{equation}\label{EEE: def BZ}
    B^{\;\alpha}_\lambda\equiv\mathcal{P}^{\rho\alpha}\partial_\rho\beta_\lambda=\mathcal{P}^{\rho\alpha}\left(\xi_{\rho\lambda}-\varpi_{\rho\lambda}\right)\;,\qquad Z^\alpha_i\equiv \mathcal{P}^{\rho\alpha}\partial_\rho\mathfrak{Z}_i\;.
\end{equation}
Now being $\vartheta_T=-\Delta\beta_\nu$ and $\vartheta_{\mathcal{V}_3}=\Delta\mathfrak{Z}_3$ we get:
\begin{equation*}
    \mathfrak{D}^\alpha\mathcal{R}^T_{\mu\nu}\big|_x=-\nu_\mu\,B_\nu^{\;\alpha}\;,\qquad  \mathfrak{D}^\alpha\mathcal{R}^{\mathcal{V}_3}_{\mu}\big|_x=\nu_\mu\,Z^\alpha_3\;,
\end{equation*}
implying:
\begin{equation*}
    h^{\mu\nu}_T\mathfrak{D}^\alpha\mathcal{R}^T_{\mu\nu}+(\mathsf{t}_3)^{\mm\mm}h^\mu_\mathcal{V}\mathfrak{D}^\alpha\mathcal{R}^{\mathcal{V}_3}_{\mu}=-2k^\nu B_\nu^{\;\alpha}+2\mm\,Z^\alpha_3\;.
\end{equation*}
For the diagonal pion component, the longitudinal term $i=3$ has $(\mathsf{t}_3)^{\mm\mm}=\mm$\;, hence the sum of the stress-energy contraction and longitudinal isochemical potential combine in:
\begin{equation}\label{EEE: def U}
    U^\alpha_\mm\equiv  k^\nu B_\nu^{\;\alpha}-\mm\,Z^\alpha_3
\end{equation}
Overall, using \eqref{EEE: def U} the longitudinal (in the isospin space) correction to the Wigner function reads:
\begin{equation}\label{eq: Wquad par}
    \begin{split}
        \Delta^{(2)}_{\rm quad,\|}W^{\mm\mm}(x,k)&=\frac{1}{4(2\pi)^3}\Big\{\delta(k^2-m^2)\Big[\frac{\chi^{(2)}_\mm}{2}\left(B^{\;\alpha}_\beta B_{\alpha}^{\;\beta}-2\nu^\alpha U_{\mm\gamma}B^{\;\gamma}_\alpha+(\nu\cdot U_\mm)^2\right)\\
        &+4L^{\mm\mm}_{[2]}\left((\beta\cdot\nu)(U_\mm\cdot U_\mm)-\beta_\alpha B_{\beta}^{\;\alpha}U^{\beta}_\mm\right)\Big]-\chi^{(2)}_{\mm}\delta'(k^2-m^2)(U_\mm\cdot U_\mm)\Big\}\;.
    \end{split}
\end{equation}
The other contribution to the diagonal part, which is transverse, comes from the combination of the vertices $\mathcal{V}_1\mathcal{V}_1+\mathcal{V}_2\mathcal{V}_2$. The corresponding isospin factor comes from the combined contraction of $\mathsf{t}_1$ and $\mathsf{t}_2$ with themselves in \eqref{eq: kernel list scalar}:
\begin{equation*}
    w^{\mm\rr}\equiv (\mathsf{t}_1)^{\mm\rr}(\mathsf{t}_1)^{\rr\mm}=(\mathsf{t}_2)^{\mm\rr}(\mathsf{t}_2)^{\rr\mm}=\begin{cases}
        \frac{1}{2}&{\rm if}\, |\mm-\rr|=1\;,\\
        0&{\rm otherwise}\;.
    \end{cases}
\end{equation*}
We refer to this correction as \emph{transverse} (in the isospin space) correction and it reads:
\begin{equation}\label{eq: Wquad perp}
    \begin{split}
        &\Delta^{(2)}_{\rm quad,\bot}W^{\mm\mm}(x,k)=\frac{1}{4(2\pi)^3}\sum_{i=1}^{2}\sum_{\rr=-1}^{+1}w^{\mm\rr}\Big\{\delta(k^2-m^2)\times\\
        &\times\Big[\mathcal{N}^{\mm\rr}_0(\nu\cdot Z_i)^2-4\mathcal{N}^{\mm\rr}_{a}(\beta\cdot\nu)Z_i\cdot Z_i-4\mathcal{N}^{\mm\rr}_{ab}(\beta\cdot Z_i)^2\Big]-2\mathcal{N}^{\mm\rr}_0\delta'(k^2-m^2)Z_i\cdot Z_i\Big\}\;,
    \end{split}
\end{equation}
with the various derivative of off-diagonal elements the statistical factor given in \eqref{EEE: der N}.
The above combined with \eqref{EEE eq: Wscal2 final} gives the total local equilibrium correction to the diagonal components of the Wigner function up to the second order in the gradients:
\begin{equation}\label{EEE: SCALAR W quad1}
    \begin{split}
       \Delta^{(2)}W^{\mm\mm}(x,k)&=\frac{1}{4(2\pi)^3}\Bigg\{\delta(k^2-m^2)\Big[\frac{\chi^{(2)}_\mm}{2}\left(B^{\;\alpha}_\beta B_{\alpha}^{\;\beta}-2\nu^\alpha U_{\mm\gamma}B^{\;\gamma}_\alpha+(\nu\cdot U_\mm)^2\right)-\chi^{(1)}_\mm\mathcal{C}\\
        &\qquad\qquad\qquad+4L^{\mm\mm}_{[2]}\left(\frac{1}{2}\beta_\alpha\beta_\beta\mathcal{Z}^{\alpha\beta}_\mm+(\beta\cdot\nu)(U_\mm\cdot U_\mm)-\beta_\alpha B_{\beta}^{\;\alpha}U^{\beta}_\mm\right)\Big]\\
        &+\delta'(k^2-m^2)\left[\chi^{(1)}_\mm g_{\alpha\beta}\mathcal{Z}^{\alpha\beta}_\mm-\chi^{(2)}_{\mm}(U_\mm\cdot U_\mm)\right]\Bigg\}+\Delta^{(2)}_{\rm quad,\bot} W^{\mm\mm}(x,k)\;.
    \end{split}
\end{equation}
Using the definitions \eqref{EEE: def U}, \eqref{EEE eq: C e Z final} and \eqref{def: chim} we can prove the following identity:
\begin{equation*}
g_{\alpha\beta}\mathfrak{D}^\alpha\circ\mathfrak{D}^\beta\,n_{\rm B}(k,\mm)=-\left(\chi^{(1)}_\mm g_{\alpha\beta}\mathcal{Z}^{\alpha\beta}_\mm-\chi^{(2)}_{\mm}U^\alpha_\mm\cdot U_{\alpha\mm}\right)\;,
\end{equation*}
which is exactly minus the coefficient multiplying the off-shell contribution in the combined quadratic longitudinal and second order linear term in \eqref{EEE: SCALAR W quad1}. Then, introducing the \emph{on-shell} coefficient:
\begin{equation}\label{EEE: Konshel}
    \begin{split}
        \mathcal{K}^\mm&\equiv \frac{\chi^{(2)}_\mm}{2}\left(B^{\;\alpha}_\beta B_{\alpha}^{\;\beta}-2\nu^\alpha U_{\mm\gamma}B^{\;\gamma}_\alpha+(\nu\cdot U_\mm)^2\right)-\chi^{(1)}_\mm\mathcal{C}\\
    &+4L^{\mm\mm}_{[2]}\left(\frac{1}{2}\beta_\alpha\beta_\beta\mathcal{Z}^{\alpha\beta}_\mm+(\beta\cdot\nu)(U_\mm\cdot U_\mm)-\beta_\alpha B_{\beta}^{\;\alpha}U^{\beta}_\mm\right)\;,
    \end{split}
\end{equation}
the quadratic correction can be finally written as:
\begin{equation}\label{EEE: SCALAR W quad FINAL}
    \begin{split}
        \Delta^{(2)}W^{\mm\mm}(x,k)&=\frac{\delta(k^2-m^2)}{4(2\pi)^3}\mathcal{K}^\mm(x,k)-\frac{\delta'(k^2-m^2)}{4(2\pi)^3}g_{\alpha\beta}\mathfrak{D}^\alpha\circ\mathfrak{D}^\beta n_{\rm B}(k,\mm)+\Delta^{(2)}_{\rm quad,\bot} W^{\mm\mm}(x,k)\;.
    \end{split}
\end{equation}
Note that the off-shell term is proportional to a total derivative. Hence, once integrated over the decoupling hypersurface, it does not contribute, modulo a boundary term, to the total spectra of the pions.

In order to perform this integration we consider the particular case where  the isovector chemical potential is aligned everywhere on the hypersurface along the third direction. This allows to simplify the computations significantly. So consider $\ZZ_{1,2}=0$, $\ZZ_3=\ZZ$. This immediately implies:
\begin{equation*}
    \Delta^{(2)}_{\rm quad,\bot} W^{\mm\mm}(x,k)=0\;.
\end{equation*}
The spectrum \eqref{eq: CF in terms of W} involves the integral $\int_{\Sigma_{\rm D}}\di\Sigma(x)\cdot k$ of the Wigner function at fixed $k$. We can prove that, for any vector field $V^\mu(x)$ vanishing at the boundary of $\Sigma_{\rm D}$, and for any $k$, not necessarily on-shell:
\begin{equation}\label{eq: IBP}
    \int_{\Sigma_{\rm D}}\di\Sigma(x)\cdot k\;\mathfrak{D}_\mu V^\mu(x)=0\;,\qquad\mathfrak{D}_\mu V^\mu\equiv\mathcal{P}^{\beta}_{\;\,\mu}(x)\,\partial_\beta V^\mu(x)\;.
\end{equation}
The above identity allows to simplify considerably the resulting expression for the spectra \eqref{eq: CF pions}.

To prove it consider the antisymmetric tensor $A^{\beta\mu}\equiv k^\mu V^\beta-k^\beta V^\mu$, built with $V^\mu$ and $k^\nu$. Then the integral of its divergence, according to Stokes theorem is just a boundary term:
\begin{equation*}
    \di\Sigma_\mu\,\partial_\beta A^{\beta\mu}=\di\Sigma\cdot k\left(\partial_\beta V^\beta-\nu_\mu k^\beta\partial_\beta V^\mu\right)=\di\Sigma\cdot k\;\mathcal{P}^\beta_{\;\,\mu}\partial_\beta V^\mu\;,
\end{equation*}
where we used d$\Sigma_\mu=$d$\Sigma\cdot k\,\nu_\mu$. The integral of the left hand side over $\Sigma_{\rm D}$ is a boundary term, which vanishes, and this proves \eqref{eq: IBP}.

The off-shell term of \eqref{EEE: SCALAR W quad FINAL} has exactly the form \eqref{eq: IBP}, with $V^\alpha=-(1/4(2\pi)^3)\mathfrak{D}^\alpha n_{\rm B}(k,\mm)$. Its integral over the hypersurface thus vanishes for every $k$, and so does its derivative with respect to $k^0$, which appears after the integration of $\delta'(k^2-m^2)$ over $k^0$ in \eqref{eq: CF in terms of W}: only the on-shell part contributes to the spectra.

The on-shell coefficient can also be simplified once integrated over the decoupling hypersurface. According to the definitions \eqref{def: chim} and \eqref{eq: derivatives of L}:
\begin{equation}\label{EEE: derivate intermedie}
    \mathfrak{D}^\alpha\chi^{(1)}_\mm=-\chi^{(2)}_\mm U^\alpha_\mm\;,\qquad\mathfrak{D}^\alpha L^{\mm\mm}_{[2]}\equiv\dot{L}^{\mm\mm}_{[2]}U^\alpha_\mm=\frac{\di L^{\mm\mm}_{[2]}}{\di(k\cdot\beta-\mm\mathfrak{Z}_3)}U^\alpha_\mm\;.
\end{equation}
Now consider the term of \eqref{EEE: Konshel} containing the tensor \eqref{EEE def: matZ}:
\begin{equation}\label{EEE: IZ}
    \mathcal{I}_\mathcal{Z}\equiv2 L^{\mm\mm}_{[2]}\beta_\alpha\beta_\beta\mathcal{Z}^{\alpha\beta}_\mm=2 L^{\mm\mm}_{[2]}\beta_\alpha\beta_\beta\mathfrak{D}^\alpha U^\beta_\mm\;.
\end{equation}
Now let us define the following vector:
\begin{equation*}
    V_\mathcal{Z}^\alpha\equiv 2L^{\mm\mm}_{[2]}\beta^\alpha(\beta\cdot U_\mm)\;.
\end{equation*}
Taking the projected divergence of the above vector and using \eqref{EEE: derivate intermedie} we obtain:
\begin{equation*}
    \mathfrak{D}_\alpha V^\alpha_\mathcal{Z}=\mathcal{I}_\mathcal{Z}+2\dot{L}^{\mm\mm}_{[2]}(\beta\cdot U_\mm)^2+2L^{\mm\mm}_{[2]}\left(B_\alpha^{\;\alpha}(\beta\cdot U_\mm)+\beta_\alpha B_\beta^{\;\alpha}U_{\beta\mm}\right)\;.
\end{equation*}
Hence, being the left hand side a total divergence, according to the Stokes theorem the integral on the left hand side is a boundary term and we obtain:
\begin{equation*}
    \int_{\Sigma_{\rm D}}\di\Sigma(x)\cdot k\,\mathcal{I}_\mathcal{Z}=-\int_{\Sigma_{\rm D}}\di\Sigma(x)\cdot k\left[2\dot{L}^{\mm\mm}_{[2]}(\beta\cdot U_\mm)^2+2L^{\mm\mm}_{[2]}\left(B_\alpha^{\;\alpha}(\beta\cdot U_\mm)+\beta_\alpha B_\beta^{\;\alpha}U^\beta_{\;\mm}\right)\right]\;.
\end{equation*}
Note that, after integration, a term containing second derivatives and curvature $\mathfrak{D}U$ has been replaced by terms involving squares of first derivatives.

A similar work can be carried out for the scalar coefficient $\mathcal{C}$ in equation \eqref{EEE def: matC}. We can split it in the sum of four terms:
\begin{equation}\label{EEE: C1234}
    \mathcal{C}_1\equiv \nu^\mu\mathfrak{D}_\mu B_\alpha^{\;\alpha}\;,\qquad\mathcal{C}_2\equiv-\nu^\nu\mathfrak{D}_\alpha B_\nu^{\;\alpha}\;,\qquad\mathcal{C}_3\equiv K^\mu_{\;\mu} B_\alpha^{\;\alpha}\;,\qquad\mathcal{C}_4\equiv-K_{\mu\nu}B^{\mu\nu}\;,
\end{equation}
One can show that, integrating by parts $\chi^{(1)}_\mm\mathcal{C}_{1,2}$ their curvature terms cancel those of $\mathcal{C}_{3,4}$. Indeed, for arbitrary, regular enough $F$:
\begin{equation}\label{EEE: int by part}
    \int_{\Sigma_{\rm D}}\di\Sigma(x)\cdot k\, F(x,k)\mathfrak{D}_\alpha V^\alpha(x,k)=-\int_{\Sigma_{\rm D}}\di\Sigma(x)\cdot k\,V^\alpha(x,k)\mathfrak{D}_\alpha\,F(x,k)\;.
\end{equation}
Applying this to the two terms $\mathcal{C}_1$ and $\mathcal{C}_2$ in \eqref{EEE: C1234} and using \eqref{EEE: int by part} we get:
\begin{equation*}
    \begin{split}
        \int_{\Sigma_{\rm D}}\di\Sigma(x)\cdot k\,\left(-\chi^{(1)}_\mm\mathcal{C}_1\right)&=\int_{\Sigma_{\rm D}}\di\Sigma(x)\cdot k\left(-\chi^{(2)}_\mm(\nu\cdot U_\mm)B_\alpha^{\;\alpha}+\chi^{(1)}_\mm K^\mu_{\;\mu}B_{\alpha}^{\;\alpha}\right)\;,\\
        \int_{\Sigma_{\rm D}}\di\Sigma(x)\cdot k\,\left(-\chi^{(1)}_\mm\mathcal{C}_2\right)&=\int_{\Sigma_{\rm D}}\di\Sigma(x)\cdot k\left(\chi^{(2)}_\mm U_{\alpha\mm}\nu^\nu\,B_\nu^{\;\alpha}-\chi^{(1)}_\mm K_\alpha^{\;\nu}B_\nu^{\;\alpha}\right)\;.
    \end{split}
\end{equation*}
Now combining the above with \eqref{EEE: C1234} and \eqref{EEE def: matC} we find:
\begin{equation}\label{EEE: int C intermedio}
    \begin{split}
        \int_{\Sigma_{\rm D}}\di\Sigma(x)\cdot k\,\chi^{(1)}_\mm\left(\mathcal{C}_1+\mathcal{C}_3\right)&=\int_{\Sigma_{\rm D}}\di\Sigma(x)\cdot k\,\chi^{(2)}_\mm\,(\nu\cdot U_\mm)B_\alpha^{\;\alpha}\;,\\
        \int_{\Sigma_{\rm D}}\di\Sigma(x)\cdot k\,\chi^{(1)}_\mm\left(\mathcal{C}_2+\mathcal{C}_4\right)&=-\int_{\Sigma_{\rm D}}\di\Sigma(x)\cdot k\,\chi^{(2)}_\mm\,U_{\alpha\mm}\nu^\gamma B_\gamma^{\;\alpha}\;,
    \end{split}
\end{equation}
where in the second equality we used $K_{\gamma\lambda}=K_{\lambda\gamma}$. Overall, by means of \eqref{EEE: int C intermedio} the integral of the coefficient $\mathcal{C}$ in equation \eqref{EEE: Konshel} simplifies considerably:
\begin{equation}\label{EEE: int C}
    \int_{\Sigma_{\rm D}}\di\Sigma(x)\cdot k\,(-\chi^{(1)}_\mm\mathcal{C})=\int_{\Sigma_{\rm D}}\di\Sigma(x)\cdot k\,\chi^{(2)}_\mm\left(-(\nu\cdot U_\mm)B_\alpha^{\;\alpha}+\nu^\alpha U_{\gamma\mm}B_\alpha^{\;\gamma}\right)\;.
\end{equation}
Note that, again there are no second derivatives of the gradient, and moreover the explicit dependence on the curvature $K$ has disappeared. Now, collecting \eqref{EEE: IZ} and \eqref{EEE: int C}, the on-shell coefficient \eqref{EEE: Konshel}, after integration over the hypersurface can be written as:
\begin{equation*}
    \int_{\Sigma_{\rm D}}\di\Sigma(x)\cdot k\,\mathcal{K}^\mm(k)=\int_{\Sigma_{\rm D}}\di\Sigma(x)\cdot k\,\widetilde{\mathcal{K}}^\mm(k)\;,
\end{equation*}
where:
\begin{equation}\label{EEE Ktilde 1}
    \begin{split}
        \widetilde{\mathcal{K}}^\mm(k)&=-2\dot{L}^{\mm\mm}_{[2]}(\beta\cdot U_\mm)^2-2L^{\mm\mm}_{[2]}\left(B_\alpha^{\;\alpha}(\beta\cdot U_\mm)+3\beta_\alpha B_\beta^{\;\alpha}U^\beta_\mm-2(U_\mm\cdot U_\mm)\beta\cdot\nu\right)\\
        &+\frac{\chi^{(2)}_\mm}{2}\left(B_\beta^{\;\alpha}B_\alpha^{\;\beta}+(\nu\cdot U_\mm)^2\right)-\chi^{(2)}_\mm B_\alpha^{\;\alpha}(\nu\cdot U_\mm)\;.
    \end{split}
\end{equation}
The above can be expressed, equivalently, in terms of the Bose-Einstein factor and its derivatives using the definitions \eqref{def: chim} and \eqref{eq: derivatives of L}:
\begin{equation}\label{EEE eq: relazione tra chi L e nb}
    \chi^{(1)}_\mm=-n_{\rm B}'(\chi_\mm)\;,\quad\chi^{(2)}_\mm=n_{\rm B}''(\chi_\mm)\;,\quad L^{\mm\mm}_{[2]}=-\frac{1}{12}n'''_{\rm B}(\chi_\mm)\;,\quad \dot{L}^{\mm\mm}_{[2]}=-\frac{1}{12}n_{\rm B}''''(\chi_\mm)\;,
\end{equation}
where the ``prime" denote derivatives with respect to the argument $\chi_\mm=k\cdot\beta-\mm\mathfrak{Z}_3$. Plugging \eqref{EEE eq: relazione tra chi L e nb} in \eqref{EEE Ktilde 1} we obtain:
\begin{equation}\label{EEE: Ktilde 2}
    \begin{split}
        \widetilde{\mathcal{K}}^\mm(k)&=\frac{1}{6}n''''_{\rm B}(\beta\cdot U_\mm)^2+\frac{1}{6}n'''_{\rm B}\left(B_\alpha^{\;\alpha}(\beta\cdot U_\mm)+3\beta_\alpha B_\beta^{\;\alpha}U^\beta_\mm-2(U_\mm\cdot U_\mm)\beta\cdot\nu\right)\\
        &+\frac{n''_{\rm B}}{2}\left[B_\beta^{\;\alpha}B_\alpha^{\;\beta}+(\nu\cdot U_\mm)^2-2B_\alpha^{\;\alpha}(\nu\cdot U_\mm)\right]\;,
    \end{split}
\end{equation}
where the argument of the Bose-Einstein distribution and the derivatives are understood. 

We conclude this appendix by studying the global equilibrium limit of \eqref{EEE: SCALAR W quad FINAL}. In global equilibrium the vorticity and the chemical potentials are constant while the shear tensor exactly vanishes:
\begin{equation}\label{EEE: GE limite}
    \varpi_{\rho\lambda}={\rm const.}\;,\qquad \xi_{\rho\lambda}=0\;,\qquad\mathfrak{Z}_i={\rm const.}
\end{equation}
In this limit \eqref{EEE: def U} and \eqref{EEE: def BZ} simplify:
\begin{equation}\label{EEE: B e U GE}
    B_\lambda^{\;\alpha}=-\mathcal{P}^{\rho\alpha}\varpi_{\rho\lambda}\;,\qquad U^\alpha_\mm=-\varpi^{\alpha}_{\;\lambda}k^\lambda\;.
\end{equation}
Moreover, being $\varpi$ constant and antisymmetric the tensor $\mathcal{Z}^{\alpha\beta}_\mm$ defined in equation \eqref{EEE def: matZ} vanishes altogether. The same conclusion holds for the coefficient $\mathcal{C}$ defined in \eqref{EEE def: matC}. To see this consider:
\begin{equation*}
    \nu_\mu\mathfrak{D}^\mu(B^{\,\nu}_\nu)=-\nu_\mu K^{\mu\nu}\varpi_{\nu\sigma}k^\sigma\;,\qquad \nu^\nu\mathfrak{D}_\alpha(B^{\,\alpha}_\nu)=- K^\alpha_{\;\,\alpha}\nu^\gamma\varpi_{\gamma\sigma}k^\sigma\;,\qquad K^{\mu\nu}B_{\mu\nu}=- K^{\mu\nu}\nu_\mu \varpi_{\nu\sigma}k^\sigma\;,
\end{equation*}
which implies that the four terms in \eqref{EEE def: matC} cancel in pairs.
 Finally, for the remaining term in \eqref{EEE: SCALAR W quad FINAL}, using \eqref{EEE: B e U GE}, we obtain:
\begin{equation*}
    \beta_\alpha B_\rho^{\;\alpha}U^\rho_\mm-(\beta\cdot\nu)U^\alpha_\mm U_{\alpha\mm}=\beta^\sigma\varpi_{\sigma\rho}\varpi^{\rho}_{\;\nu}k^\nu\;,\qquad\frac{1}{2}\left[B^{\,\alpha}_\beta B^{\,\beta}_\alpha+(\nu\cdot U_\mm)^2\right]-U_{\mm\gamma}B^{\,\gamma}_\alpha\nu^\alpha=\frac12\varpi^\alpha_{\;\;\beta}\varpi^\beta_{\;\;\alpha}\;.
\end{equation*}
Therefore, in the global equilibrium limit, the quadratic correction to the diagonal Wigner function \eqref{EEE: SCALAR W quad FINAL} reads:
\begin{equation}\label{EEE eq: pion W2 GE}
    \begin{split}
        \Delta^{(2)}W^{\mm\mm}_{\rm GE}(x,k)=\frac{1}{4(2\pi)^3}\bigg\{\delta(k^2-m^2)\left[-4L^{\mm\mm}_{[2]}\beta_\sigma\varpi^{\sigma}_{\;\rho}\varpi^\rho_{\;\lambda}k^\lambda+\frac{\chi^{(2)}_\mm}{2}\varpi^\sigma_{\;\rho}\varpi^{\rho}_{\;\sigma}\right]+\\
        -\delta'(k^2-m^2)\chi^{(2)}_\mm\varpi_{\sigma\lambda}k^\lambda\varpi^{\sigma}_{\;\gamma}k^\gamma\bigg\}\;.
    \end{split}
\end{equation}
All the dependence on $\nu$, namely on the geometry of the hypersurface, has disappeared, as expected. This cancellation involves both the linear and the quadratic response.

The above expression solves the generalized on-shell equation for the Wigner function:
\begin{equation}\label{EEE: free eq WGE}
   \left(k^2-m^2-\frac{1}{4}\Box\right)\left[W^{\mm\mm}_{\rm GE}(x,k)+\Delta^{(2)}_\varpi W^{\mm\mm}_{\rm GE}\right]=\mathcal{O}(\varpi^3)\;,
\end{equation}
up to second order in the vorticity as expected. To see this it is sufficient to consider that, at global equilibrium:
\begin{equation*}
    \partial_\rho\left(\beta\cdot k-\mm\,\mathfrak{Z}_3\right)=-\varpi_{\rho\sigma}k^\sigma\;,
\end{equation*}
so that:
\begin{equation*}
    \Box_xn^\mm_{\rm B}(\beta(x)\cdot k)=\chi^{(2)}_\mm(k,x)\varpi_{\rho\lambda}k^\lambda\varpi^{\rho}_{\;\sigma}k^\sigma\;.
\end{equation*}
Now using that $(k^2-m^2)\delta'(k^2-m^2)=-\delta(k^2-m^2)$\;, when the d'Alembert operator $\Box_x$ acts on the leading order term $W^{\mm\mm}_{\rm GE}$ the contribution is exactly canceled by the term proportional to $\delta'$ in \eqref{EEE eq: pion W2 GE} implying \eqref{EEE: free eq WGE}.

The global equilibrium expression \eqref{EEE eq: pion W2 GE} can be written in a more interesting way. At global equilibrium the thermal vorticity, being a constant antisymmetric tensor, can be expanded using the four-velocity $u^\mu=T\beta^\mu$  as follows:
\begin{equation}
    \varpi_{\lambda\rho}=\alpha_\lambda u_\rho-\alpha_\rho u_\lambda+\epsilon_{\lambda\rho\sigma\gamma}\omega^\sigma u^\gamma\;;
\end{equation}
where $\alpha^\mu=a^\mu/T$ and $\omega^\mu=w^\mu/T$ with $a^\mu$ and $w^\mu$ four-acceleration and four-angular velocity respectively and $T$ proper temperature. Then we have:
\begin{equation*}
    \beta_\sigma\varpi^{\sigma}_{\;\rho}\varpi^{\rho}_{\;\nu}k^\nu=-\frac{a_\rho}{T^2}\varpi^{\rho\nu} k_\nu\;.
\end{equation*}
Consequently, using that $\varpi^\sigma_{\;\rho}\varpi^{\rho}_{\;\sigma}=-\varpi_{\sigma\rho}\varpi^{\sigma\rho}\equiv\varpi:\varpi$, and expressing the coefficients $\chi_\mm$ and $L^{\mm\mm}$ in terms of the Bose-Einstein distribution, we get:
\begin{equation}\label{EEE eq: pion W2 GE2 final}
    \begin{split}
       \Delta^{(2)}W^{\mm\mm}_{\rm GE}(x,k)&=\frac{\delta(k^2-m^2)}{4(2\pi)^3}n^\mm_{\rm B}(k)(1+n^\mm_{\rm B}(k))\\
        &\times\left(\frac{1+2n^\mm_{\rm B}(k)}{2}\varpi:\varpi+\frac{1+6n^\mm_{\rm B}(k)+6(n^{\mm}_{\rm B}(k))^2}{3T^2}a_\mu k_\nu\varpi^{\mu\nu}\right)\\
        &-\frac{\delta'(k^2-m^2)}{4(2\pi)^3}n^\mm_{\rm B}(k)(1+n^\mm_{\rm B}(k))\left(1+2n_{\rm B}^\mm(k)\right)\varpi_{\sigma\lambda}k^\lambda\varpi^{\sigma}_{\;\gamma}k^\gamma\;.
    \end{split}
\end{equation}
%

\bibliographystyle{JHEP}
\bibliography{main}

\end{document}